\documentclass[%
 aip,
 amsmath,amssymb,
 reprint,%
]{revtex4-1}

\usepackage{graphicx}
\usepackage{dcolumn}
\usepackage{bm}

\usepackage[utf8]{inputenc}
\usepackage[T1]{fontenc}
\usepackage[dvipsnames]{xcolor}
\usepackage{mathptmx}
\usepackage{etoolbox}
\usepackage{kotex}
\usepackage{ulem}
\usepackage{hyperref} 

\usepackage{bm}
\usepackage{color}
\usepackage{caption}
\usepackage{subcaption}
\usepackage[export]{adjustbox}
\usepackage{orcidlink}
\definecolor{lgrey}{HTML}{d3c9cc}
\definecolor{ngrey}{HTML}{808080}
\definecolor{dgrey}{HTML}{595959}

\graphicspath{{Figures/}}

\def\sizeV{N}
\def\sizeW{M}
\def\I{\mathbf{I}}
\def\V{\mathcal{V}}
\def\W{\mathcal{W}}
\def\E{\mathcal{E}}
\def\H{\mathcal{H}}

\newcommand{\blue}{\textcolor{blue}}
\newcommand{\bI}{{\bf I}}

\makeatletter
\def\@email#1#2{%
 \endgroup
 \patchcmd{\titleblock@produce}
  {\frontmatter@RRAPformat}
  {\frontmatter@RRAPformat{\produce@RRAP{*#1\href{mailto:#2}{#2}}}\frontmatter@RRAPformat}
  {}{}
}%
\makeatother
\begin{document}

\preprint{AIP/123-QED}

\title[Clustering coefficients for  networks with higher order interactions]{Clustering coefficients for  networks with higher order interactions}
\author{Gyeong-Gyun Ha(하경균)\,\orcidlink{0009-0009-9298-1806}}
 \email{gyeong-gyun.ha@kcl.ac.uk}
\author{Izaak Neri\,\orcidlink{0000-0001-9529-5742}}%
\author{Alessia Annibale\,\orcidlink{0000-0003-4010-6742}}
\affiliation{Department of Mathematics, King's College London, Strand, London, WC2R 2LS, UK}%

\date{\today}

\begin{abstract}
We introduce a clustering coefficient for nondirected and directed hypergraphs, which we call the {\it quad clustering coefficient}.   We determine the average quad clustering coefficient and its distribution in real-world hypergraphs and compare its value with those of random hypergraphs drawn from the configuration model.  
We find that  real-world hypergraphs exhibit a nonnegligible fraction of nodes with a maximal value of the quad clustering coefficient, while we do  not find such nodes in random hypergraphs.   Interestingly,  these highly clustered nodes can have large degrees and can be incident to hyperedges of large cardinality.   Moreover,  highly clustered nodes are not observed in an analysis based on the pairwise clustering coefficient  of the associated projected graph that has binary interactions, and hence higher order interactions are required to identify nodes with a large quad  clustering coefficient. 
\end{abstract}

\maketitle

\begin{quotation}
 Real-world networks   exhibit, so-called, higher  order interactions, which are  relations  that involve more than two parties.   Such higher order interactions can be represented by hyperedges, and a collection of nodes and hyperedges is called a hypergraph.  The question arises what are the topological properties of real-world systems that have higher order interactions, such as,   social collaboration networks or product composition networks.    This problem is challenging as real-world networks can consist of a  large number of nodes and hyperedges.   Moreover, hyperedges in real-world networks can connect up to hundreds of nodes.   To address  the topological properties of hypergraphs, we introduce in this Paper a clustering coefficient that determines the density of quads incident to a node, and which we call the quad clustering coefficient.   Comparing the  quad clustering coefficients of nodes in real-world networks with those in random networks, we find that real-world systems have topological properties that are significantly different from those of random systems.   Notably, real-world hypergraphs have a large fraction of nodes with a maximal value of the quad clustering coefficient.  This  feature is only observed when accounting for the higher order interactions and is not seen in a classical network analysis based on binary interactions.  We believe that these results are interesting for developing  more accurate null models for real-world networks with higher order interactions.

\end{quotation}

\section{Introduction} 
Networks consist of  nodes, representing components of a system, and relations between those nodes.     When the relations are binary, they can be represented as links in a graph~\cite{newman2006structure, barabasi,dorogovtsev2022nature}. However in real-world systems  relations  often include  three or more vertices, and these are called higher order interactions~\cite{battiston2020networks}.     For example, a protein-protein interaction network can be seen as a network of binary relations, where two proteins are connected when they bind to each other, or it can be seen as a network with higher order interactions where a protein complex of $\chi$ proteins corresponds to a higher order interaction of cardinality $\chi$. 

Although in a first approximation real-world networks appear to be random,  random networks have a smaller number of cliques than what is observed in real-world networks~\cite{newman2006structure, barabasi,dorogovtsev2022nature}.  Indeed, the average clustering coefficient of a random graph, measuring the density of triangles~\cite{watts1998collective} (the smallest possible clique),  decreases linearly as a function of  the number of nodes in the graph.  On the other hand,  the average clustering coefficient of real-world networks is larger and  approximately independent of $N$~\cite{albert2002statistical}.  Because of this observation,  more realistic models for real-world networks have been developed that are based on a hierarchical network~\cite{ravasz2003hierarchical} or a small-world network structure~\cite{barrat2000properties, watts1998collective}.

For systems with higher order interactions, Refs.~\cite{opsahl2013triadic,brunson2015triadic,kartun2019beyond,serrano2020centrality}  define  a clustering coefficient that measures the degree of local  transitivity, and corresponds with quantifying clustering of nodes in the projected graph associated with a higher order network.    However, contrarily to the case of simple graphs, the clustering coefficients of Refs.~\cite{opsahl2013triadic,brunson2015triadic,yin2018higher,kartun2019beyond,serrano2020centrality} do not capture the density of the shortest cycles in  hypergraphs.    

In this Paper,   we propose  an alternative observable for  clustering in  hypergraphs that quantifies  the density of the shortest possible simple cycle.  The shortest simple cycle of a hypergraph is a quad. In a bipartite representation of a hypergraph, where nodes and hyperedges represent the two parties of the bipartite graph, a quad 
is a closed path of length four consisting of an alternating sequence of two nodes and two hyperedges.   The quad clustering coefficient that we introduce in this paper quantifies the density of  quads and it  is reminiscent of clustering coefficients that quantify densities of squares in  bipartite graphs, see Refs.~\cite{lind2005cycles,zhang2008clustering,kitsak2011hidden}, but there are also some notable distinctions.   For example, as we show here,   the quad clustering coefficient is more effective in quantifying the density of quads in a hypergraph than coefficients defined previously in the literature.     After a comparison with these previous works, we study clustering of quads in random graphs and real-world networks.

The paper is structured as follows. In Sec. \ref{ch:Preliminaries}, we define  hypergraphs and introduce the notation  used in this paper. In Sec.~\ref{sec:defNot}, we define the quad clustering coefficient and compare this coefficient with similar coefficients studied in the context of bipartite graphs. In Sec.~\ref{ch:avg_quad_random}, we derive exact expressions of the ensemble average of the quad clustering coefficient in a random hypergraph model. In Sec.~\ref{ch:real_hypergraph}, we compare the results of Sec.~\ref{ch:avg_quad_random} with real-world hypergraphs and discuss notable distinctions between real-world networks and random graphs. In Sec.~\ref{ch:directed_hypergraph}, we extend the quad clustering coefficient to directed hypergraphs, and make a corresponding study for real-world networks.   Conclusions are given in Sec.~\ref{ch:discussion}, and the Papers ends with several Appendices containing technical details on the calculations in this Paper.

\section{Preliminaries on hypergraphs} \label{ch:Preliminaries}

A nondirected, hypergraph is a triplet $\H=(\V, \W, \E)$ consisting of a set $\V$ of $\sizeV=|\V|$ nodes, a set of $\W$ of $\sizeW=|\W|$ hyperedges, and a set $\E$ of links.
We denote nodes by roman indices, $i,j\in \V$, and hyperedges by Greek indices $\alpha,\beta\in \W$. 
The set of links $\E$ consists of pairs $(i,\alpha)$  with $i\in \V$ and $\alpha\in \W$.   We say that the hypergraph is {\it simple} when each pair $(i,\alpha)$  occurs at most once in the set $\E$.   

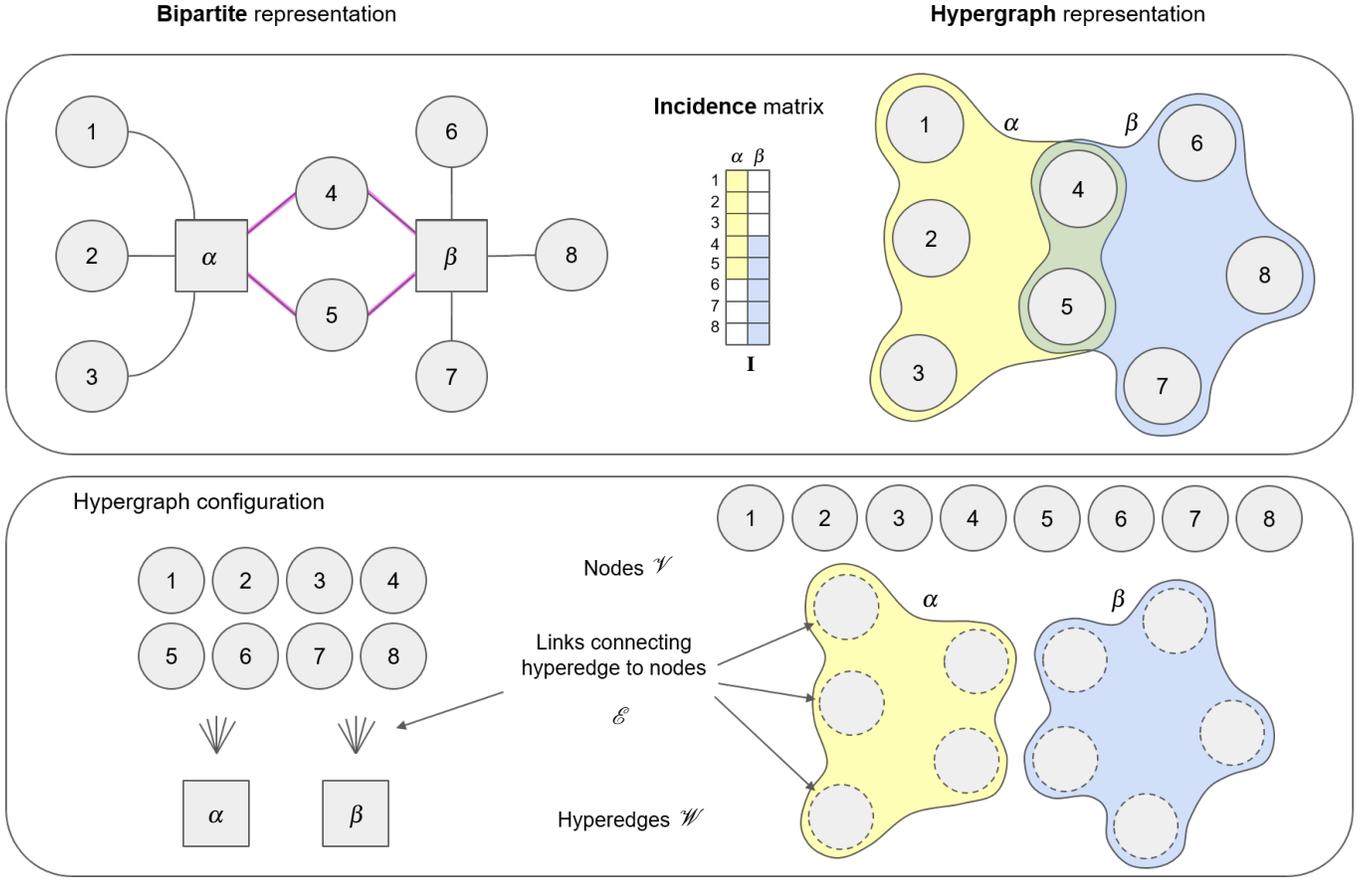
\begin{figure*}
 \centering
 \setlength{\unitlength}{0.1\textwidth}
 \includegraphics[width=\textwidth]{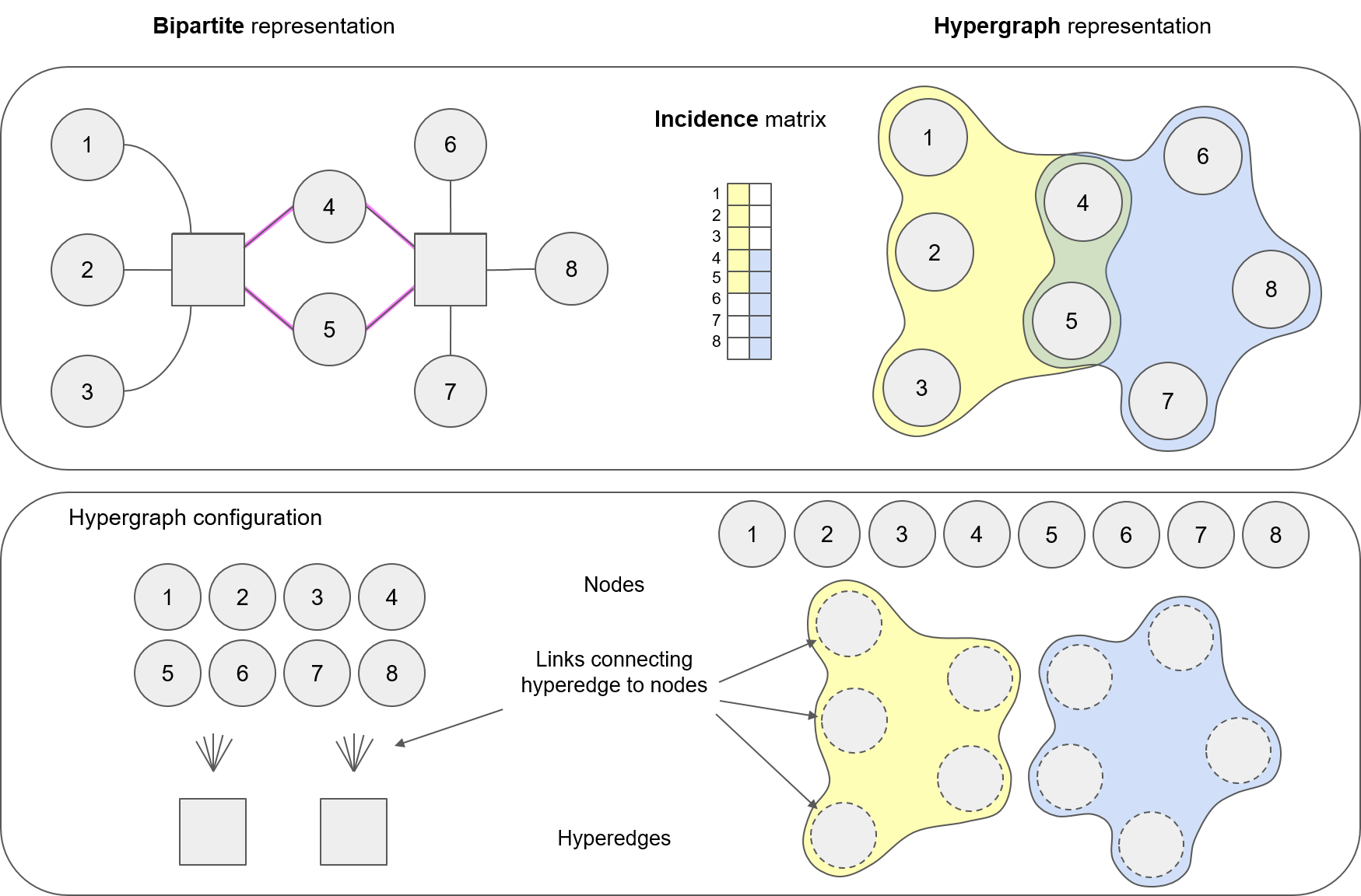}
 \put(-2.6,5.54){\small$\alpha$}
 \put(-1.7,5.54){\small$\beta$}
 \put(-8.55,4.54){\small$\alpha$}
 \put(-6.75,4.54){\small$\beta$}
 \put(-4.5,3.75){\footnotesize$\I$}
 \put(-4.62,5.3){\scriptsize$\alpha$}
 \put(-4.45,5.3){\scriptsize$\beta$}
 \put(-5.2,2.25){\small$\V$}
 \put(-3.2,2.){\small$\alpha$}
 \put(-1.8,2.){\small$\beta$}
 \put(-5.0,0.38){\small$\W$}
 \put(-5.5,1.13){\small$\E$}
 \put(-8.5,0.4){\small$\alpha$}
 \put(-7.45,0.4){\small$\beta$}
 \caption{{\it Illustration of a hypergraph, its different representations, and the quad motif.} The upper panel shows the three ways of representing a hypergraph, namely, as a bipartite graph, as an incidence matrix, and as a graph with higher order interactions. The illustrated hypergraph in the left top panel has one quad, highlighted in magenta, consisting of the hyperedges $\alpha$ and $\beta$ and the nodes $4$ and $5$.  The lower panel visualises the three different components of a hypergraph, namely, the set of nodes $\V$, the set of links $\E$, and the set of hyperedges $\W$.}
 \label{fig:description of the notation}
\end{figure*}

A simple, nondirected hypergraph can be represented by an {\it incidence matrix}  of dimensions $N\times M$ that is defined by 
\begin{equation}
[\I]_{i\alpha} \equiv \left\{\begin{array}{ccc} 1 &{\rm if}& (i,\alpha)\in \E ,\\ 0 &{\rm if} & (i,\alpha)\notin \E.\end{array}\right.
\end{equation}
Consequently, a hypergraph can also be represented as a bipartite graph whose vertices are the nodes and the hyperedges of the hypergraph.   
Figure~\ref{fig:description of the notation} shows an example of a hypergraph represented as a bipartite graph and an incidence matrix.  For simplicity,  we often make no distinction between the hypergraph $\mathcal{H}$ and its representation $\mathbf{I}$.

We define the network observables that we use   in this Paper.  
The {\it degree} of node $i\in\V$ is defined by 
\begin{equation}
k_i(\mathbf{I})  \equiv \sum^\sizeW_{\alpha=1}I_{i\alpha} , 
\end{equation}
and we use the vector notation 
\begin{equation}
\vec{k}(\mathbf{I}) \equiv (k_1(\mathbf{I}), k_2(\mathbf{I}), \ldots, k_N(\mathbf{I}))
\end{equation}
to denote the sequence of degrees of the hypergraph $\mathbf{I}$.
Analogously, we define the {\it cardinality}  of a hyperedge $\alpha$ by 
\begin{equation}
\chi_\alpha(\mathbf{I})  \equiv \sum^\sizeV_{i=1} I_{i \alpha } ,
\end{equation}      
and the sequence of cardinalities is 
\begin{equation}
\vec{\chi}(\mathbf{I}) \equiv (\chi_1(\mathbf{I}), \chi_2(\mathbf{I}), \ldots, \chi_M(\mathbf{I})).
\end{equation}      
As a hypergraph is a graph with higher order interactions, we also consider the degrees  
\begin{equation}
k_i(\mathbf{I};\chi) = \sum^\sizeW_{\alpha=1}I_{i\alpha}\delta_{\chi_{\alpha}(\mathbf{I}),\chi}\label{def:dec_k}
\end{equation}
that determine the number of hyperedges of  cardinality $\chi$ that are incident to node $i$.   In (\ref{def:dec_k}) $\delta_{n,m}$, with $n,m\in \mathbb{N}$, represents the Kronecker-delta function.     
We denote the number of  hyperedges  incident to node $i$, excluding those with cardinality $1$, by the so-called {\it modified degree}
\begin{equation}
k^\ast_i(\mathbf{I}) \equiv \sum^{\infty}_{\chi=2}k_i(\mathbf{I};\chi). \label{def:kAst}
\end{equation}
Lastly, we define the neighbourhood set 
\begin{equation}
\partial_{i\alpha}(\mathbf{I})  \equiv \{j\in \V| I_{i\alpha}I_{j\alpha}\neq0 \} 
\end{equation} 
consisting of nodes that are incident to the hyperedge $\alpha$ that is  connected to the node $i$.  

When $\chi_{\alpha}(\mathbf{I})=2$ for all $\alpha\in \W$, then $\mathbf{I}$ represents a  graph.   In this case, we can also represent the graph in terms of the adjacency matrix $\mathbf{A}$ with off-diagonal entries
\begin{equation}
A_{ij} = \sum^\sizeW_{\alpha=1}I_{i\alpha}I_{j\alpha} 
\end{equation}
and zero-valued diagonal entries, $A_{ii} = 0$.   
We say that  the graph is simple when $A_{ij}\in\left\{0,1\right\}$.    

Given a  hypergraph, we can define the so-called {\it projected  graph} by the adjacency matrix $\mathbf{A}^{\rm proj}$ with entries  
\begin{equation}
A^{\rm proj}_{ij} = \Theta\left(\sum^\sizeW_{\alpha=1}I_{i\alpha}I_{j\alpha}\right) \label{eq:AProj}
\end{equation}
where $\Theta(x)$ is the Heaviside function, i.e., $\Theta(x)=1$ when $x>0$ and $\Theta(x)=0$ when $x\leq 0$.   Note that this map is surjective, as a projected graph can correspond with multiple hypergraphs.

\section{Quad clustering coefficient: definition and motivation}   \label{sec:defNot}

For  simple graphs with pairwise interactions   determined by the adjacency matrix $\mathbf{A}$, the clustering coefficient  of a node with degree $k_i(\mathbf{A})\geq 2$ is given by~\cite{watts1998collective}
\begin{equation}
C^{\rm pi}_i(\mathbf{A}) \equiv \frac{T_i(\mathbf{A})}{t_{\rm max}(k_i(\mathbf{A}))} , \label{eq:cGraph}
\end{equation} 
where $T_i(\mathbf{A})$ is the number of triangles incident to node $i$, and 
\begin{equation}
t_{\rm max}(k_i(\mathbf{A})) = \frac{k_i(\mathbf{A})(k_i(\mathbf{A})-1)}{2}
\end{equation} 
is the maximum possible number of triangles incident to a node with degree $k_i(\mathbf{A})$.     Hence, the clustering coefficient $C^{\rm pi}_i$ determines the density of triangles incident to node $i$.  If $C^{\rm pi}_i=1$, then all possible triangles  incident to node $i$ are present, and if  $C^{\rm pi}_i=0$ then none of the triangles are present.      If $k_i\leq 1$, then by convention we set $C^{\rm pi}_i=0$.   

Since a triangle is the shortest cycle in a simple graph, the clustering coefficient $C^{\rm pi}_i$ is the density of shortest cycles incident to a node $i$, and we use this property of the clustering coefficient  for graphs with pairwise interactions to derive  a  clustering coefficient valid for hypergraphs.    To this aim, we represent a hypergraph as a bipartite graph, see Fig.~\ref{fig:description of the notation}.   In this bipartite representation, there exist no triangles, and instead  the  cycle of shortest length is a {\it quad}  consisting of two nodes and two hyperedges, see the motif illustrated in magenta in Fig.~\ref{fig:description of the notation}  for an illustration of the quad.    Specifically, the quad is a simple cycle of four links forming an alternating sequence of nodes  and hyperedges.

\subsection{Definition of the quad clustering coefficient}\label{sec:quadclust}
   
\begin{figure*}[t]
     \centering
     \setlength{\unitlength}{0.1\textwidth}
     \includegraphics[width=0.8\textwidth]{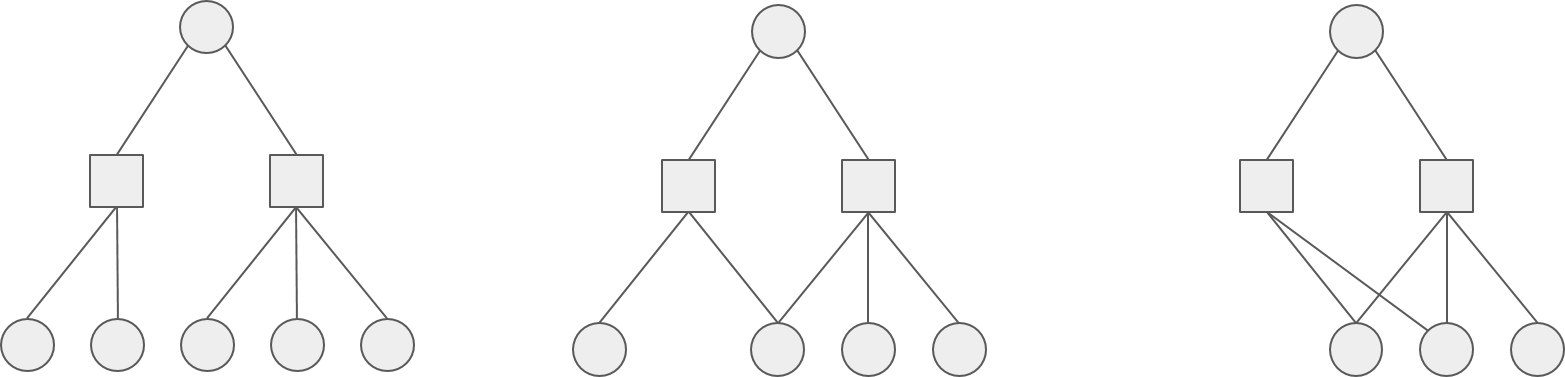}
     \put(-8.3,2){\normalsize$(a)$}
     \put(-6.97,1.73){\small$i$}
     \put(-7.47,0.95){\small$\alpha$}
     \put(-6.55,0.95){\small$\beta$}
     \put(-5.4,2){\normalsize$(b)$}
     \put(-4.05,1.7){\small$i$}
     \put(-4.55,0.93){\small$\alpha$}
     \put(-3.62,0.93){\small$\beta$}
     \put(-2,2){\normalsize$(c)$}
     \put(-1.1,1.7){\small$i$}
     \put(-1.6,0.93){\small$\alpha$}
     \put(-0.67,0.93){\small$\beta$}
     \caption{{\it  The quad clustering coefficient of a node in a few simple examples.} Node $i$ is connected to hyperedges $\alpha$ and $\beta$ with cardinalities $3$ and $4$, respectively. Depending on the number of quads,  $C^{\rm q}_{i}$ equals  $0$ (a),  $1/2$ (b), and  $1$ (c), respectively.}
     \label{fig:bipar_clustering}
\end{figure*}
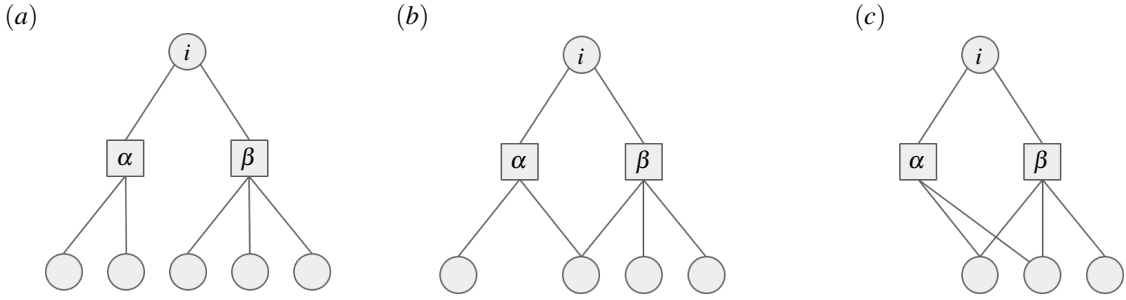

In this Section, we define the quad clustering coefficient  $C^{\rm q}_i(\mathbf{I})$ of a node $i$ in  a hypergraph.  
Let $i$ be a node that is connected to two or more hyperedges of cardinality two or higher, i.e., $k^\ast_i(\mathbf{I})\geq 2$.    We define the quad clustering coefficient of $i$ by   
\begin{equation}
C^{\rm q}_i(\mathbf{I}) \equiv \frac{Q_i(\mathbf{I})}{q_{\rm max}(\left\{k_i(\mathbf{I};\chi)\right\}_{\chi\in \mathbb{N}})}, \label{def:CQuad}
\end{equation}

where 
\begin{equation}
Q_i(\mathbf{I}) \equiv \sum^M_{\alpha<\beta}q_{i\alpha\beta}(\textbf{I})
\end{equation}
is the number of quads incident to node $i$, 
with  $\sum_{\alpha<\beta}=\sum^M_{\alpha=1}\sum^M_{\beta=\alpha+1}$ and 
\begin{equation}
q_{i\alpha\beta}(\textbf{I})\equiv\sum^N_{j=1;j\neq i}I_{j\alpha}I_{j\beta}I_{i\alpha}I_{i\beta}\blue{,} \label{eq:qialphabeta}
\end{equation} 
and where
\begin{equation}
q_{\rm max}(\left\{k_i(\mathbf{I};\chi)\right\}_{\chi\in \mathbb{N}})\equiv \sum_{\alpha<\beta}{\rm min}\left\{\chi_{\alpha}(\textbf{I})-1,\chi_{\beta}(\textbf{I})-1\right\}I_{i\alpha}I_{i\beta} \label{eq:qimax1}
\end{equation} 
is the maximal possible number of quads that a node with degrees $\left\{k_i(\mathbf{I};\chi)\right\}_{\chi\in \mathbb{N}}$ can have.   In Appendix~\ref{app:C} we show that the maximal number of quads can also be expressed by
\begin{equation}
q_{\rm max} =  \frac{1}{2} \sum^{\infty}_{\chi=2} (\chi-1) k_{i}(\mathbf{I};\chi) \left( \sum^{\infty}_{\chi'=\chi}k_i(\mathbf{I};\chi')-1\right), \label{eq:qimax2}
\end{equation}
which makes it evident that $q_{\rm max}$  is fully determined by the set $\left\{k_i(\mathbf{I};\chi)\right\}_{\chi\in \mathbb{N}}$ of degrees associated with node $i$.   If $k^\ast_i(\mathbf{I})<2$, then $C^{\rm q}_i(\mathbf{I}) = 0$, as the number of quads incident to a node with a degree less than two equals zero.      
Note that the formula for the maximal possible number of quads, $q_{\rm max}$, assumes that both the degree of node $i$  and the cardinalities of the hyperedges connected to $i$ are given.    
Also, note  that the quad clustering coefficient is a density, i.e., $C^{\rm q}_i(\mathbf{I})\in [0,1]$, and in the example of Fig.~\ref{fig:bipar_clustering}, $C^{\rm q}_i(\mathbf{I})=0$, $C^{\rm q}_i(\mathbf{I})=1/2$ and $C^{\rm q}_i(\mathbf{I})=1$.

The quad clustering coefficient $C^{\rm q}_i$  has two useful properties.  First, for fixed degrees $k_i(\mathbf{I};\chi)$, the quad clustering coefficient is a {\it linear function} of $Q_i$.   Second, the proportionality factor is such that $C^{\rm q}_i\in[0,1]$, and  $C^{\rm q}_i=1$  is attained when the number of quads around the node $i$ is maximal.   As will become evident, these properties do not hold for clustering coefficients of bipartite graphs considered previously in the literature.

Note that quads quantify the multitude of ways  neighbouring  nodes interact with each other, and in simple graphs we need higher order interactions to have multiple interaction paths.     
In the case of simple graphs (i.e., all hyperedges have cardinality $2$ and for each pair of nodes there is at most one hyperedge connecting them) the quad clustering coefficient is zero, as the only way to create multiple interactions between two nodes is through multiple edges, which are absent when the graph is simple.

In the next two Subsections,  we compare the quad clustering coefficient with two other clustering coefficients for bipartite graphs, namely, Lind's clustering coefficient~\cite{lind2005cycles} in Sec.~\ref{sec:Lind} and Zhang's clustering coefficient~\cite{zhang2008clustering} in Sec.~\ref{sec:Zhang}.     As we will see, Lind's and Zhang's clustering coefficients are not functions of $Q_i$, except when $k_i=2$, and in the latter case Lind's and Zhang's clustering coefficients are nonlinear functions in $Q_i$.    In addition to   Lind's and Zhang's clustering coefficients, other clustering coefficients have been defined in the litureature, see Refs.~\cite{jeong2022effects,jia2021measuring,aksoy2017measuring,malizia2023hyperedge,lee2021hyperedges}, but since these are significantly different from the quad clustering coefficient we do not discuss them here. Specifically, the clustering coefficients in Refs.~\cite{jeong2022effects,jia2021measuring} apply to nodes in standard networks without higher order interactions, the clustering coefficient in Ref.~\cite{aksoy2017measuring} has a denominator that does not depend on  the cardinalities of the  hyperedges incident to the considered node, and the coefficients in Ref.~\cite{malizia2023hyperedge,lee2021hyperedges} do not   count the number of quads.

\subsection{Lind's clustering coefficient}\label{sec:Lind} 
In Ref.~\cite{lind2005cycles}, Lind, Gonz\'{a}lez, and Herrmann  define a clustering coefficient by 
\begin{eqnarray} 
 C^{{\rm Lind}}_i(\textbf{I})\equiv 
\frac{Q_i(\mathbf{I})}
{q^{\rm Lind}_{i, \rm max}(\mathbf{I})},\label{def:CLind}
\end{eqnarray} 
where 
\begin{multline}
q^{\rm Lind}_{i, \rm max}(\mathbf{I}) \equiv \sum_{\alpha<\beta}\bigl[(\chi_{\alpha}(\textbf{I})-\eta_{i\alpha\beta}(\textbf{I}))(\chi_{\beta}(\textbf{I})-\eta_{i\alpha\beta}(\textbf{I}))\\
+q_{i\alpha\beta}(\textbf{I})\bigr]I_{i\alpha}I_{i\beta}
\end{multline}
with
\begin{equation}
\eta_{i\alpha\beta}(\textbf{I})\equiv1+q_{i\alpha\beta}(\textbf{I}).
\end{equation}
 For simplicity we call  $C^{{\rm Lind}}_i(\textbf{I})$ Lind's  clustering coefficient.     In the example of Fig.~\ref{fig:bipar_clustering}, $C^{\rm Lind}_i(\mathbf{I})=0$ for $(a)$, $C^{\rm Lind}_i(\mathbf{I})=1/3$ for $(b)$ and $C^{\rm Lind}_i(\mathbf{I})=1$ for $(c)$.

The difference between the formulas for  $C^{\rm Lind}_i(\mathbf{I})$ and $C^{\rm q}_i(\mathbf{I})$, given by Eqs.~(\ref{def:CQuad}) and (\ref{def:CLind}), respectively,  is in the definition of the maximal possible number of quads.     For Lind's clustering coefficient, $q^{\rm Lind}_{i, \rm max}$ is the sum of the existing quads $q_i$ and the number of ways $(\chi_{\alpha}(\textbf{I})-\eta_{i\alpha\beta}(\textbf{I}))(\chi_{\beta}(\textbf{I})-\eta_{i\alpha\beta}(\textbf{I}))$ that the remaining edges can be combined to form quads.   In general, the number $q^{\rm Lind}_{i, \rm max}$  overcounts significantly the number of possible quads.  For example, in Fig.~\ref{fig:bipar_clustering}  $q^{\rm Lind}_{i, \rm max} = 3$, even though $q_{\rm max} = 2$. 

\begin{figure}
     \centering
     \setlength{\unitlength}{0.1\textwidth}
     \begin{subfigure}[b]{0.4\textwidth}
         \centering
         \includegraphics[width=\textwidth]{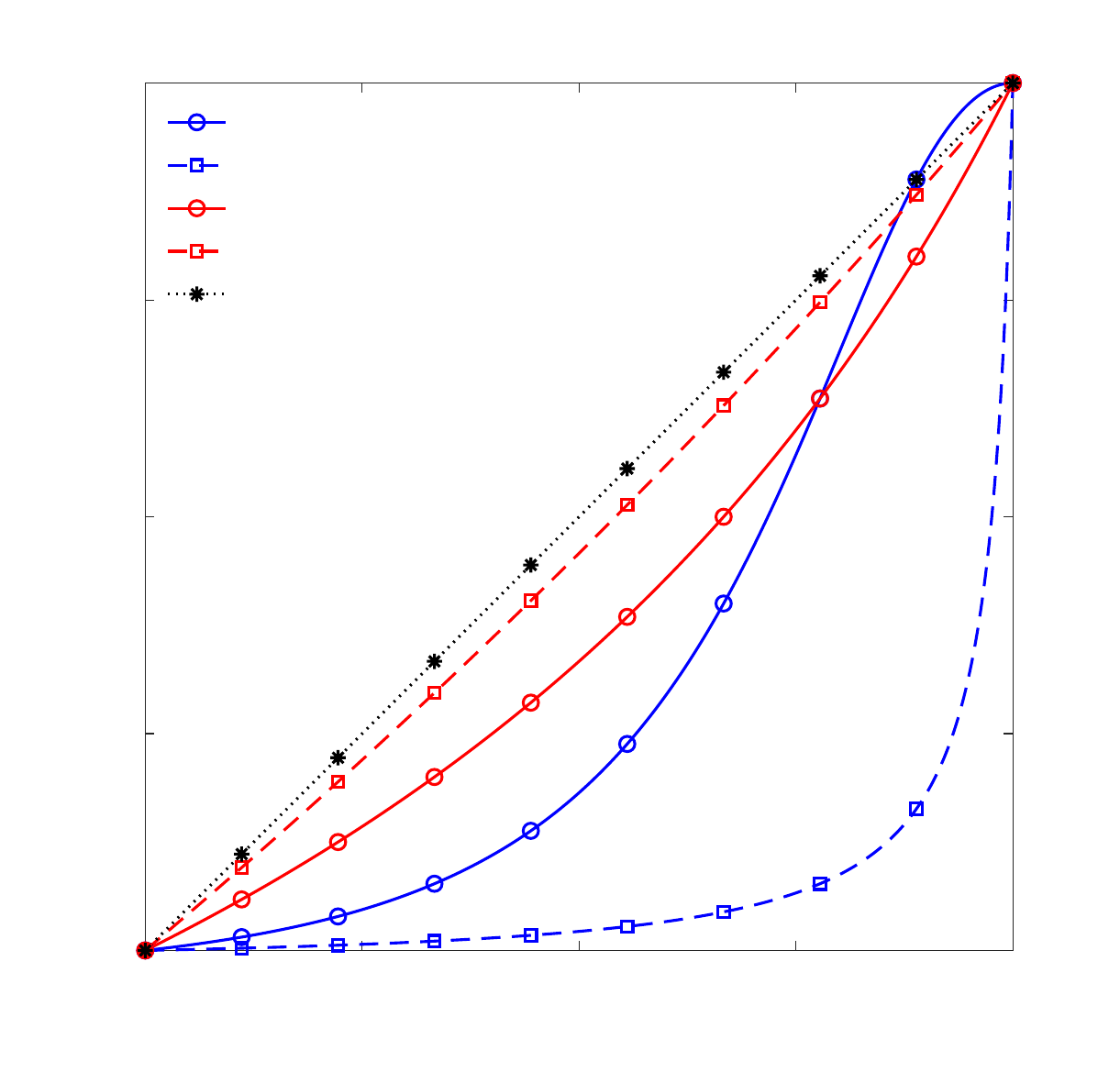}
         \label{fig:two_hyperedges}
     \end{subfigure}
      \put(-3.15,3.62){\tiny\color{blue}$C=C_{i}^{\text{Lind}}$ with $\chi_{\alpha}/\chi_{\beta}=1$}
      \put(-3.15,3.45){\tiny\color{blue}$C=C_{i}^{\text{Lind}}$ with $\chi_{\alpha}/\chi_{\beta}=0.2$}
      \put(-3.15,3.28){\tiny\color{red}$C=\Gamma_{i} C_{i}^{\text{Zhang}}$ with $\chi_{\alpha}/\chi_{\beta}=1$}
      \put(-3.15,3.11){\tiny\color{red}$C=\Gamma_{i} C_{i}^{\text{Zhang}}$ with $\chi_{\alpha}/\chi_{\beta}=0.2$}
      \put(-3.15,2.94){\tiny$C=C_{i}^{\text{Quad}}$ universal}
      \put(-4.2,3.79){\normalsize$(a)$}
      \put(-4.1,2.15){\small$C$}
      \put(-3.7,0.45){\footnotesize$0$}
      \put(-2.85,0.45){\footnotesize$0.25$}
      \put(-2.03,0.45){\footnotesize$0.5$}
      \put(-1.27,0.45){\footnotesize$0.75$}
      \put(-0.42,0.45){\footnotesize$1$}
      \put(-3.9,1.37){\footnotesize$0.25$}
      \put(-3.84,2.15){\footnotesize$0.5$}
      \put(-3.9,2.91){\footnotesize$0.75$}
      \put(-3.7,3.7){\footnotesize$1$}
      \put(-2.1,0.15){\small	$q=\frac{Q_i(\mathbf{I})}{\chi_{\alpha}-1}$}
     \hfill
     \begin{subfigure}[b]{0.4\textwidth}
         \centering
         \includegraphics[width=\textwidth]{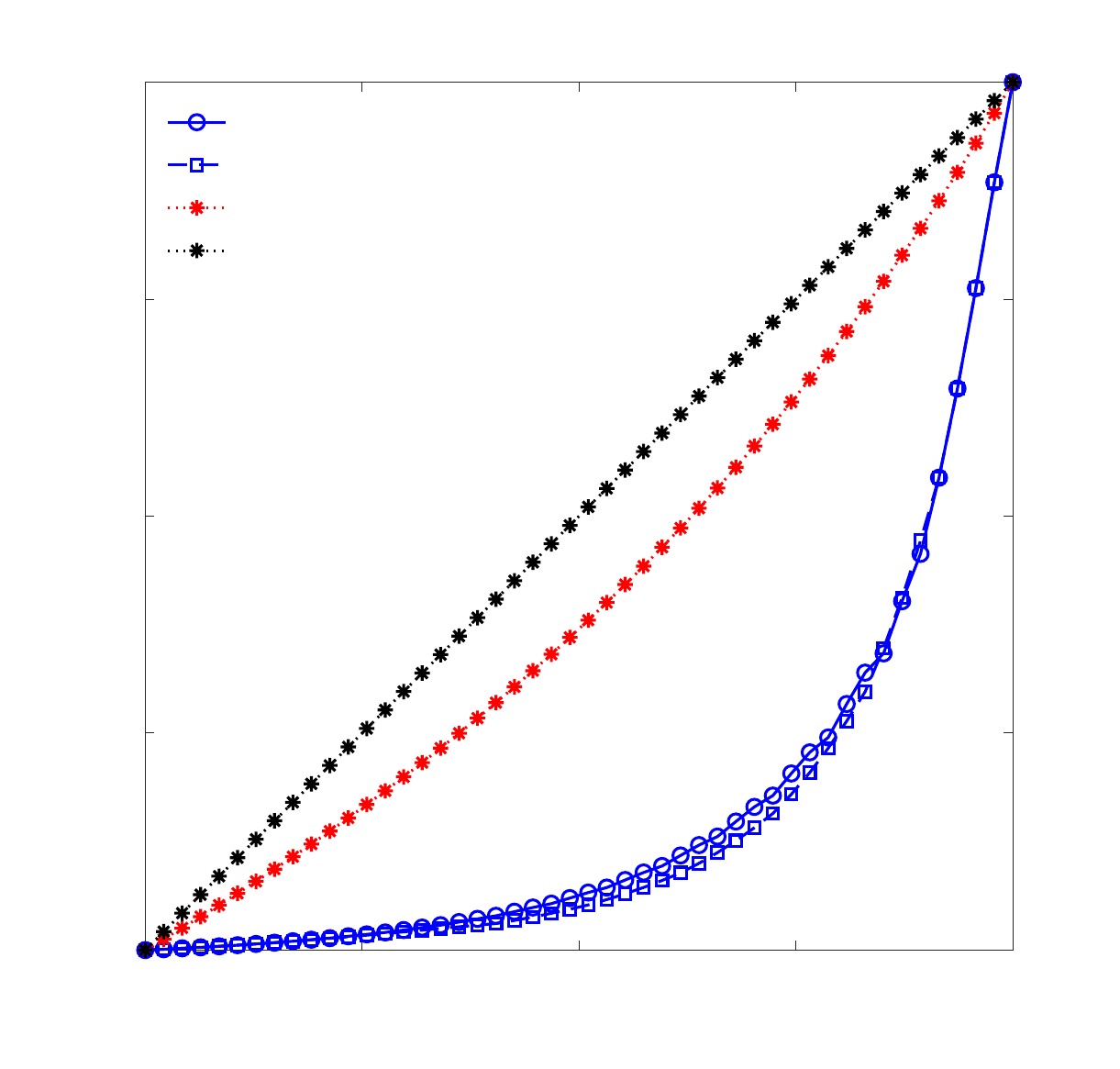}
         \label{fig:three_hyperedges}
     \end{subfigure}
      \put(-4.2,4.2){\normalsize$(b)$}
      \put(-4.1,2.15){\small$C$}
      \put(-3.15,3.58){\tiny\color{blue}$C=C_{i}^{\rm Lind}$ with uniform case}
      \put(-3.15,3.42){\tiny\color{blue}$C=C_{i}^{\rm Lind}$ with biased case}
      \put(-3.15,3.26){\tiny\color{red}$C=\Gamma_{i} C_{i}^{\rm Zhang}$}
      \put(-3.15,3.1){\tiny$C=C_{i}^{\rm q}$}
      \put(-3.7,0.45){\footnotesize$0$}
      \put(-2.85,0.45){\footnotesize$0.25$}
      \put(-2.03,0.45){\footnotesize$0.5$}
      \put(-1.27,0.45){\footnotesize$0.75$}
      \put(-0.42,0.45){\footnotesize$1$}
      \put(-3.9,1.37){\footnotesize$0.25$}
      \put(-3.84,2.15){\footnotesize$0.5$}
      \put(-3.9,2.91){\footnotesize$0.75$}
      \put(-3.7,3.7){\footnotesize$1$}
      \put(-2.27,0.15){\small	$\frac{Q_i(\mathbf{I})}{2\chi_\alpha+\chi_\beta-3}$}
        \caption{{\it Comparison among the different clustering coefficients of hypergraphs.}   The quad clustering coefficient $C^{\rm q}_i$, Lind's clustering coefficient $C^{\rm Lind}_i$, and Zhang's clustering coefficient $C^{\rm Zhang}_i$ (markers) are plotted as a function of the number of quads $Q_i$ incident to a node $i$ (the scaling factor on the $x$-axis is chosen to make the variable's range $[0,1]$). Upper Panel: node $i$ has degree $k_i=2$ and is connected to two hyperedges $\alpha$ and $\beta$, with cardinalities $\chi_{\alpha}=10$ and $\chi_\beta$ as indicated in the legend.   Lines denote the functions given by Eqs.~(\ref{eq:CLindCont}) and (\ref{eq:ZhangCont})  for $C^{\rm Lind}_i$ and $C^{\rm Zhang}_i$, respectively. 
        Lower Panel: node $i$ has degree $k_i=3$ and interacts with hyperedges $\alpha$, $\beta$ and $\gamma$, of cardinalities $\chi_{\alpha}=15$, $\chi_{\beta}=20$, and  $\chi_{\gamma}=25$, respectively.  Circles and squares represent  values of  $C^{\rm Lind}$  for different values of $q_{i\alpha\beta}$, $q_{i\alpha\gamma}$, $q_{i\beta\gamma}$, obtained from two different prescriptions, i.e. uniform and biased, as explained in Appendix~\ref{app:QDetail}).  In the lower panel, lines are a guide to the eye. 
      }
        \label{fig:clustering_landscape}
\end{figure}

Another notable difference between the quad clustering coefficient and Lind's clustering coefficient is  that the former is a linear function of $Q_i$, while the latter is, in general, not a function of $Q_i$.     An exception is when $k_i=2$, in which case Lind's clustering coefficient is a function of $Q_i$, but this function is nonlinear.   
This feature is illustrated in the upper panel of   Fig.~\ref{fig:clustering_landscape} that plots   Lind's clustering coefficient as a function of the quad clustering coefficient for a node of degree $2$ that is connected to a hyperedge with cardinality $\chi_\alpha$ and a hyperedge with cardinality $\chi_\beta$.  The solid lines in   Fig.~\ref{fig:clustering_landscape} are obtained by taking the limit  $\chi_{\alpha}\rightarrow \infty$  with the ratio $r= \chi_{\beta}/\chi_{\alpha}>1$ fixed, yielding the function
\begin{equation}
 C^{{\rm Lind}}(q) = \lim_{\chi_{\alpha}\rightarrow \infty}C^{{\rm Lind}}_i(\mathbf{I}) = 
\frac{q}
{\chi_{\alpha}(1-q)(r-q)+q}, \label{eq:CLindCont}
\end{equation}
where $q = Q_i/(\chi_{\alpha}-1)\in [0,1]$ (see Appendix~\ref{app:Clustering_inf}).   We observe a strong nonlinearity in $C^{{\rm Lind}}(q)$ for large values of  $\chi_\alpha$. Indeed, as shown in Fig.~\ref{fig:clustering_landscape}(a),  for $q$ below one   and
 large  enough values of $\chi_{\alpha}$, it holds that  $C^{{\rm Lind}}(q) \approx  0$, and for  $q= 1$ it holds that   $C^{{\rm Lind}}(q) =1$, which can be recovered from Eq.~(\ref{eq:CLindCont}) by taking the limit  $\chi_\alpha\rightarrow \infty$.

For nodes with a degree  $k_i>2$, Lind's clustering coefficient,  is not  a function of $Q_i$,  contrarily to the quad clustering coefficient,    as $q^{\rm Lind}_{i, \rm max}$ depends on   all $q_{i\alpha\beta}$, with $\alpha,\beta\in \W$.    For the simplest case of $k_i=3$,   we illustrate this feature in the lower panel of   Fig.~\ref{fig:clustering_landscape}.    The circles and squares denote $C^{\rm Lind}_i$ for two different assignments for 
$q_{i\alpha\beta}$, $q_{i\alpha\gamma}$, and $q_{i\beta\gamma}$, as detailed in Appendix~\ref{app:QDetail}.     As Fig.~\ref{fig:clustering_landscape}(b) shows, the two curves for $C^{\rm Lind}_i$ are different for different prescriptions on the $q$'s indicating that $C^{\rm Lind}_i$ is not a function of $Q_i$.

\subsection{Zhang's clustering coefficient}\label{sec:Zhang} 
In Ref.~\cite{zhang2008clustering}, Zhang et al.  introduce the clustering coefficient
\begin{eqnarray} 
\begin{split}
C^{\rm Zhang}_{i}(\textbf{I})\equiv&
\frac{Q_i(\mathbf{I})}
{q^{\rm Zhang}_{i,\rm max}(\mathbf{I})}, \label{def:CZhang}
\end{split}
\end{eqnarray}
where 
\begin{multline}
q^{\rm Zhang}_{i,\rm max}(\mathbf{I}) = \sum_{\alpha<\beta}\bigl[(\chi_{\alpha}(\textbf{I})-\eta_{i\alpha\beta}(\textbf{I}))+(\chi_{\beta}(\textbf{I})-\eta_{i\alpha\beta}(\textbf{I}))\\
+q_{i\alpha\beta}(\textbf{I})\bigr]I_{i\alpha}I_{i\beta}
\end{multline}
is the maximal possible number of quads.    We call $C^{\rm Zhang}_{i}$ Zhang's clustering coefficient.       Note that Zhang's clustering coefficient can also be written as~\cite{kitsak2011hidden} 
\begin{eqnarray} 
\begin{split}
C^{\rm Zhang}_{i}(\textbf{I})
=&
\frac{\sum_{\alpha,\beta;\alpha<\beta}|\partial_{i\alpha}(\textbf{I})\cap\partial_{i\beta}(\textbf{I})|}
{\sum_{\alpha,\beta;\alpha<\beta}|\partial_{i\alpha}(\textbf{I})\cup\partial_{i\beta}(\textbf{I})|},
\end{split}
\end{eqnarray}
which is known as the Jaccard similarity coefficient~\cite{jaccard1901etude}.

Comparing $C^{\rm Zhang}_i(\mathbf{I})$ with $C^{\rm Lind}_i(\mathbf{I})$ and $C^{\rm q}_i(\mathbf{I})$, we see that Zhang et al. considered yet another way of counting the maximal, possible  number of quads.   In the example of Fig.~\ref{fig:bipar_clustering}, we get $C^{\rm Zhang}_i(\mathbf{I})=0$ for $(a)$, $C^{\rm Zhang}_i(\mathbf{I})=1/4$ for $(b)$ and $C^{\rm Zhang}_i(\mathbf{I})=2/3$ for $(c)$.

Like Lind's clustering coefficient, for nodes with degree $k_i=2$ Zhang's clustering coefficient is a nonlinear function of $Q_i$.   Indeed, taking the limit $\chi_{\alpha}\rightarrow \infty$ while keeping $r=\chi_{\alpha}/\chi_{\beta}>1$ fixed, we get  
\begin{equation}
 C^{{\rm Zhang}}(q) = \lim_{\chi_{\alpha}\rightarrow \infty}C^{{\rm Zhang}}_i(\mathbf{I}) = 
\frac{q}
{1+r-q}, \label{eq:ZhangCont}
\end{equation}
for $q \in[0,1]$ (see Appendix~\ref{app:Clustering_inf}).  
We illustrate this function in the upper panel of Fig.~\ref{fig:clustering_landscape}. 
  Note that  Zhang's clustering coefficient is not normalised, as  $C^{{\rm Zhang}}(1) = 1/r$, and more generally  $C^{\rm Zhang}_{i}(\mathbf{I})\in [0,1/\Gamma_i(\mathbf{I})]$, with
\begin{equation}
\Gamma_{i}(\textbf{I})=\frac{\sum_{\alpha<\beta}\text{max}\{\chi_{\alpha}(\textbf{I})-1,\chi_{\beta}(\textbf{I})-1\}I_{i\alpha}I_{i\beta}}{\sum_{\alpha<\beta}\text{min}\{\chi_{\alpha}(\textbf{I})-1,\chi_{\beta}(\textbf{I})-1\}I_{i\alpha}I_{i\beta}}.
\end{equation} 

For nodes with degrees $k_i>2$,  $ C^{{\rm Zhang}}_i$ is not a function of $Q_i$, as $q^{\rm Zhang}_{i, \rm max}$ depends on $q_{i\alpha\beta}$ for all $\alpha,\beta\in \W$.

\section{Average quad clustering coefficient for random hypergraphs} \label{ch:avg_quad_random}
  
In this Section, we determine the average quad clustering coefficients for random 
hypergraphs.  First, in Sec.~\ref{sec:regCard} we derive  the ensemble averaged clustering coefficient in random hypergraph models with regular cardinalities, i.e., $\chi_\alpha(\mathbf{I})=\chi$  for all $\alpha\in \W$.   For these models we obtain compact expressions   for  the ensemble averaged quad clustering coefficient in terms of the model parameters.    Subsequently, in Sec.~\ref{sec:birreg} we deal with models that are biregular in the cardinalities, i.e., $\chi_\alpha(\mathbf{I})\in \left\{\chi_1,\chi_2\right\}$, and, as will become evident, the calculations in biregular models are significantly more difficult than those in models with regular cardinalities.     

\subsection{Regular cardinalities}\label{sec:regCard}
We consider three random hypergraph models with regular cardinalities, i.e., for which $\chi_\alpha(\mathbf{I})=\chi$ for all $\alpha\in \W$.   The three models are distinguished by the fluctuations in their degrees $k_i(\mathbf{I})$.   In the $\chi$-regular ensemble, considered in Sec.~\ref{sec:chireg}, the degrees are unconstrained; in the $(k,\chi)$-regular ensemble, considered in Sec.~\ref{sec:chikreg}, the degrees are regular, i.e., $k_i(\mathbf{I})=k$ for all $i\in \V$; lastly, in the $(\vec{k},\chi)$-regular ensemble, considered in Sec.~\ref{sec:chikvecreg}, the degrees are prescribed by the sequence $\vec{k}$, i.e., $k_i(\mathbf{I})=k_i$ for all $i\in \V$. $\\$

\subsubsection{$\chi$-regular ensemble} \label{sec:chireg}
In the $\chi$-regular ensemble the probability of drawing a hypergraph with  incidence matrix $\mathbf{I}\in \left\{0,1\right\}^{NM}$ is given by 
\begin{eqnarray} 
P_{\chi}(\textbf{I}) \equiv \frac{1}{\mathcal{N}_{\chi}}\prod^M_{\alpha=1}\delta_{\chi,\chi_{\alpha}(\mathbf{I})},
\label{Eq:rand_model}
\end{eqnarray}
with the normalisation constant $\mathcal{N}_{\chi}$
as derived in~Appendix~\ref{app:chireg}.   

The average quad clustering coefficient 
\begin{equation}
\langle C^{\rm q}_i(\textbf{I}) \rangle_{\chi} \equiv \sum_{\mathbf{I}}P_{\chi}(\mathbf{I})C^{\rm q}_i(\textbf{I}),  \label{eq:defensembleAverageCQ}
\end{equation}
 where $\sum_{\mathbf{I}}$ is a sum over all possible incidence matrices $\mathbf{I}\in \left\{0,1\right\}^{NM}$, 
is given by (see~Appendix~\ref{app:chireg} for a derivation)
\begin{equation}
\langle C^{\rm q}_i(\textbf{I}) \rangle_{\chi}=\frac{\chi-1}{N-1}\left[1-\left(1-\frac{\chi}{N}\right)^{M}\left(1+\frac{M\chi}{N-\chi}\right)\right].
\label{eq:chi_regular_cq}
\end{equation}
Taking the limit of large $N$ while keeping  the mean node degree 
\begin{equation}
c \equiv \frac{M}{N} \chi
\end{equation}
fixed, and thus finite, we obtain   
\begin{equation}
\langle C^{\rm q}_i(\textbf{I}) \rangle_{\chi} =\frac{\chi-1}{N}\left[1-e^{- c}\left(1+ c\right)\right] + \mathcal{O}\left(\frac{1}{N^2}\right).
\label{eq:clustering_finite_limit}
\end{equation}
Note that the average quad clustering coefficient decreases as  $1/N$ with the order of the hypergraph, implying that the density of quads vanishes in the limit of infinitely large, sparse, hypergraphs.   For large values of $\chi$, but still $\chi\ll N$, we get the simple formula 
\begin{equation}
\langle C^{\rm q}_i(\textbf{I}) \rangle_{\chi} = \frac{\chi}{N} + \mathcal{O}\left(\frac{1}{N^2}\right) +   \mathcal{O}\left(\frac{1}{\chi}\right) \label{eq:universal}
\end{equation}
stating that the average density of quads equals the cardinality $\chi$ divided by the number $N$ of nodes.  
$\\$

\subsubsection{$(c,\chi)$-regular ensemble} \label{sec:chikreg}

In the $(c,\chi)$-regular ensemble the probability assigned to a hypergraph with incidence matrix $\mathbf{I}$ is defined by 
\begin{eqnarray} 
P_{c,\chi}(\textbf{I})\equiv \frac{1}{\mathcal{N}_{k,\chi}}\prod^N_{j=1}\delta_{c,k_{j}(\mathbf{I})}\prod^M_{\alpha=1}\delta_{\chi,\chi_{\alpha}(\mathbf{I})},
\label{Eq:regular_model}
\end{eqnarray}
where $\mathcal{N}_{k,\chi} = \sum_{\mathbf{I}}\prod^N_{j=1}\delta_{c,k_{j}(\mathbf{I})}\prod^M_{\alpha=1}\delta_{\chi,\chi_{\alpha}(\mathbf{I})}$ is the  normalization constant.

In Appendix~\ref{app:chiveckreg} we derive the average quad clustering coefficient for this model in the limit $N\gg 1$ with fixed values of $c$ and $\chi$, and with $M = (c/\chi) N$.   
Neglecting subleading order corrections, we find for the  average quad clustering coefficient the expression 
\begin{align}
\langle C^{\rm q}_i(\textbf{I}) \rangle_{c,\chi}= \frac{c-1}{c} \frac{\chi-1}{N}+\mathcal{O}\left(\frac{1}{N^2}\right). \label{eq:CAvkchi}
\end{align}
In the limit of large values of $k$ and $\chi$, we recover Eq.~(\ref{eq:universal}), indicating that in this limit the average clustering coefficient   is independent of the degree distribution.  However, at finite $k$ and $\chi$ the average clustering coefficient depends on the degree fluctuations, as (\ref{eq:CAvkchi})   differs from (\ref{eq:clustering_finite_limit}). 

\subsubsection{$(\vec{k},\chi)$-regular ensemble}\label{sec:chikvecreg}
In the $(\vec{k},\chi)$-regular ensemble the probability assigned to incidence matrices  $\mathbf{I}$ is given by
\begin{eqnarray} 
P_{\vec{k},\chi}(\textbf{I})=\frac{1}{\mathcal{N}_{\vec{k},\chi}}\prod^N_{j=1}\delta_{k_j,k_{j}(\mathbf{I})}\prod^M_{\alpha=1}\delta_{\chi,\chi_{\alpha}(\mathbf{I})},
\label{Eq:irregular_model}
\end{eqnarray}
where $\mathcal{N}_{\vec{k},\chi} = 
\sum_{\mathbf{I}}\prod^N_{j=1}\delta_{k_j,k_{j}(\mathbf{I})}\prod^M_{\alpha=1}\delta_{\chi,\chi_{\alpha}(\mathbf{I})}$ is the normalization constant.

Neglecting subleading order terms, the average quad clustering coefficient is given by (see  Appendix~\ref{app:chiveckreg})
\begin{eqnarray}
\lefteqn{\langle C^{\rm q}_i(\textbf{I}) \rangle_{\vec{k},\chi} }&& \nonumber\\  
&=&\frac{\chi-1}{\overline{k}^{2}N} 
\overline{ k(k-1)}\left(1-p_{\rm deg}(0)-p_{\rm deg}(1) \right)
+ \mathcal{O}\left(\frac{1}{N^2}\right) ,
\label{eq:CqkPresc}
\end{eqnarray}
where the overline denotes the mean value  
\begin{equation}
\overline{f(k)} \equiv \sum^{\infty}_{k=0}p_{\rm deg}(k)f(k),
\end{equation}
with 
\begin{equation}
p_{\rm deg}(k) \equiv  \lim_{N\rightarrow \infty}\frac{\sum^N_{j=1}\delta_{k,k_j}}{N} 
\end{equation}
and $f(k)$ is an arbitrary real-valued function defined on $k\in\mathbb{N}\cup \left\{0\right\}$.
Using $p_{\rm deg}(k) = \delta_{k,c}$ and $p_{\rm deg}(k) = e^{-c}c^k/k!$ in Eq.~(\ref{eq:CqkPresc}), we find, respectively, the Eqs.~(\ref{eq:CAvkchi}) and (\ref{eq:clustering_finite_limit}).  Hence, the formula (\ref{eq:CqkPresc}) generalises Eqs.~(\ref{eq:CAvkchi}) and  (\ref{eq:clustering_finite_limit}).

Notice  that the first term in  Eq.~(\ref{eq:CqkPresc}) diverges when the degree distribution $p_{\rm deg}(k)$ has a diverging second moment,  indicating that the average clustering coefficient of random hypergraphs with diverging second moments decreases slower than $1/N$  as a function of $N$.   This results is compatible with what is known for random graphs, as the average number of cycles of finite length  diverges with the second moment of the degree distribution (see Equation (9) in Ref.~\cite{bianconi2005loops}).

\subsection{Biregular cardinalities }\label{sec:birreg}
Having studied in detail the case with regular cardinalities, including the effect of degree fluctuations, we now analyze how fluctuations in the cardinality affect the average quad clustering coefficient.  We focus on the simplest case of biregular ensembles, where $M_1$ hyperedges have cardinality $\chi_1$ and the remaining $M-M_1$ have cardinality $\chi_2$. In this case, the probability of incidence matrices  $\mathbf{I}\in \left\{0,1\right\}^{NM}$ takes the form 
\begin{equation} 
P_{\chi_1,\chi_2}(\textbf{I})=\frac{1}{\mathcal{N}_{\chi_1,\chi_2}}\prod^{M_{1}}_{\alpha=1}\delta_{\chi_{1},\chi_{\alpha}(\mathbf{I})}\prod^{M}_{\beta=M_{1}+1}\delta_{\chi_{2},\chi_{\beta}(\mathbf{I})},
\label{Eq:rand_model2}
\end{equation}
where as before $\mathcal{N}_{\chi_1,\chi_2}$ is the normalisation constant.

In  Appendix~\ref{App:E}, we show that the average clustering coefficient, defined by 
\begin{eqnarray}
\langle C^{\rm q}_{i}(\textbf{I}) \rangle_{\chi_1,\chi_2}&\equiv & \sum_{\mathbf{I}} C^{\rm q}_{i}(\textbf{I}) P_{\chi_1,\chi_2}(\mathbf{I}) ,\label{eq:DefEnsemAvChi1Chi2}
\end{eqnarray}
is given by 
\begin{eqnarray}
\langle C^{\rm q}_{i}(\textbf{I}) \rangle_{\chi_1,\chi_2}&=&\sum_{u=2}^{M_{1}}\sum_{v=0}^{M_{2}}\frac{\Lambda_{2,0}(u,v)}{\Phi(u,v)} \nonumber\\
&& +\sum_{u=1}^{M_{1}}\sum_{v=1}^{M_{2}}\frac{\Lambda_{1,1}(u,v)}{\Phi(u,v)}+\sum_{u=0}^{M_{1}}\sum_{v=2}^{M_{2}}\frac{\Lambda_{0,2}(u,v)}{\Phi(u,v)}  \nonumber\\ \label{eq:biregAverage}
\end{eqnarray}
where $M_2=M-M_1$ and we introduced the functions
\begin{multline}
\Lambda_{a,b}(u,v)\equiv 2(N-1)\binom{M_{1}}{a}\binom{M_{2}}{b}\\
\times\left[\binom{N-2}{\chi_{1}-2}\right]^{a}\left[\binom{N-2}{\chi_{2}-2}\right]^{b}\\
\times\binom{M_{1}-a}{u-a}\left[\binom{N-1}{\chi_{1}-1}\right]^{u-a}\left[\binom{N-1}{\chi_{1}}\right]^{M_{1}-u}\\
\times\binom{M_{2}-b}{v-b}\left[\binom{N-1}{\chi_{2}-1}\right]^{v-b}\left[\binom{N-1}{\chi_{2}}\right]^{M_{2}-v}, \label{eq:lambdaab}
\end{multline}
and
\begin{multline}
\Phi(u,v)\equiv \binom{N}{\chi_{1}}^{M_{1}}\binom{N}{\chi_{2}}^{M_{2}}\bigl[(\chi_{1}-1)(u+v)(u+v-1)\\
+v(\chi_{2}-\chi_{1})(v-1)\bigr]. \label{eq:phiuv}
\end{multline}

We have not been able to simplify the expression  (\ref{eq:biregAverage})-(\ref{eq:phiuv}) further, not even in the sparse limit.  Hence, although models with degree fluctuations are analytical tractable, as shown in Sec.~\ref{sec:chikvecreg}, it is significantly more difficult to  deal with models with heterogeneous cardinalities.

Setting $\chi_1=\chi_2=\chi$ in Eq.~(\ref{eq:biregAverage}), we find the Eq.~(\ref{eq:chi_regular_cq}).  Hence, the formula (\ref{eq:biregAverage}) generalises Eq.~(\ref{eq:chi_regular_cq}). 

We understand each term in Eq.~(\ref{eq:biregAverage}) as follows: the first and last terms consider quads consisting of two hyperedges with the same cardinality, and the middle term considers the case where the two hyperedges have different cardinalities.

\section{Quad clustering coefficient in real world hypergraphs} \label{ch:real_hypergraph}
Having established a theoretical understanding of quad clustering coefficients in random hypergraphs, we focus now our attention on  the quad clustering coefficient in real-world hypergraphs.     To this aim, we  build hypergraphs out of  six datasets, which are related to    Github, Youtube,   NDC-subtances,  food recipes, Wallmart, and crime involvement.    As detailed in Table~\ref{tb:qC_value}, the real-world hypergraphs have diverse topologies:     their order ranges from $N\approx 10^{3}$ to $N\approx 10^5$, their mean degree ranges from $\overline{k}\approx 3$  to $\overline{k}\approx 60$, and their mean cardinality ranges from $\overline{\chi}\approx 3$ to $\overline{\chi}\approx 10$  [see Appendix~\ref{app:data} for more detailed information about these data sets].

\begin{table*}
\caption{Characteristics of the  real-world hypergraphs considered in this Paper: number of nodes $N$ and  hyperedges $M$,  mean degree $\overline{k}$ and mean cardinality $\overline{\chi}$, mean quad clustering coefficient $\overline{C}^{\rm q}(\mathbf{I}_{\rm real})$, mean Lind's clustering coefficient $\overline{C}^{\rm Lind}(\mathbf{I}_{\rm real})$, mean Zhang's clustering coefficient $\overline{C}^{\rm Zhang}(\mathbf{I}_{\rm real})$, the average mean quad clustering coefficient $\langle \overline{C}^{\rm q}(\mathbf{I})\rangle$, the average mean Lind's clustering coefficient $\langle \overline{C}^{\rm Lind}(\mathbf{I})\rangle$ and the average mean Zhang's clustering coefficient $\langle \overline{C}^{\rm Zhang}(\mathbf{I})\rangle$ of the corresponding configuration model with fixed degree sequence $\vec{k}(\mathbf{I}_{\rm real})$ and cardinality sequence  $\vec{\chi}(\mathbf{I}_{\rm real})$.  For more details see Appendix~\ref{app:data}.}\label{tb:qC_value}
\centering
\begin{tabular}{cccccccccccc}
\hline\hline
Dataset & $N$ & $M$ & $\overline{k}$ & $\overline{\chi}$ & $\overline{C}^{\rm q}(\textbf{I}_{\rm real})$ & $\overline{C}^{\rm Lind}(\textbf{I}_{\rm real})$ & $\overline{C}^{\rm Zhang}(\textbf{I}_{\rm real})$ & $\langle\overline{C}^{\rm q}(\textbf{I})\rangle$ & $\langle\overline{C}^{\rm Lind}(\textbf{I})\rangle$ & $\langle\overline{C}^{\rm Zhang}(\textbf{I})\rangle$\\
\hline
NDC-substances & 5,556 & 112,919 & 12.2 & 2.0 & 0.2760 & 0.1418 & 0.1792 & 0.0252 & 0.0012 & 0.0093\\
Youtube & 94,238 & 30,087 & 3.1 & 9.8 & 0.0920 & 0.0094 & 0.0225 & 0.0142 & 0.0001 & 0.0043\\
Food recipe & 6,714 & 39,774 & 63.8 & 10.8 & 0.1118 & 0.0178 & 0.0501 & 0.0658 & 0.0054 & 0.0271\\
Github & 56,519 & 120,867 & 7.8 & 3.6 & 0.1129 & 0.0408 & 0.0329 & 0.0084 & 0.0001 & 0.0029\\
Crime involvement & 829 & 551 & 1.8 & 2.7 & 0.0369 & 0.0243 & 0.0169 & 0.0037 & 0.0010 & 0.0013\\
Wallmart & 88,860 & 69,906 & 5.2 & 6.6 & 0.0120 & 0.0046 & 0.0046 & 0.0010 & 0.0001 & 0.0003\\
\hline\hline
\end{tabular}
\end{table*}

\subsection{Mean quad clustering coefficient}  

 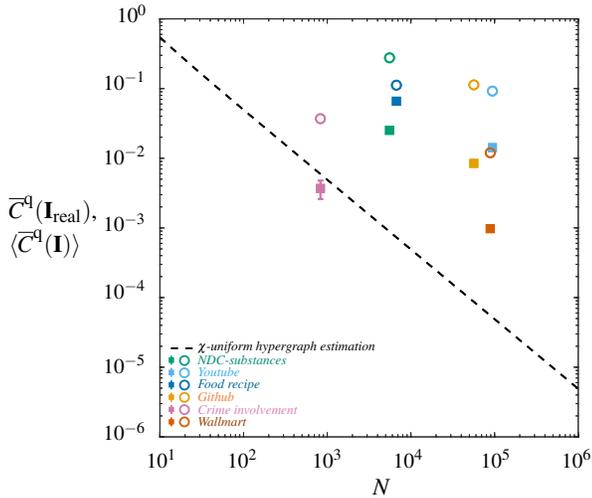
\begin{figure}
     \centering
     \setlength{\unitlength}{0.1\textwidth}
     \hspace*{0.8cm}
     \includegraphics[width=0.4\textwidth]{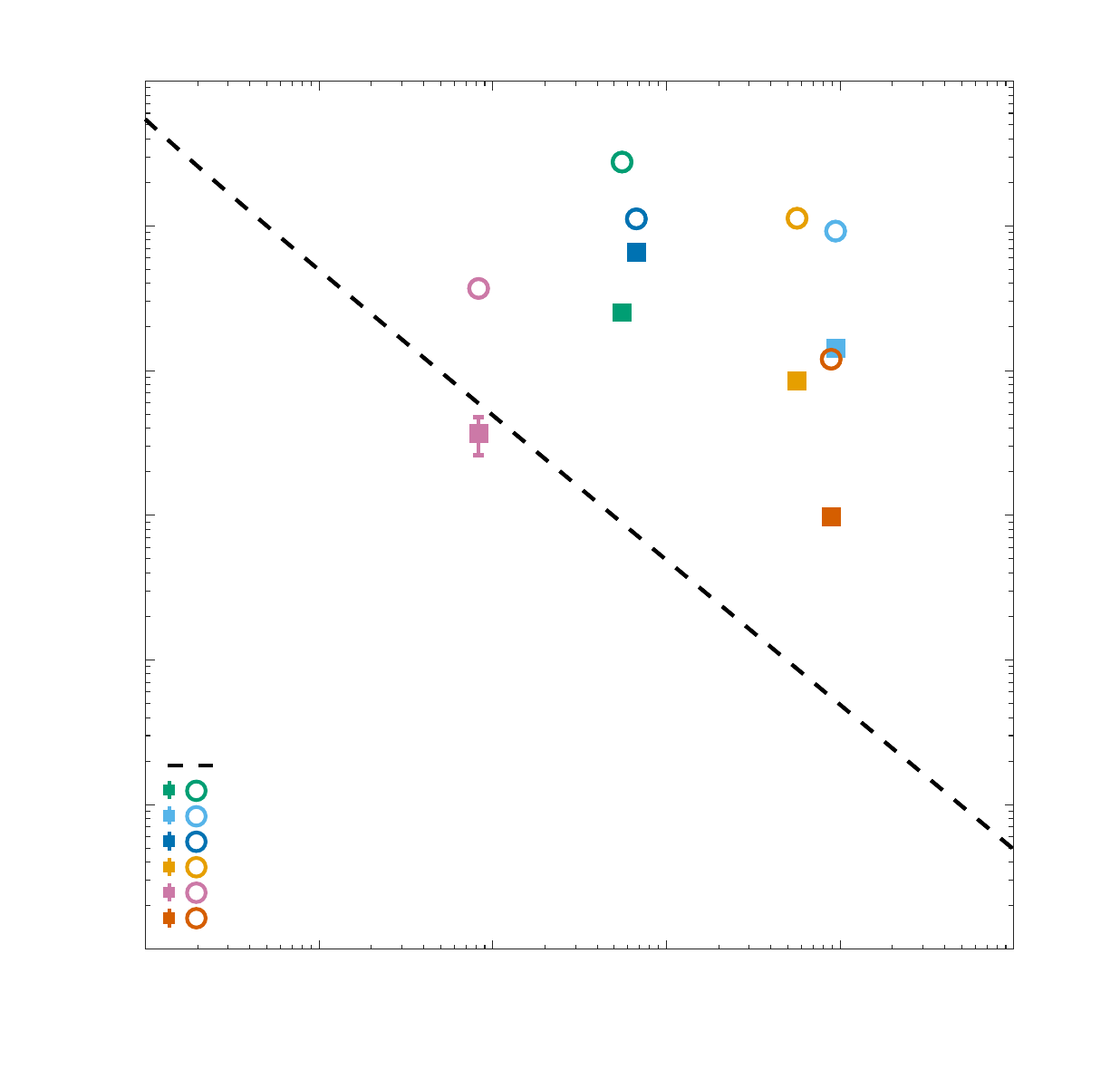}
     \put(-4.6,2.05){\small $\overline{C}^{\rm q}(\mathbf{I}_{\rm real})$,}
    \put(-4.6,1.8){\small$\langle \overline{C}^{\rm q}(\mathbf{I}) \rangle$}
      \put(-1.9,0.){\small $N$}
      \put(-3.9,0.4){\footnotesize$10^{-6}$}
      \put(-3.58,0.2){\footnotesize$10^{1}$}
      \put(-2.98,0.2){\footnotesize$10^{2}$}
      \put(-2.35,0.2){\footnotesize$10^{3}$}
      \put(-1.75,0.2){\footnotesize$10^{4}$}
      \put(-1.10,0.2){\footnotesize$10^{5}$}
      \put(-0.5,0.2){\footnotesize$10^{6}$}
      \put(-3.9,0.9){\footnotesize$10^{-5}$}
      \put(-3.9,1.45){\footnotesize$10^{-4}$}
      \put(-3.9,1.95){\footnotesize$10^{-3}$}
      \put(-3.9,2.45){\footnotesize$10^{-2}$}
      \put(-3.9,3.0){\footnotesize$10^{-1}$}
      \put(-3.8,3.5){\footnotesize$10^{0}$}
      \put(-3.2,1.07){\tiny {\it $\chi$-uniform hypergraph estimation}}
      \put(-3.2,0.9625){\tiny\color{PineGreen} {\it NDC-substances}}
      \put(-3.2,0.875){\tiny\color{CornflowerBlue} {\it Youtube}}
      \put(-3.2,0.7875){\tiny\color{MidnightBlue} {\it Food recipe}}
      \put(-3.2,0.69){\tiny\color{Orange} {\it Github}}
      \put(-3.2,0.5925){\tiny\color{Thistle} {\it Crime involvement}}
      \put(-3.2,0.505){\tiny\color{RawSienna} {\it Wallmart}}
     \caption{Comparison between   mean, quad clustering coefficients $\overline{C}^{\rm q}(\mathbf{I}_{\rm real})$  (unfilled, circles) in real-world hypergraphs, and   average, mean clustering coefficients $\langle \overline{C}^{\rm q}(\mathbf{I}) \rangle$   (filled, squares) in random hypergraphs with prescribed degree and cardinality sequences $\vec{k}(\mathbf{I}_{\rm real})$ and $\vec{\chi}(\mathbf{I}_{\rm real})$.  Estimates of $\langle \overline{C}^{\rm q}(\mathbf{I}) \rangle$  are based on $100$  hypergraph realisations, and error bars show the error on the mean, whenever they are larger than the marker size.   The dashed line represents the prediction  Eq.~(\ref{eq:clustering_finite_limit}) for $\chi$-regular hypergraphs with $\chi=5.9$ and $c=20/\chi$, which are, respectively, the average cardinality and mean degree of all hyperedges and nodes in all real-world datasets.
     }  
     \label{fig:avgquadCcompare}
\end{figure}

The mean quad clustering coefficient 
\begin{eqnarray} 
\overline{C}^{\rm q}(\textbf{I})\equiv\frac{1}{N}\sum^N_{i=1}C^{{\rm q}}_i(\textbf{I})
\end{eqnarray}
is  a real number $\overline{C}^{\rm q}(\textbf{I})\in [0,1]$ that quantifies the density of quads in the hypergraph represented by $\mathbf{I}$.   In Figure~\ref{fig:avgquadCcompare}, we compare the mean clustering coefficients $\overline{C}^{\rm q}(\textbf{I}_{\rm real})$ for the six canonical hypergraphs under study, represented by $\textbf{I}_{\rm real}$, with  those of the configuration model~\cite{Newman2001} with a prescribed degree sequence $\vec{k}(\mathbf{I}_{\rm real})$ and cardinality sequence $\vec{\chi}(\mathbf{I}_{\rm real})$ (see Appendix~\ref{app:conf_model} for a description of the algorithm used to generate hypergraphs from the configuration model).  The results in Fig.~\ref{fig:avgquadCcompare} reveal that the  quad clustering coefficients of real-world networks are significantly larger than the average clustering coefficient  $\langle \overline{C}^{\rm q}(\mathbf{I})\rangle$ of the  corresponding  configuration models ($\langle \overline{C}^{\rm q}(\mathbf{I})\rangle \approx 0.10 \:\overline{C}^{\rm q}_i(\mathbf{I}_{\rm real})$, see Table~\ref{tb:qC_value}).  Hence, the density of quads in real-world networks is higher than what  is expected in the configuration  model, similarly to  previous findings for clustering coefficients in networks with pairwise interactions, see, e.g., Ref.~\cite{barabasi}. {Similar conclusions can be drawn from comparing Lind's and Zhang's clustering coefficients between real-world and random networks (see Table~\ref{tb:qC_value}).  However, the corresponding values of Lind's and Zhang's clustering coefficients are one order of magnitude smaller than the quad clustering coefficient, consistent with the  behaviour of the clustering coefficients as a function of the number of quads as shown in Fig.~\ref{fig:clustering_landscape} and discussed in Sec.~\ref{sec:defNot}.}

 \begin{figure*}[th!]
     \centering
     \setlength{\unitlength}{0.1\textwidth}
     \hspace*{0.8cm}
     \includegraphics[width=\textwidth]{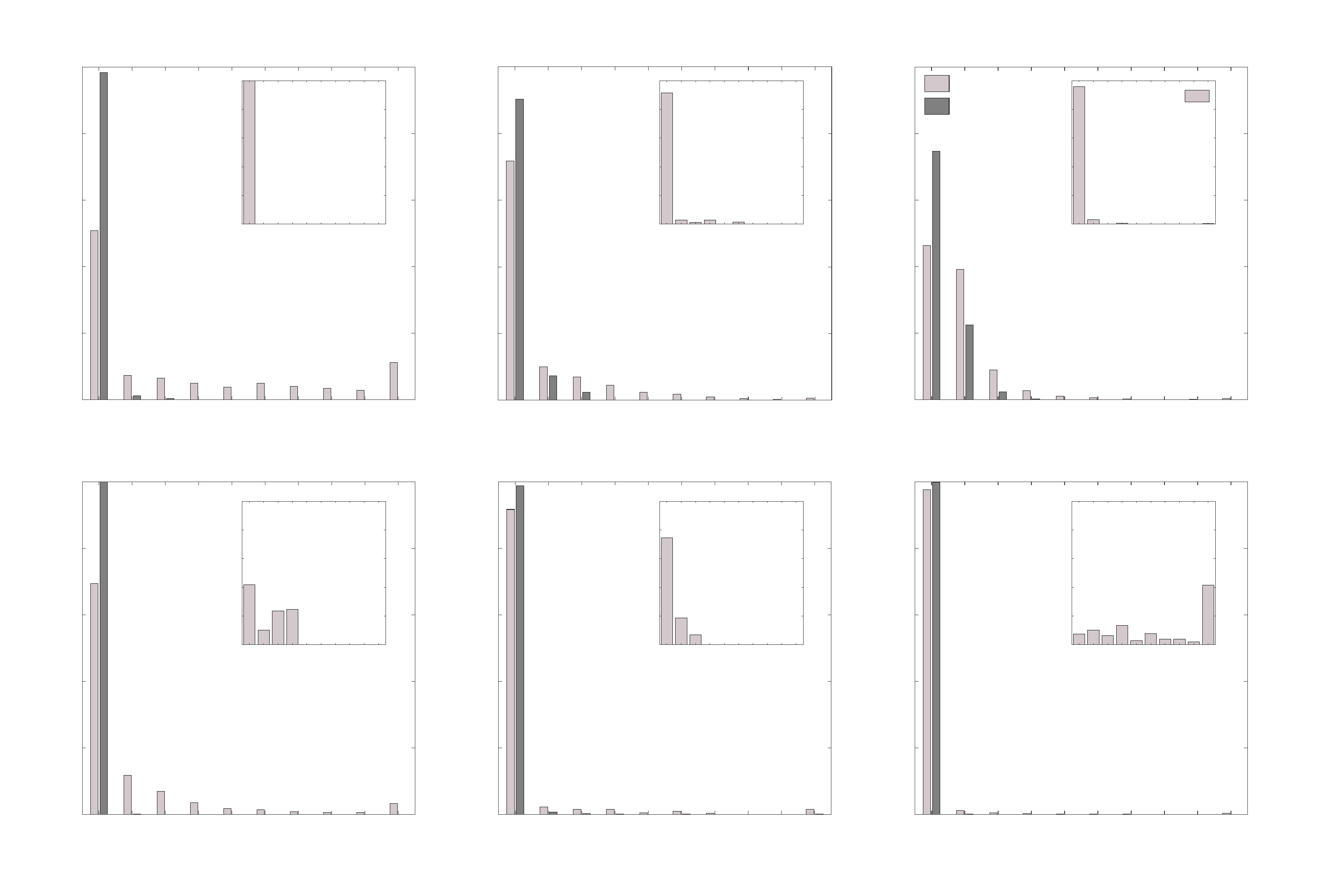}
      \put(-2.84,6.06){\scriptsize $C^{\rm q}(\mathbf{I}_{\rm real})$}
      \put(-2.84,5.9){\scriptsize $C^{\rm q}(\mathbf{I}_{\rm real})$}
      \put(-1.62,5.98){\tiny $C^{\rm pi}(\mathbf{A}^{\rm proj}_{\rm real})$}
      \put(-10.4,6.15){\small $(a)$}
      \put(-6.7,6.15){\small $(b)$}
      \put(-3.5,6.15){\small $(c)$}
      \put(-10.4,3.05){\small $(d)$}
      \put(-6.7,3.05){\small $(e)$}
      \put(-3.5,3.05){\small $(f)$}
   \put(-10.4,5.0){\small $P(C^{\rm q};\mathbf{I}_{\rm real})$,}
      \put(-10.4,4.8){\small $\langle P(C^{\rm q};\mathbf{I})\rangle$}
  \put(-10.4,1.9){\small $P(C^{\rm q};\mathbf{I}_{\rm real})$,}
      \put(-10.4,1.7){\small $\langle P(C^{\rm q};\mathbf{I})\rangle$}
      \put(-5,0.15){\small $C^{\rm q}$}
      \put(-7.8,4.76){\footnotesize$C^{\rm pi}$}
      \put(-9,5.65){\footnotesize$P(C^{\rm pi};\mathbf{A}^{\rm proj}_{\rm real})$}
      \put(-8.18,4.92){\scriptsize$0.1$}
      \put(-7.75,4.92){\scriptsize$0.5$}
      \put(-7.18,4.92){\scriptsize$1$}
      \put(-8.3,5.02){\scriptsize$0$}
      \put(-8.4,5.425){\scriptsize$0.4$}
      \put(-8.4,5.875){\scriptsize$0.8$}
      \put(-8.3,6.08){\scriptsize$1$}
      \put(-5.05,4.92){\scriptsize$0.1$}
      \put(-4.62,4.92){\scriptsize$0.5$}
      \put(-4.03,4.92){\scriptsize$1$}
      \put(-5.17,5.02){\scriptsize$0$}
      \put(-5.27,5.425){\scriptsize$0.4$}
      \put(-5.27,5.875){\scriptsize$0.8$}
      \put(-5.17,6.08){\scriptsize$1$}
      \put(-1.94,4.92){\scriptsize$0.1$}
      \put(-1.51,4.92){\scriptsize$0.5$}
      \put(-0.93,4.92){\scriptsize$1$}
      \put(-2.06,5.02){\scriptsize$0$}
      \put(-2.16,5.425){\scriptsize$0.4$}
      \put(-2.16,5.875){\scriptsize$0.8$}
      \put(-2.06,6.08){\scriptsize$1$}
      \put(-8.18,1.75){\scriptsize$0.1$}
      \put(-7.75,1.75){\scriptsize$0.5$}
      \put(-7.18,1.75){\scriptsize$1$}
      \put(-8.3,1.85){\scriptsize$0$}
      \put(-8.4,2.255){\scriptsize$0.4$}
      \put(-8.4,2.705){\scriptsize$0.8$}
      \put(-8.3,2.91){\scriptsize$1$}
      \put(-5.05,1.75){\scriptsize$0.1$}
      \put(-4.62,1.75){\scriptsize$0.5$}
      \put(-4.03,1.75){\scriptsize$1$}
      \put(-5.17,1.85){\scriptsize$0$}
      \put(-5.27,2.255){\scriptsize$0.4$}
      \put(-5.27,2.705){\scriptsize$0.8$}
      \put(-5.17,2.91){\scriptsize$1$}
      \put(-1.94,1.75){\scriptsize$0.1$}
      \put(-1.51,1.75){\scriptsize$0.5$}
      \put(-0.93,1.75){\scriptsize$1$}
      \put(-2.06,1.85){\scriptsize$0$}
      \put(-2.16,2.255){\scriptsize$0.4$}
      \put(-2.16,2.705){\scriptsize$0.8$}
      \put(-2.06,2.91){\scriptsize$1$}
      \put(-9.35,0.4){\footnotesize$0.1$}
      \put(-8.62,0.4){\footnotesize$0.4$}
      \put(-7.86,0.4){\footnotesize$0.7$}
      \put(-7.04,0.4){\footnotesize$1$}
      \put(-6.21,0.4){\footnotesize$0.1$}
      \put(-5.48,0.4){\footnotesize$0.4$}
      \put(-4.72,0.4){\footnotesize$0.7$}
      \put(-3.9,0.4){\footnotesize$1$}
      \put(-3.08,0.4){\footnotesize$0.1$}
      \put(-2.34,0.4){\footnotesize$0.4$}
      \put(-1.57,0.4){\footnotesize$0.7$}
      \put(-0.77,0.4){\footnotesize$1$}
      \put(-9.55,3.7){\footnotesize$0$}
      \put(-9.7,4.15){\footnotesize$0.2$}
      \put(-9.7,4.65){\footnotesize$0.4$}
      \put(-9.7,5.15){\footnotesize$0.6$}
      \put(-9.7,5.65){\footnotesize$0.8$}
      \put(-9.55,6.15){\footnotesize$1$}
      \put(-9.55,0.6){\footnotesize$0$}
      \put(-9.7,1.05){\footnotesize$0.2$}
      \put(-9.7,1.55){\footnotesize$0.4$}
      \put(-9.7,2.05){\footnotesize$0.6$}
      \put(-9.7,2.55){\footnotesize$0.8$}
      \put(-9.55,3.05){\footnotesize$1$}
      \put(-9.35,3.5){\footnotesize$0.1$}
      \put(-8.62,3.5){\footnotesize$0.4$}
      \put(-7.86,3.5){\footnotesize$0.7$}
      \put(-7.04,3.5){\footnotesize$1$}
      \put(-6.21,3.5){\footnotesize$0.1$}
      \put(-5.48,3.5){\footnotesize$0.4$}
      \put(-4.72,3.5){\footnotesize$0.7$}
      \put(-3.9,3.5){\footnotesize$1$}
      \put(-3.08,3.5){\footnotesize$0.1$}
      \put(-2.34,3.5){\footnotesize$0.4$}
      \put(-1.57,3.5){\footnotesize$0.7$}
      \put(-0.77,3.5){\footnotesize$1$}
     \caption{{\it Distribution of quad clustering coefficients in nondirected hypergraphs.}
     Comparison between the distributions $P(C^{\rm q};\mathbf{I}_{\rm real})$   of  quad clustering coefficients in real-world hypergraphs (light grey histograms) and the  average distribution $\langle P(C^{\rm q};\mathbf{I})\rangle$ (dark grey histograms) of the corresponding configuration model with a prescribed degree sequence $\vec{k}(\mathbf{I}_{\rm real})$ and cardinality sequence  $\vec{\chi}(\mathbf{I}_{\rm real})$. The estimate of $\langle P(C^{\rm q};\mathbf{I})\rangle$ has been obtained from   $100$ graph realisations.    The inset  shows the distribution  $P(C^{\rm pi};\mathbf{A}^{\rm proj}_{\rm real})$  of pairwise clustering coefficients in the projected network $\mathbf{A}^{\rm proj}_{\rm real}$ formed from   pairwise interactions obtained with the formula (\ref{eq:AProj}). Note that the distributions  $P(C^{\rm q};\mathbf{I}_{\rm real})$  show a peak at $C^{\rm q}=1$, while the distributions $P(C^{\rm pi};\mathbf{A}^{\rm proj}_{\rm real})$ do not show a peak at $C^{\rm pi}=1$ [except for Hypergraph $(f)$]. The real-world hypergraphs considered are: $(a)$ {\it NDC-substances}, $(b)$ {\it Youtube}, $(c)$ {\it Food recipe}, $(d)$ {\it Github}, $(e)$ {\it Crime involvement} and $(f)$ {\it Wallmart}; see Table~\ref{tb:qC_value}. }
     \label{fig:undquadCdist}
\end{figure*}
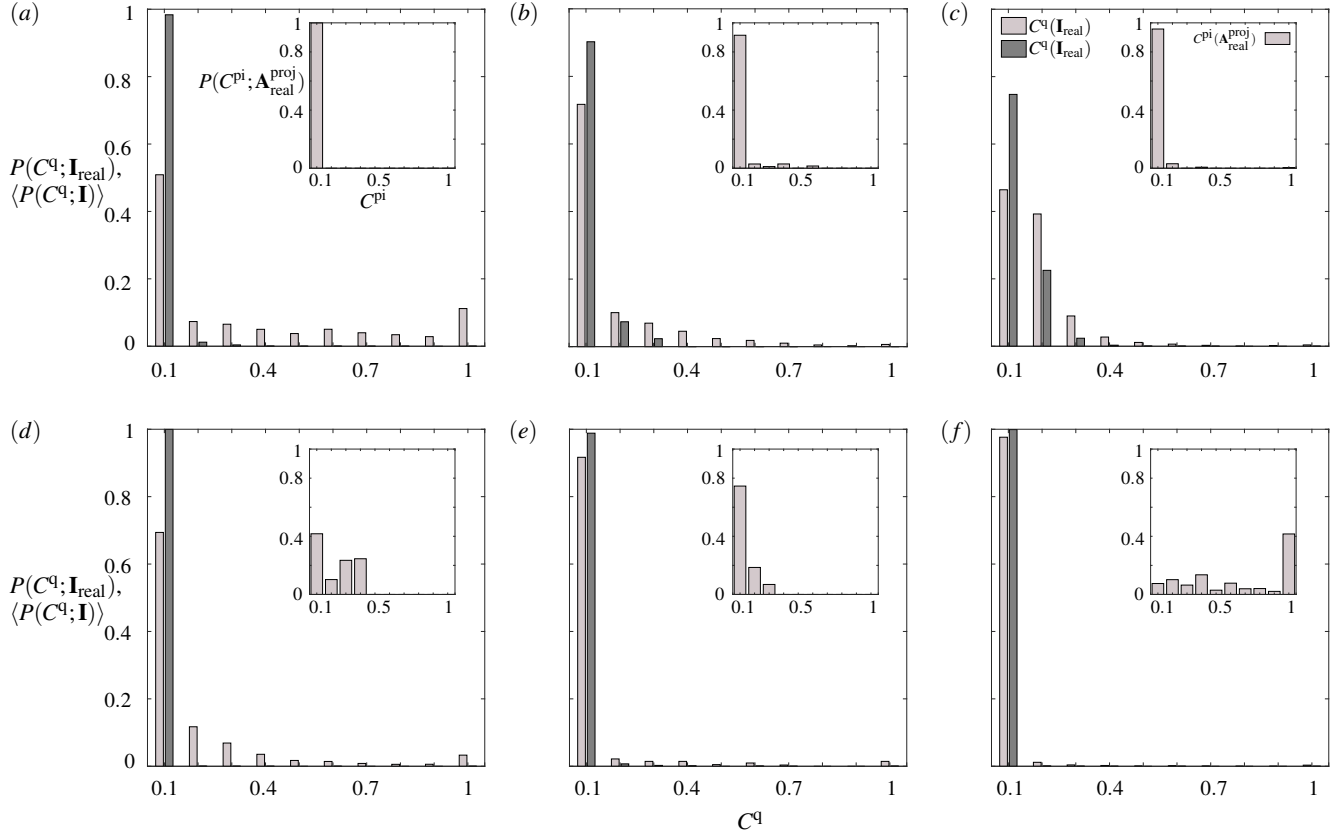

\subsection{Distribution of quad clustering coefficients}

\begin{figure*}
     \centering
     \setlength{\unitlength}{0.1\textwidth}
     \hspace*{0.8cm}
     \begin{subfigure}[b]{0.3\textwidth}
         \centering
         \includegraphics[width=\textwidth]{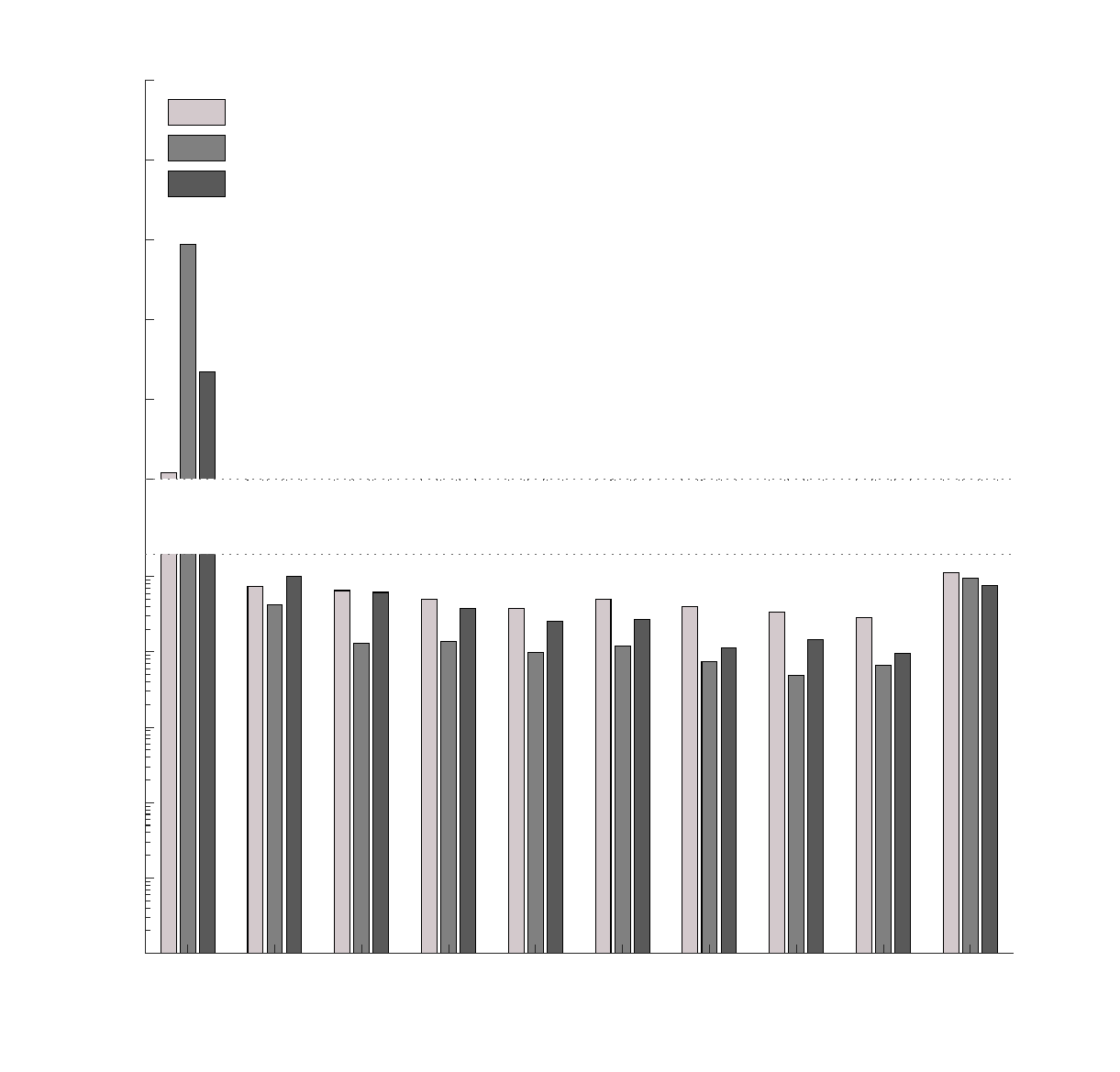}
     \end{subfigure}
      \put(-3.45,2.8){\normalsize$(a)$}
      \put(-2.77,2.6){\footnotesize$1$}
      \put(-2.9,2.39){\footnotesize$0.9$}
      \put(-2.9,1.97){\footnotesize$0.7$}
      \put(-2.9,1.55){\footnotesize$0.5$}
      \put(-2.35,2.32){\scriptsize\color{dgrey} $C=C^{\rm Zhang}$}
      \put(-2.35,2.44){\scriptsize\color{ngrey} $C=C^{\rm Lind}$}
      \put(-2.35,2.56){\scriptsize\color{lgrey} $C=C^{\rm q}$}
      \put(-2.97,1.25){\footnotesize$10^{-1}$}
      \put(-2.97,0.85){\footnotesize$10^{-3}$}
      \put(-2.97,0.45){\footnotesize$10^{-5}$}
      \put(-2.97,0.25){\footnotesize$10^{-6}$}
      \put(-3.45,2.2){\small $P\left(C;\mathbf{I}_{\rm real}\right)$}
      \put(-2.6,0.15){\footnotesize$0.1$}
      \put(-1.9,0.15){\footnotesize$0.4$}
      \put(-1.2,0.15){\footnotesize$0.7$}
      \put(-0.43,0.15){\footnotesize$1$}
     \hfill
     \begin{subfigure}[b]{0.3\textwidth}
         \centering
         \includegraphics[width=\textwidth]{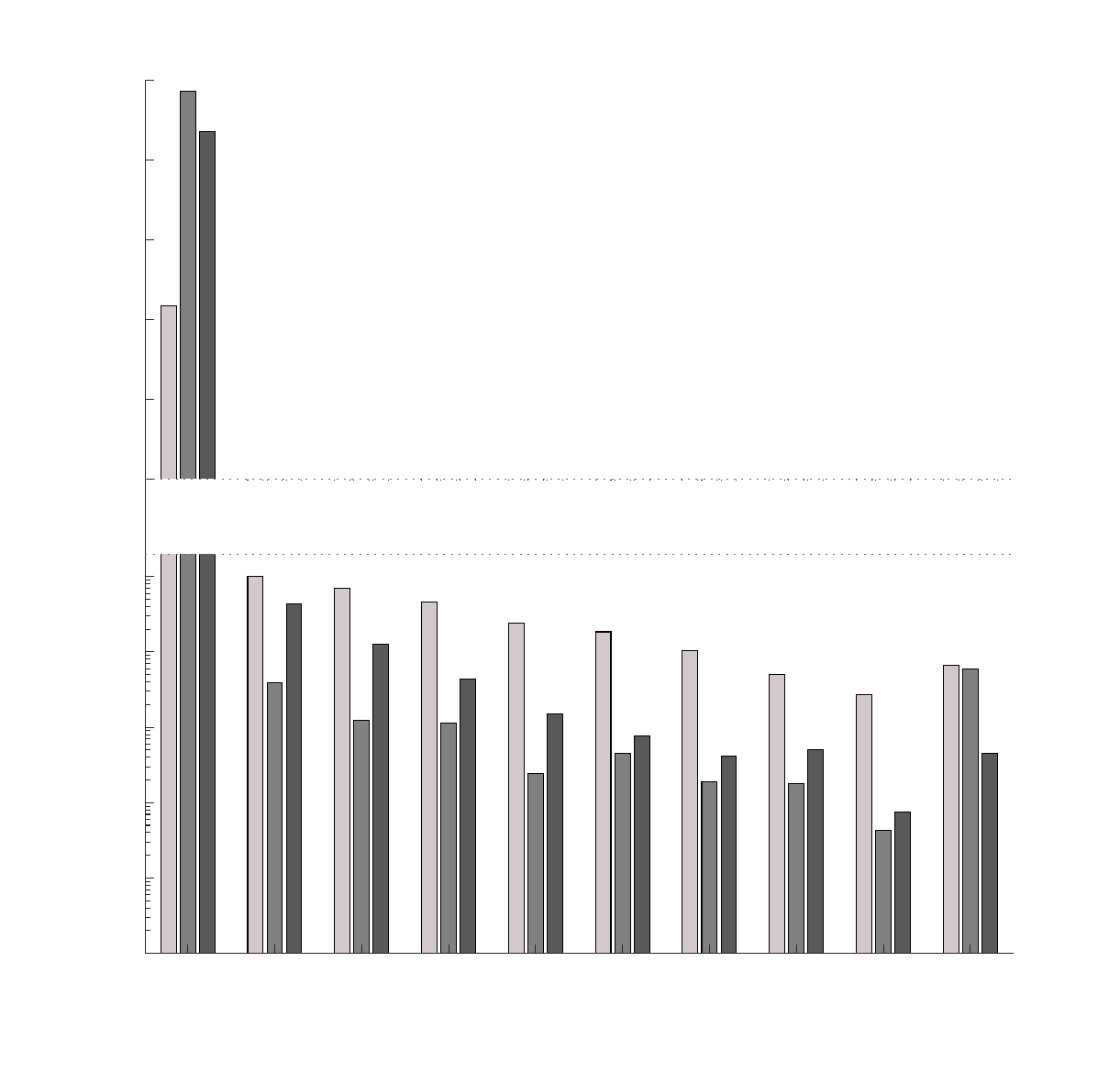}
     \end{subfigure}
      \put(-3.2,2.8){\normalsize$(b)$}
      \put(-2.77,2.6){\footnotesize$1$}
      \put(-2.9,2.39){\footnotesize$0.9$}
      \put(-2.9,1.97){\footnotesize$0.7$}
      \put(-2.9,1.55){\footnotesize$0.5$}
      \put(-2.97,1.25){\footnotesize$10^{-1}$}
      \put(-2.97,0.85){\footnotesize$10^{-3}$}
      \put(-2.97,0.45){\footnotesize$10^{-5}$}
      \put(-2.97,0.25){\footnotesize$10^{-6}$}
      \put(-2.6,0.15){\footnotesize$0.1$}
      \put(-1.9,0.15){\footnotesize$0.4$}
      \put(-1.2,0.15){\footnotesize$0.7$}
      \put(-0.43,0.15){\footnotesize$1$}
     \hfill
     \begin{subfigure}[b]{0.3\textwidth}
         \centering
         \includegraphics[width=\textwidth]{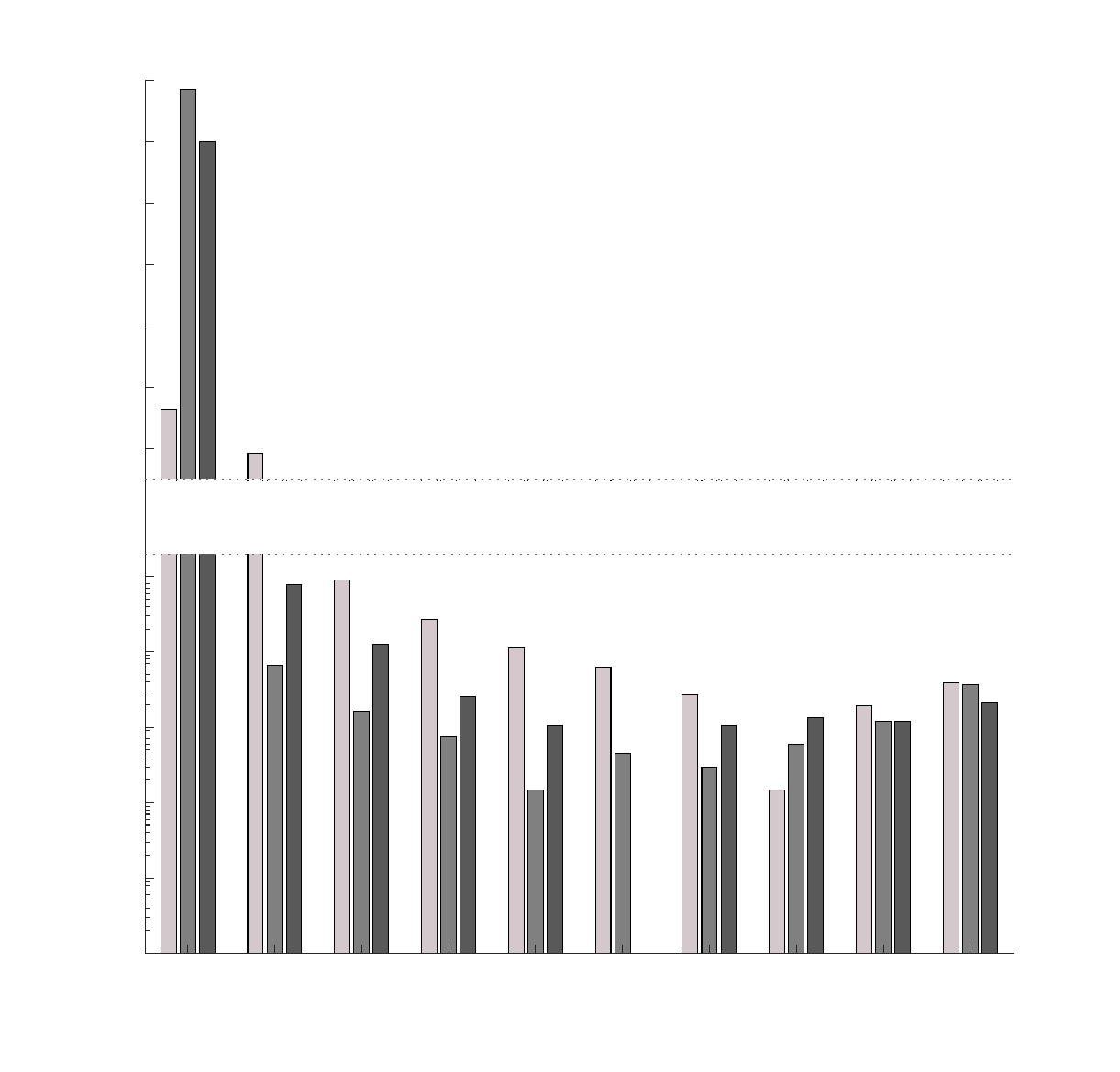}
     \end{subfigure}
      \put(-3.2,2.8){\normalsize$(c)$}
      \put(-2.77,2.6){\footnotesize$1$}
      \put(-2.9,2.29){\footnotesize$0.8$}
      \put(-2.9,1.96){\footnotesize$0.6$}
      \put(-2.9,1.63){\footnotesize$0.4$}
      \put(-2.97,1.25){\footnotesize$10^{-1}$}
      \put(-2.97,0.85){\footnotesize$10^{-3}$}
      \put(-2.97,0.45){\footnotesize$10^{-5}$}
      \put(-2.97,0.25){\footnotesize$10^{-6}$}
      \put(-2.6,0.15){\footnotesize$0.1$}
      \put(-1.9,0.15){\footnotesize$0.4$}
      \put(-1.2,0.15){\footnotesize$0.7$}
      \put(-0.43,0.15){\footnotesize$1$}
     \hfill
     \hspace*{0.8cm}
     \begin{subfigure}[b]{0.3\textwidth}
         \centering
         \includegraphics[width=\textwidth]{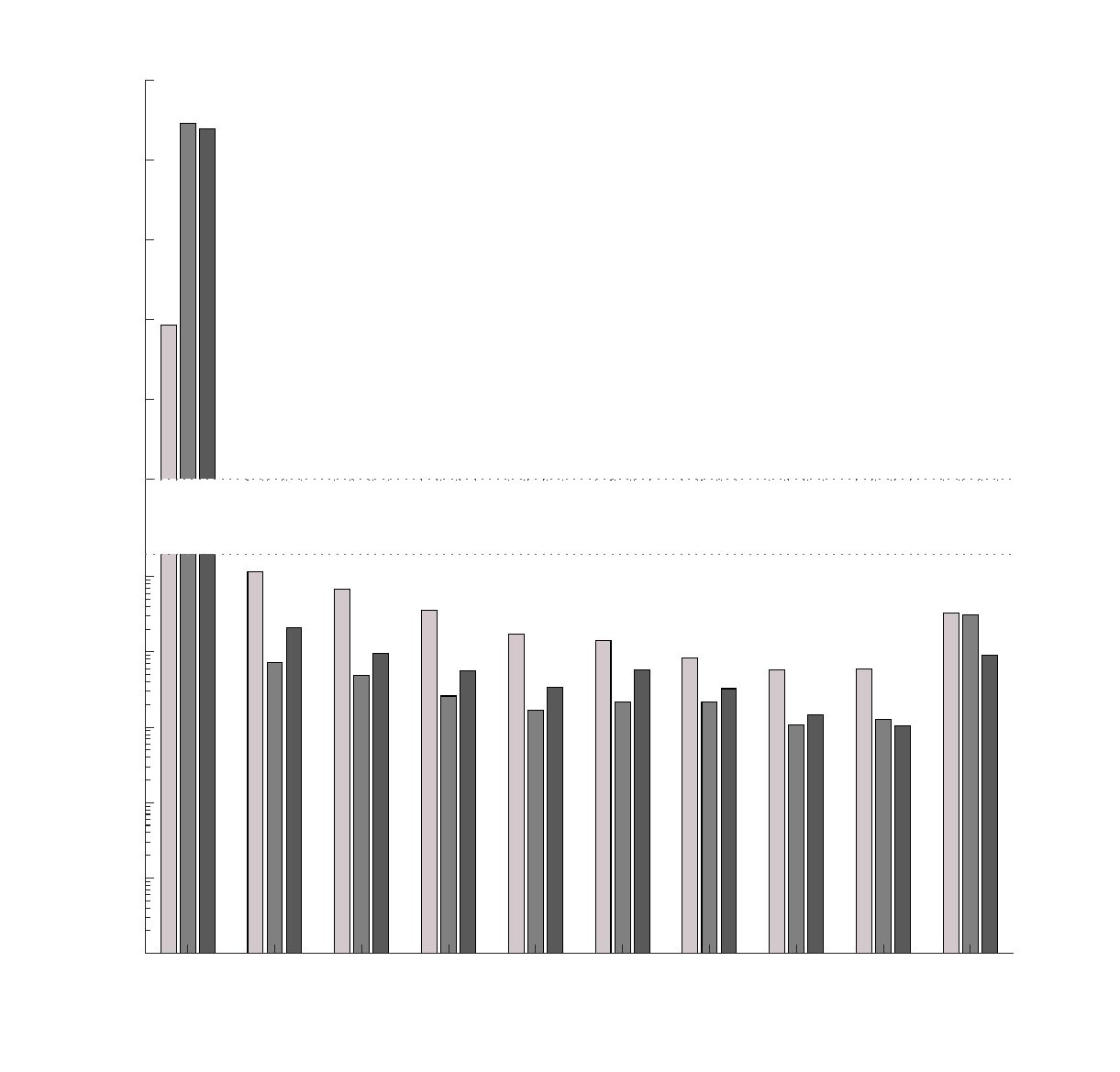}
     \end{subfigure}
      \put(-3.45,2.8){\normalsize$(d)$}
      \put(-2.77,2.6){\footnotesize$1$}
      \put(-2.9,2.39){\footnotesize$0.9$}
      \put(-2.9,1.97){\footnotesize$0.7$}
      \put(-2.9,1.55){\footnotesize$0.5$}
      \put(-2.97,1.25){\footnotesize$10^{-1}$}
      \put(-2.97,0.85){\footnotesize$10^{-3}$}
      \put(-2.97,0.45){\footnotesize$10^{-5}$}
      \put(-2.97,0.25){\footnotesize$10^{-6}$}
      \put(-3.45,2.2){\small $P\left(C;\mathbf{I}_{\rm real}\right)$}
      \put(-2.6,0.15){\footnotesize$0.1$}
      \put(-1.9,0.15){\footnotesize$0.4$}
      \put(-1.2,0.15){\footnotesize$0.7$}
      \put(-0.43,0.15){\footnotesize$1$}
     \hfill
     \begin{subfigure}[b]{0.3\textwidth}
         \centering
         \includegraphics[width=\textwidth]{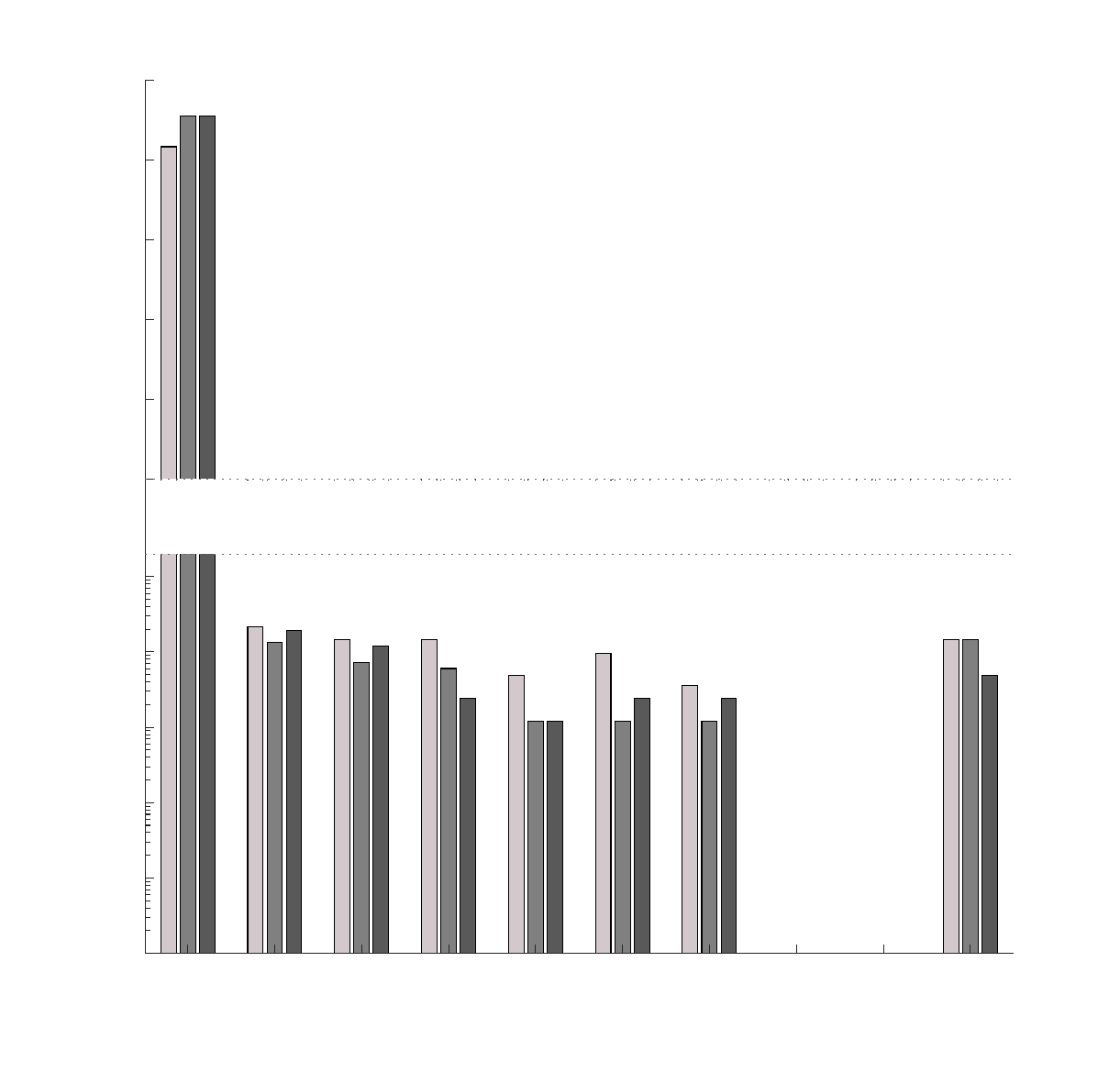}
     \end{subfigure}
      \put(-3.2,2.8){\normalsize$(e)$}
      \put(-2.77,2.6){\footnotesize$1$}
      \put(-2.9,2.39){\footnotesize$0.9$}
      \put(-2.9,1.97){\footnotesize$0.7$}
      \put(-2.9,1.55){\footnotesize$0.5$}
      \put(-2.97,1.3){\footnotesize$10^{-1}$}
      \put(-2.97,0.88){\footnotesize$10^{-3}$}
      \put(-2.97,0.45){\footnotesize$10^{-5}$}
      \put(-2.97,0.25){\footnotesize$10^{-6}$}
      \put(-2.6,0.15){\footnotesize$0.1$}
      \put(-1.9,0.15){\footnotesize$0.4$}
      \put(-1.2,0.15){\footnotesize$0.7$}
      \put(-0.43,0.15){\footnotesize$1$}
      \put(-1.5,0){\small$C^{\rm q}$}
     \hfill
     \begin{subfigure}[b]{0.3\textwidth}
         \centering
         \includegraphics[width=\textwidth]{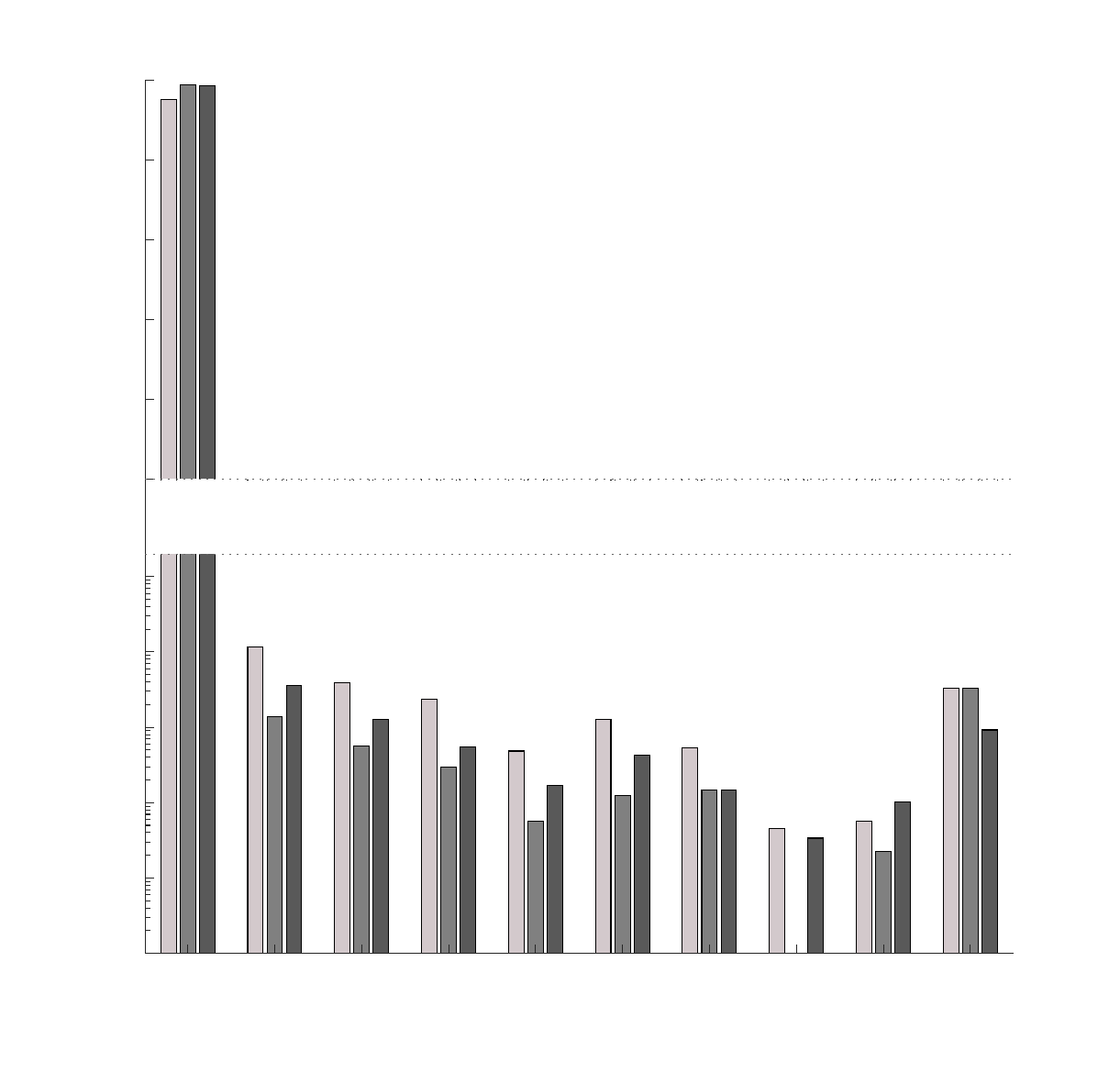}
     \end{subfigure}
      \put(-3.2,2.8){\normalsize$(f)$}
      \put(-2.77,2.6){\footnotesize$1$}
      \put(-2.9,2.39){\footnotesize$0.9$}
      \put(-2.9,1.97){\footnotesize$0.7$}
      \put(-2.9,1.55){\footnotesize$0.5$}
      \put(-2.97,1.3){\footnotesize$10^{-1}$}
      \put(-2.97,0.88){\footnotesize$10^{-3}$}
      \put(-2.97,0.45){\footnotesize$10^{-5}$}
      \put(-2.97,0.25){\footnotesize$10^{-6}$}
      \put(-2.6,0.15){\footnotesize$0.1$}
      \put(-1.9,0.15){\footnotesize$0.4$}
      \put(-1.2,0.15){\footnotesize$0.7$}
      \put(-0.43,0.15){\footnotesize$1$}
        \caption{{\it Comparison of distributions of three clustering coefficients examined in the real-world hypergraphs.}  The light grey histograms represent the distributions of the quad clustering coefficient $P(C^{\rm q};\mathbf{I}_{\rm real})$. The grey bar graphs show the distributions of Lind's clustering coefficient $P(C^{\rm Lind};\mathbf{I}_{\rm real})$. And the dark grey histograms denote the distributions of Zhang's clustering coefficient $P(C^{\rm Zhang};\mathbf{I}_{\rm real})$.   Panels represent  different real-world hypergraphs, as explained in the caption of Fig.~\ref{fig:undquadCdist}.   Note  the discontinuous scale on the y-axis, with a linear scale for $y>0.5$ and a logarithmic scale for $y<0.5$.   }
        \label{fig:compare_quadC_dist}
\end{figure*}

As real-world hypergraphs exhibit a larger  number of quads than expected from random models, we investigate the fluctuations in the quad clustering coefficient.   We quantify the fluctuations of the quad clustering coefficient   by its distribution

\begin{equation}
P(C^{\rm q};
\mathbf{I}) \equiv \frac{1}{N}\sum^N_{i=1}\delta(C^{\rm q}-C^{\rm q}_i(\mathbf{I})).
\end{equation}

Figure~\ref{fig:undquadCdist} shows the distribution $P(C^{\rm q};\mathbf{I}_{\rm real})$  for the six  real-world hypergraphs under study.   We highlight a few  noteworthy features of these plots.   Firstly, a significant proportion of nodes  possess a near zero quad clustering coefficient, viz.,  between 50-70 \% in the Hypergraphs (a)-(d) and over 90\% in the Hypergraphs (e)-(f).   Secondly,  for the remaining nodes the distribution of $C^{\rm q}_i$ is broad.        This latter feature stands in contrast with the average distribution $\langle P(C^{\rm q};\mathbf{I})\rangle$ in the corresponding configuration model with prescribed degree sequence $\vec{k}(\mathbf{I}_{\rm real})$ and cardinality sequence $\vec{\chi}(\mathbf{I}_{\rm real})$, generated by a standard stub-joining algorithm~\cite{coolen2017generating},  also plotted in Fig.~\ref{fig:undquadCdist}.  Thirdly, the hypergraphs in Fig.~\ref{fig:undquadCdist} exhibit a peak at 
$C^{\rm q}\approx 1$, which is most  clearly visible in the NDC-substances hypergraph (a) and the Github hypergraph (hypergraph (d)).

As discussed in Sec.~\ref{sec:defNot}, quad clustering can also be quantified with the Lind and Zhang clustering coefficients.    As shown in Fig.~\ref{fig:compare_quadC_dist}, 
the peak at $C^{\rm q}\approx 1$ also appears when quantifying quad clustering with the Lind clustering coefficient or the Zhang clustering. However, the distributions $P(C^{\rm Lind};\mathbf{I}_{\rm real})$ and $P(C^{\rm Zhang};\mathbf{I}_{\rm real})$ have a larger peak at the origin, while the number of nodes with an intermediate value (not zero or one) is smaller.   This result is consistent with the nonlinearity observed in Fig.~\ref{fig:clustering_landscape}.  Indeed, since the $C^{\rm Lind}$  and $C^{\rm Zhang}$ clustering coefficients are nonlinear, nodes accumulate at values $C^{\rm Lind}\approx 0,1$  and $C^{\rm Zhang}\approx 0,1$, and hence these clustering coefficients are less effective at discriminating nodes based on their density of quads.

Importantly,  disregarding  for now Hypergraph  (f) on which we come back later, the peak at $C^{\rm q}(\mathbf{I})\approx 1$ peak is not captured by the pairwise clustering coefficient evaluated on the corresponding projected graphs  represented by $\mathbf{A}^{\rm proj}$.    Indeed,  as shown in the inset of Figure~\ref{fig:undquadCdist}, the
\begin{equation}
P(C^{\rm pi};
\mathbf{A}^{\rm proj}) \equiv \frac{1}{N}\sum^N_{i=1}\delta(C^{\rm pi}-C^{\rm pi}_i(\mathbf{A}^{\rm proj}))
\end{equation}
where $\mathbf{A}^{\rm proj}$ is the adjacency matrix of the projected graph as defined in (\ref{eq:AProj}), does not exhibit a peak at large valuees.    Hence quad clustering captures a characteristic  distinct to  hypergraphs and that is not captured by pairwise clustering coefficients.

 \begin{figure*}
     \centering
     \setlength{\unitlength}{0.1\textwidth}
     \includegraphics[width=0.9\textwidth]{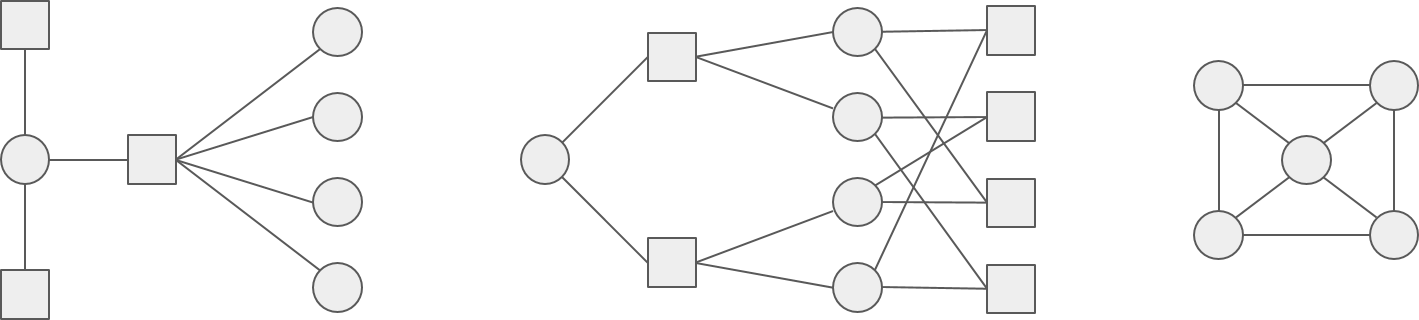}
      \put(-9.5,2.25){\small $(a)$}
      \put(-6.2,2.25){\small $(b)$}
      \put(-1.9,2.25){\small $(c)$}
      \put(-8.88,0.95){\small $i$}
      \put(-6.92,1.75){\small $1$}
      \put(-6.92,1.22){\small $2$}
      \put(-6.92,0.68){\small $3$}
      \put(-6.92,0.13){\small $4$}
      \put(-5.58,0.95){\small $i$}
      \put(-3.62,1.75){\small $1$}
      \put(-3.62,1.22){\small $2$}
      \put(-3.62,0.68){\small $3$}
      \put(-3.62,0.13){\small $4$}
      \put(-0.770,0.95){\small $i$}
      \put(-1.33,1.4){\small $1$}
      \put(-0.22,1.4){\small $2$}
      \put(-0.22,0.46){\small $3$}
      \put(-1.33,0.46){\small $4$}
     \caption{Illustration of  motifs   in the Walmart network (Hypergraph (f)) centered around nodes $i$  for which both $C^{\rm q}_i=0$ and $C^{\rm pi}_i=1$.  Panel (a): motif with degree $\sum^{\infty}_{\chi=3}k_i(\mathbf{I};\chi)=1$;  Panel (b): motif with degree $k^\ast_i>1$, but nevertheless $C^{\rm q}_i=0$ and $C^{\rm pi}_i=1$; Panel (c): projected graph for the networks illustrated in Panels (a) and (b), yielding $C^{\rm pi}_i=1$.}
     \label{fig:wallmart_pattern}
\end{figure*}
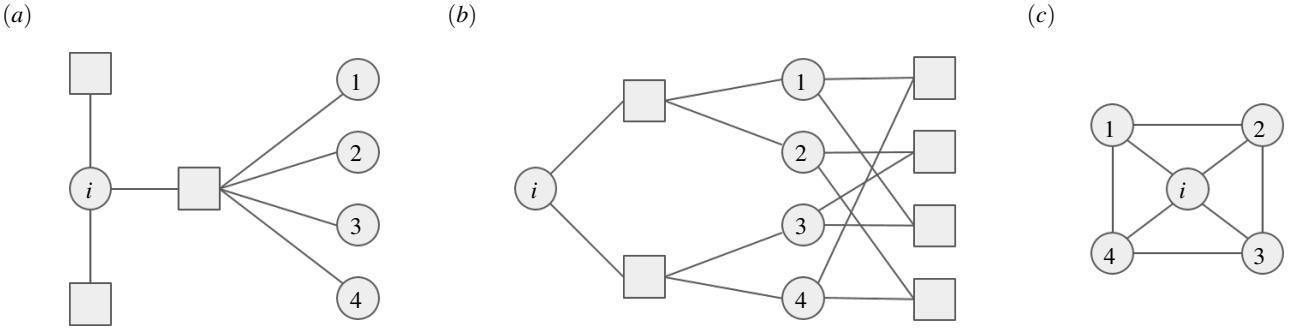

As shown in Fig.~\ref{fig:undquadCdist},  Hypergraph (f),  exhibits clustering properties that are different from those of  the other networks.    Specifically, Hypergraph (f) exhibits a peak at $1$ in the distribution of pairwise clustering coefficients of the projected graph, and does not have a peak  at $1$ observed in the distribution of quad clustering coefficients.  
To understand this peculiar property  of Hypergraph (f),  we examine the network motifs formed by the nodes $i$ for which it holds that both  $C^{\rm q}_i<0.5$ and  $C^{\rm pi}_{i}>0.8$ (a total of $38,520$ nodes out of the $88,860$ satisfy this condition).   We have found two type of structures among such  nodes: In particular, $75\%$ of the nodes have $\sum^{\infty}_{\chi=3}k_i(\mathbf{I};\chi) = 1$,
and hence their quad clustering coefficient equals zero and their pairwise clustering coefficient equals one; see   Fig.~\ref{fig:wallmart_pattern}(a) for an illustration of such a motif.  The  remaining $25\%$ of the nodes have a structure similar to those in Fig.~\ref{fig:wallmart_pattern}(b): the neighbourhoods of the hyperedges incident to the central node are disjoint when we exclude the central node.   However,  each pair of nodes $j_1,j_2$ that are incident to  hyperedges incident to the central node, are themselves directly connected by a hyperedge.    Consequently, also in this case $C^{\rm q}_i=0$ and $C^{\rm pi}_{i}=1$.     Note that in the real-world examples, the latter motifs are slightly different from those shown in Fig.~\ref{fig:wallmart_pattern}(b), and hence values of $C^{\rm q}_i\in [0,0.5]$ and  $C^{\rm pi}_{i}\in [0.8,1]$ are observed.

\subsection{Quad clustering coefficients as a function of  degree and cardinality}
In this Subsection, we make a study of the topological properties of nodes that have a large quad clustering coefficient $C^{\rm q}_i\approx 1$. 

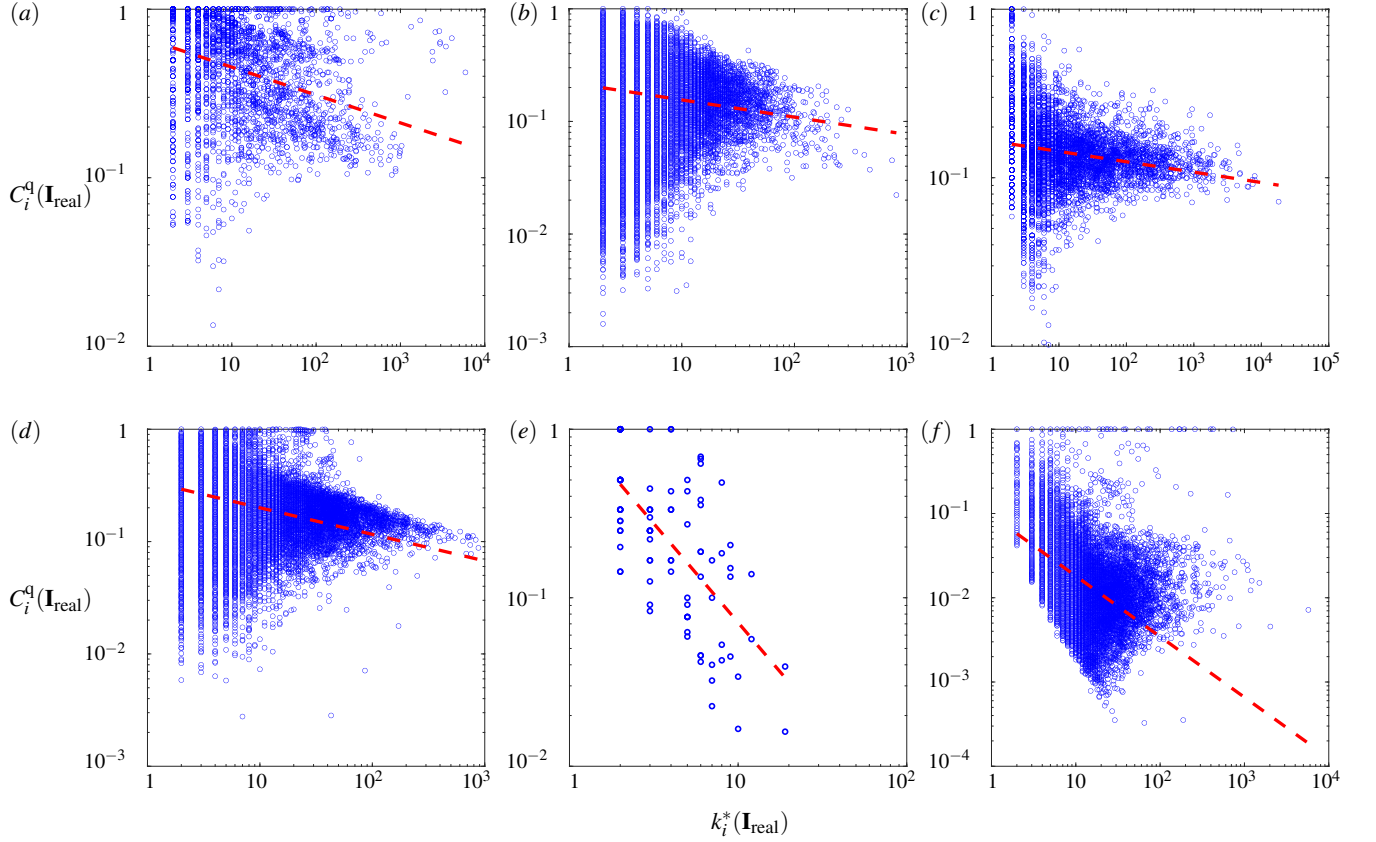
\begin{figure*}
     \centering
     \setlength{\unitlength}{0.1\textwidth}
     \hspace*{0.8cm}
     \includegraphics[width=\textwidth]{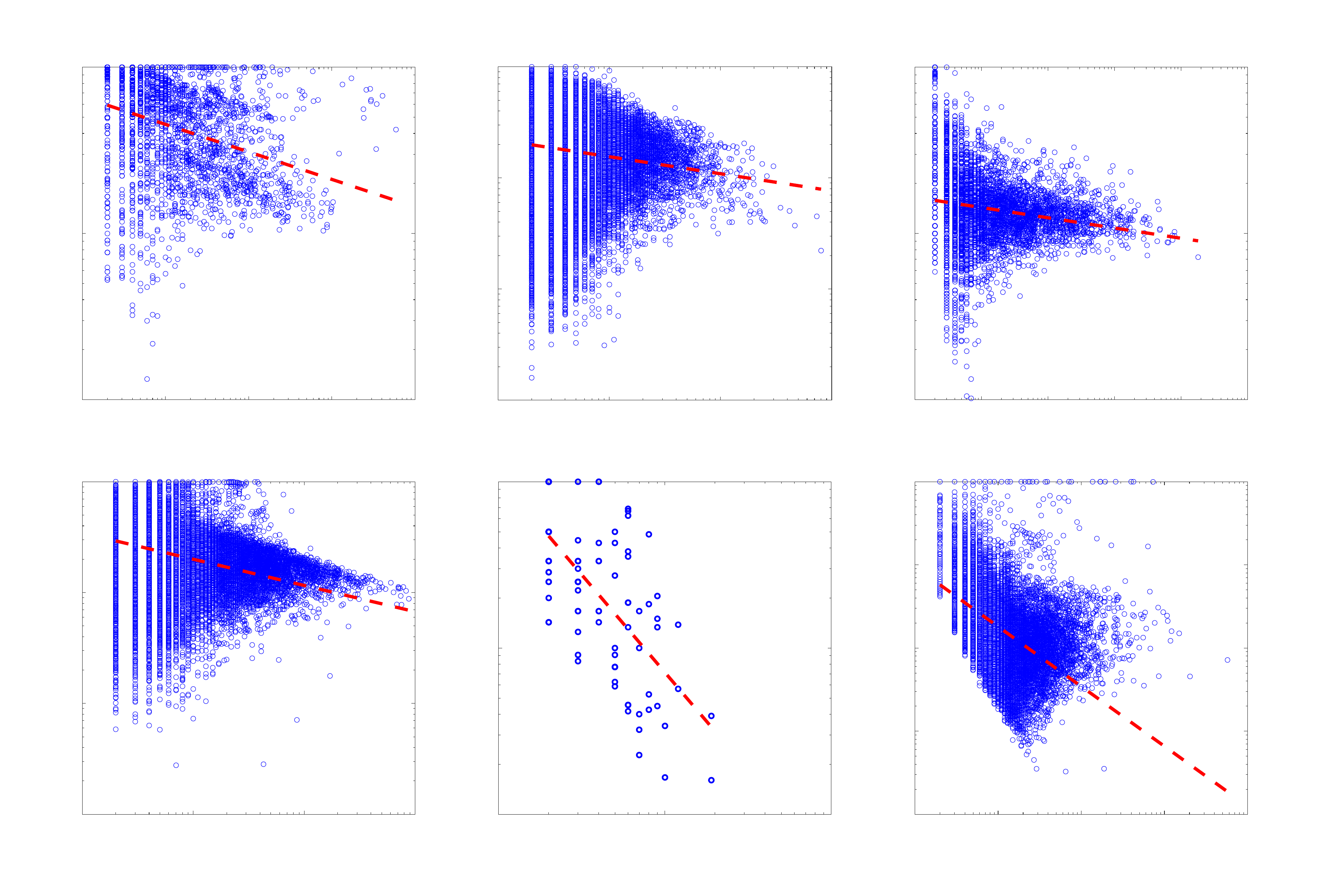}
      \put(-10.4,6.15){\small $(a)$}
      \put(-6.7,6.15){\small $(b)$}
      \put(-3.65,6.15){\small $(c)$}
      \put(-10.4,3.05){\small $(d)$}
      \put(-6.7,3.05){\small $(e)$}
      \put(-3.65,3.05){\small $(f)$}
      \put(-10.4,4.8){\small $C^{\rm q}_{i}(\mathbf{I}_{\rm real})$}
      \put(-10.4,1.8){\small $C^{\rm q}_{i}(\mathbf{I}_{\rm real})$}
      \put(-5.2,0.15){\small $k^{\ast}_{i}(\mathbf{I}_{\rm real})$}
      \put(-9.4,0.45){\footnotesize$1$}
      \put(-8.65,0.45){\footnotesize$10$}
      \put(-7.85,0.45){\footnotesize$10^{2}$}
      \put(-7.05,0.45){\footnotesize$10^3$}
      \put(-6.25,0.45){\footnotesize$1$}
      \put(-5.1,0.45){\footnotesize$10$}
      \put(-3.9,0.45){\footnotesize$10^2$}
      \put(-3.15,0.45){\footnotesize$1$}
      \put(-2.575,0.45){\footnotesize$10$}
      \put(-2.0,0.45){\footnotesize$10^2$}
      \put(-1.375,0.45){\footnotesize$10^3$}
      \put(-0.75,0.45){\footnotesize$10^4$}
      \put(-9.4,3.55){\footnotesize$1$}
      \put(-8.85,3.55){\footnotesize$10$}
      \put(-8.25,3.55){\footnotesize$10^2$}
      \put(-7.65,3.55){\footnotesize$10^3$}
      \put(-7.05,3.55){\footnotesize$10^4$}
      \put(-6.3,3.55){\footnotesize$1$}
      \put(-5.5,3.55){\footnotesize$10$}
      \put(-4.75,3.55){\footnotesize$10^2$}
      \put(-3.9,3.55){\footnotesize$10^3$}
      \put(-3.15,3.55){\footnotesize$1$}
      \put(-2.7,3.55){\footnotesize$10$}
      \put(-2.25,3.55){\footnotesize$10^2$}
      \put(-1.75,3.55){\footnotesize$10^3$}
      \put(-1.25,3.55){\footnotesize$10^4$}
      \put(-0.75,3.55){\footnotesize$10^5$}
      \put(-9.85,3.7){\footnotesize$10^{-2}$}
      \put(-9.85,4.95){\footnotesize$10^{-1}$}
      \put(-9.6,6.15){\footnotesize$1$}
      \put(-6.72,3.7){\footnotesize$10^{-3}$}
      \put(-6.72,4.5){\footnotesize$10^{-2}$}
      \put(-6.72,5.35){\footnotesize$10^{-1}$}
      \put(-6.4,6.15){\footnotesize$1$}
      \put(-3.59,3.7){\footnotesize$10^{-2}$}
      \put(-3.59,4.925){\footnotesize$10^{-1}$}
      \put(-3.3,6.15){\footnotesize$1$}
      \put(-9.85,0.6){\footnotesize$10^{-3}$}
      \put(-9.85,1.4){\footnotesize$10^{-2}$}
      \put(-9.85,2.25){\footnotesize$10^{-1}$}
      \put(-9.6,3.05){\footnotesize$1$}
      \put(-6.72,0.6){\footnotesize$10^{-2}$}
      \put(-6.72,1.8){\footnotesize$10^{-1}$}
      \put(-6.4,3.05){\footnotesize$1$}
      \put(-3.59,0.6){\footnotesize$10^{-4}$}
      \put(-3.59,1.2){\footnotesize$10^{-3}$}
      \put(-3.59,1.8){\footnotesize$10^{-2}$}
      \put(-3.59,2.45){\footnotesize$10^{-1}$}
      \put(-3.3,3.05){\footnotesize$1$}
        \caption{{\it Scatter plots  constructed from the  pairs   $(k^\ast_i(\mathbf{I}_{\rm real}), C^{\rm q}_i(\mathbf{I}_{\rm real}))$ of all nodes  $i\in \V(\mathbf{I}_{\rm real})$ in the canonical, real-world hypergraphs.}  
       The lines are a fit to $C^{\rm q}\sim \left(k^\ast\right)^{-\beta}$ with the fitted values for $\beta$ and their  $95\%$ confidence intervals  equal to   $0.17\pm0.01$ $(a)$, $0.15\pm0.01$  $(b)$, $0.06\pm0.01$  $(c)$, $0.24\pm0.01$  $(d)$, $1.2\pm0.2$  $(e)$, and $0.72\pm0.02$ $(f)$.
        Panels represent  different real-world hypergraphs, as explained in the caption of Fig.~\ref{fig:undquadCdist}.  }
\label{fig:k_quadC_scatter}
\end{figure*}

First, we address the  correlations between $C^{\rm q}_i(\mathbf{I}_{\rm real})$ and the modified degree $k^\ast_i(\mathbf{I}_{\rm real})$, as defined in Eq.~(\ref{def:kAst}).    We consider the modified degree $k^\ast_i$ instead of the degree $k_i$, as by default  hyperedges with  unit cardinality  do  not contribute  to the quad clustering coefficient.  In Fig.~\ref{fig:k_quadC_scatter} we present  scatter plots containing all the pairs   $(k^\ast_i(\mathbf{I}_{\rm real}), C^{\rm q}_i(\mathbf{I}_{\rm real}))$ for the six canonical real-world  hypergraphs that we consider in this Paper, one marker for each node in the hypergraph.     The red dashed line is a fit to the scaling relation $C^{\rm q}\sim \left(k^\ast\right)^{-\beta}$ and it shows the decreasing trend of the quad clustering with the modified degrees.  This demonstrates that  highly clustered nodes have on average lower degrees than  nodes with small quad clustering coefficients.   Nevertheless, up to modified degrees $k^\ast_i\approx 100$ there exist nodes with   $C^{\rm q}_i(\mathbf{I})\approx 1$, and hence real-world hypergraphs contain highly clustered nodes that have  large degrees.    This result is  surprising, as  the denominator of the quad clustering coefficient  increases fast as a function of $k_i$, see Eqs.~(\ref{def:CQuad}) and (\ref{eq:qimax2}), hence one may have expected that the highly clustered nodes with $C^{\rm q}_i(\mathbf{I})\approx 1$ consist exclusively of    nodes with small modified degrees.

\begin{figure*}
     \centering
     \setlength{\unitlength}{0.1\textwidth}
     \hspace*{0.8cm}
     \includegraphics[width=\textwidth]{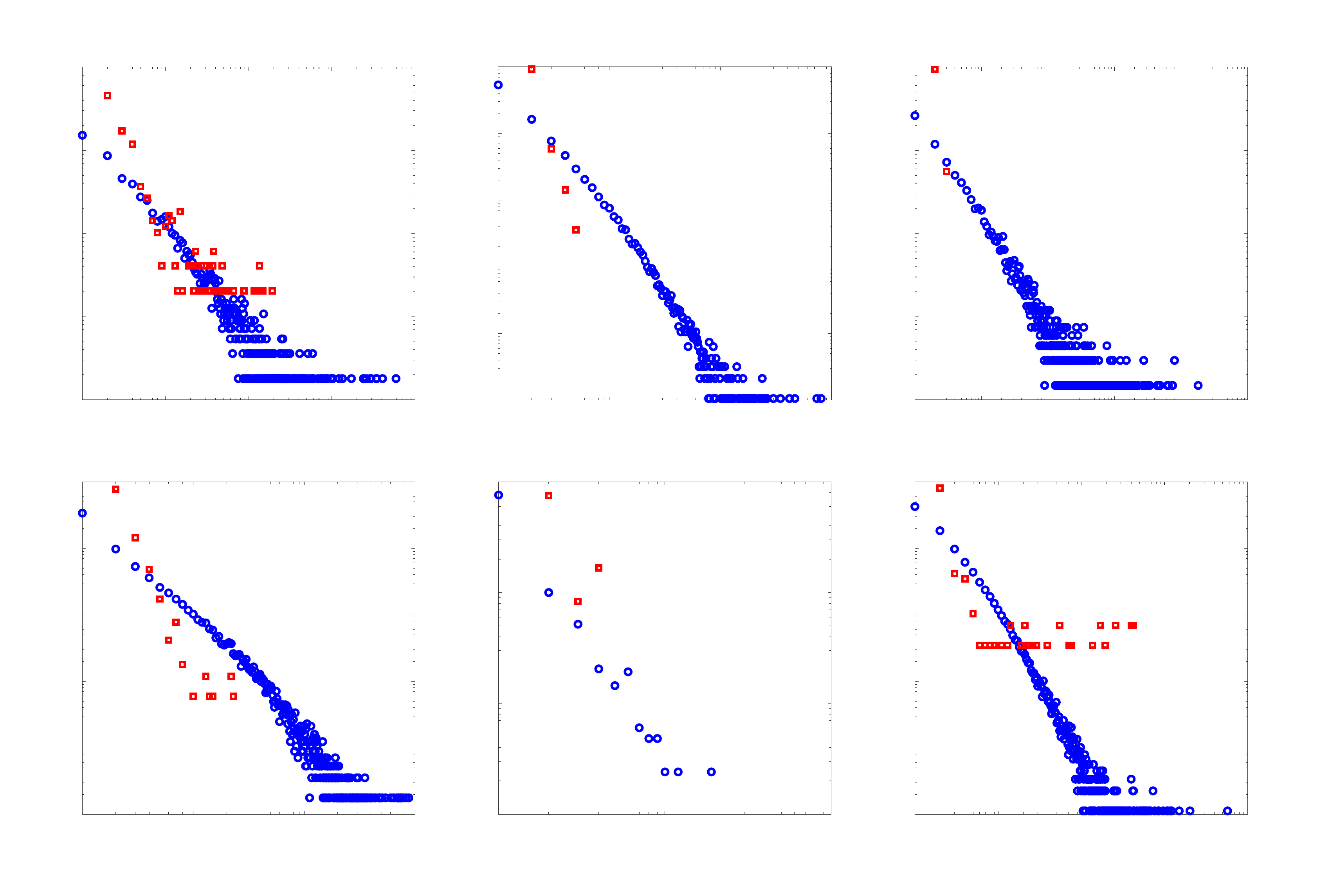}
      \put(-10.4,6.15){\small $(a)$}
      \put(-6.7,6.15){\small $(b)$}
      \put(-3.65,6.15){\small $(c)$}
      \put(-10.4,3.05){\small $(d)$}
      \put(-6.7,3.05){\small $(e)$}
      \put(-3.65,3.05){\small $(f)$}
      \put(-10.4,5.15){\small $P\left(k^{\ast};\mathbf{I}_{\rm real}\right)$}
      \put(-10.4,4.8){\small $P(k^{\ast}|C^{\rm q}=1;$}
      \put(-9.75,4.57){\small $\mathbf{I}_{\rm real})$}
      \put(-10.4,2.3){\small $P\left(k^{\ast};\mathbf{I}_{\rm real}\right)$}
      \put(-10.4,1.95){\small $P(k^{\ast}|C^{\rm q}=1;$}
      \put(-9.75,1.75){\small $\mathbf{I}_{\rm real})$}
      \put(-5,0.15){\small $k^{\ast}$}
      \put(-9.4,0.45){\footnotesize$1$}
      \put(-8.63,0.45){\footnotesize$10$}
      \put(-7.8,0.45){\footnotesize$10^{2}$}
      \put(-7.05,0.45){\footnotesize$10^3$}
      \put(-6.25,0.45){\footnotesize$1$}
      \put(-5.1,0.45){\footnotesize$10$}
      \put(-3.9,0.45){\footnotesize$10^2$}
      \put(-3.15,0.45){\footnotesize$1$}
      \put(-2.575,0.45){\footnotesize$10$}
      \put(-1.97,0.45){\footnotesize$10^2$}
      \put(-1.375,0.45){\footnotesize$10^3$}
      \put(-0.75,0.45){\footnotesize$10^4$}
      \put(-9.4,3.55){\footnotesize$1$}
      \put(-8.85,3.55){\footnotesize$10$}
      \put(-8.25,3.55){\footnotesize$10^2$}
      \put(-7.65,3.55){\footnotesize$10^3$}
      \put(-7.05,3.55){\footnotesize$10^4$}
      \put(-6.3,3.55){\footnotesize$1$}
      \put(-5.5,3.55){\footnotesize$10$}
      \put(-4.75,3.55){\footnotesize$10^2$}
      \put(-3.9,3.55){\footnotesize$10^3$}
      \put(-3.15,3.55){\footnotesize$1$}
      \put(-2.7,3.55){\footnotesize$10$}
      \put(-2.22,3.55){\footnotesize$10^2$}
      \put(-1.74,3.55){\footnotesize$10^3$}
      \put(-1.25,3.55){\footnotesize$10^4$}
      \put(-0.75,3.55){\footnotesize$10^5$}
      \put(-9.85,3.7){\footnotesize$10^{-4}$}
      \put(-9.85,4.35){\footnotesize$10^{-3}$}
      \put(-9.85,4.95){\footnotesize$10^{-2}$}
      \put(-9.85,5.55){\footnotesize$10^{-1}$}
      \put(-9.6,6.15){\footnotesize$1$}
      \put(-6.72,3.7){\footnotesize$10^{-5}$}
      \put(-6.72,4.15){\footnotesize$10^{-4}$}
      \put(-6.72,4.67){\footnotesize$10^{-3}$}
      \put(-6.72,5.15){\footnotesize$10^{-2}$}
      \put(-6.72,5.65){\footnotesize$10^{-1}$}
      \put(-6.4,6.15){\footnotesize$1$}
      \put(-3.59,3.7){\footnotesize$10^{-4}$}
      \put(-3.59,4.3125){\footnotesize$10^{-3}$}
      \put(-3.59,4.925){\footnotesize$10^{-2}$}
      \put(-3.59,5.5375){\footnotesize$10^{-1}$}
      \put(-3.3,6.15){\footnotesize$1$}
      \put(-9.85,0.6){\footnotesize$10^{-5}$}
      \put(-9.85,1.07){\footnotesize$10^{-4}$}
      \put(-9.85,1.57){\footnotesize$10^{-3}$}
      \put(-9.85,2.07){\footnotesize$10^{-2}$}
      \put(-9.85,2.58){\footnotesize$10^{-1}$}
      \put(-9.6,3.05){\footnotesize$1$}
      \put(-6.72,0.6){\footnotesize$10^{-3}$}
      \put(-6.72,1.41){\footnotesize$10^{-2}$}
      \put(-6.72,2.25){\footnotesize$10^{-1}$}
      \put(-6.4,3.05){\footnotesize$1$}
      \put(-3.59,0.6){\footnotesize$10^{-5}$}
      \put(-3.59,1.07){\footnotesize$10^{-4}$}
      \put(-3.59,1.58){\footnotesize$10^{-3}$}
      \put(-3.59,2.07){\footnotesize$10^{-2}$}
      \put(-3.59,2.57){\footnotesize$10^{-1}$}
      \put(-3.3,3.05){\footnotesize$1$}
        \caption{{\it Distributions of degrees of highly clustered nodes in real-world hypergraphs, and comparison with the full hypergraph degree distribution.}    The plot shows the   degree distributions $P(k^\ast;\mathbf{I}_{\rm real})$ (blue, circles) and $P(k^\ast|C^{\rm q}=1;\mathbf{I}_{\rm real})$ (red, squares) for the six canonical real-hypergraphs considered in this paper.   The number of nodes with $C^{\rm q}=1$ are 490  $(a)$, 560  $(b)$, 18  $(c)$, 1683  $(d)$, 12  $(e)$, and 288  $(f)$.  Panels represent the different hypergraphs, as explained in the caption of Fig.~\ref{fig:undquadCdist}.   }
        \label{fig:kstar_quadC_scatter}
\end{figure*}

This results is confirmed by 
Fig.~\ref{fig:kstar_quadC_scatter} that compares the distribution
\begin{equation}
P(k^\ast;\mathbf{I}) \equiv \frac{1}{N}\sum^N_{i=1}\delta_{k^\ast,k^\ast_i(\mathbf{I})} 
\end{equation}
of  the  modified degrees $k^\ast_i$ sampled uniformly from the set $\V$ of hypergraph nodes with the distribution 
\begin{equation}
P(k^\ast|C^{\rm q}=1;\mathbf{I}) \equiv \frac{\sum^N_{i=1}\delta_{k^\ast,k^\ast_i(\mathbf{I})} \delta_{C^{\rm q}_i(\mathbf{I}),1}}{\sum^N_{i=1}\delta_{C^{\rm q}_i(\mathbf{I}),1}}
\end{equation}
of nodes that have a clustering coefficient equal to one. As expected, the modified degree  of highly clustered nodes with $C^{\rm q}_i=1$ are concentrated on small values of the modified degrees. Surprisingly, however, in  the real-world hypergraphs (a), (d) and (f), highly clustered nodes can have modified degrees as large as $k^\ast_i\approx 100$.  As an illustration, for the {\it NDC-substances} network, Fig.~\ref{fig:kstar_quadC_scatter}$(a)$, the maximum value of $k^\ast_i$ amongst  nodes with $C^{\rm q}=1$ is $k^\ast_i=192$. This is unexpectedly large, as it implies  that the 192 hyperedges connected to node $i$ form a  fully clustered configuration.

To further describe the topological properties of the neighbourhood sets of highly clustered nodes, we analyse the  cardinalities of  the hyperedges that are incident to a highly clustered node.   We expect that strongly clustered nodes ($C^{\rm q}_i(\mathbf{I})\approx 1$) have neighbouring nodes with small cardinalities, as the denominator in the quad clustering coefficient increases fast as a function of the cardinalities of the neighbouring nodes.  To quantify fluctuations in the cardinalities of hyperedges, we define the joint distribution  
\begin{equation}
  W(k, \chi;\mathbf{I}) \equiv   \frac{\sum^N_{i=1} \sum^M_{\alpha=1}I_{i\alpha}\delta_{k,k_i(\mathbf{I})}\delta_{\chi,\chi_{\alpha}(\mathbf{I})}}{\sum^N_{i=1} \sum^M_{\alpha=1}I_{i\alpha}},
   \end{equation} 
 of degrees and cardinalities of randomly selected links connecting nodes with hyperedges.  Its marginal  distribution
  \begin{equation}
  W^\ast(\chi;\mathbf{I}) =  \frac{ \sum^M_{k=1} W(k, \chi;\mathbf{I})}{\sum^M_{k=1} \sum^{N-1}_{\chi=2} W(k, \chi;\mathbf{I})},\label{eq:Wchix}
   \end{equation}
 quantifies the fluctuations of the  cardinalities of  hyperedges  at the end point of a randomly selected link, and excluding nodes with cardinality one.   

\begin{figure*}
     \centering
     \setlength{\unitlength}{0.1\textwidth}
     \hspace*{0.8cm}
     \includegraphics[width=\textwidth]{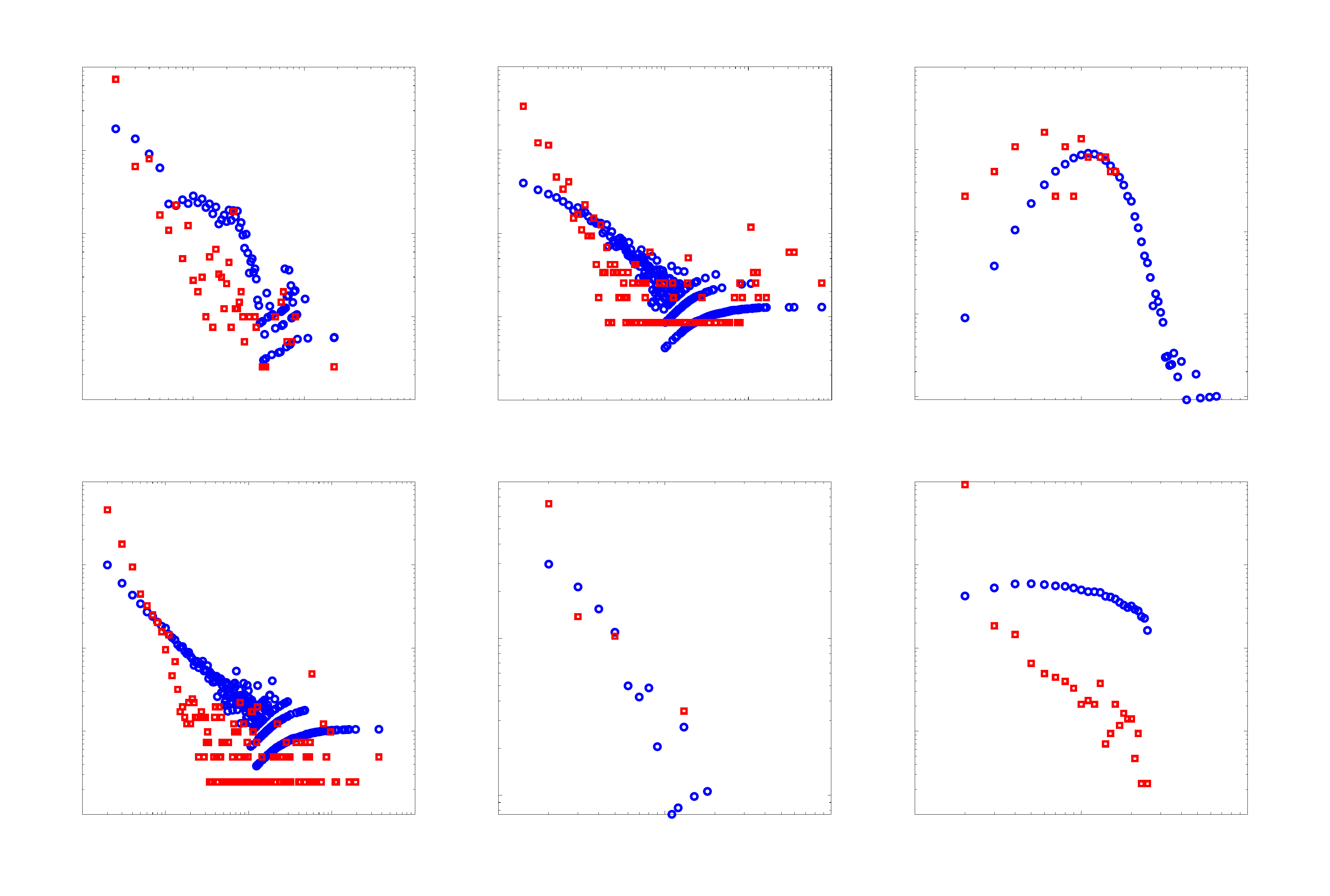}
      \put(-10.45,6.15){\small $(a)$}
      \put(-6.7,6.15){\small $(b)$}
      \put(-3.65,6.15){\small $(c)$}
      \put(-10.45,3.05){\small $(d)$}
      \put(-6.7,3.05){\small $(e)$}
      \put(-3.65,3.05){\small $(f)$}
      \put(-10.45,5.1){\small $W^{\ast}\left(\chi;\mathbf{I}_{\rm real}\right)$}
       \put(-10.45,4.75){\small $W^{\ast}(\chi|C^{\rm q}=1;$}
      \put(-9.75,4.55){\small $\mathbf{I}_{\rm real})$}
      \put(-10.45,2){\small $W^{\ast}\left(\chi;\mathbf{I}_{\rm real}\right)$}
       \put(-10.45,1.65){\small $W^{\ast}(\chi|C^{\rm q}=1;$}
      \put(-9.75,1.45){\small $\mathbf{I}_{\rm real})$}
      \put(-5,0.15){\small $\chi$}
      \put(-9.4,0.4){\footnotesize$1$}
      \put(-8.8375,0.4){\footnotesize$10$}
      \put(-8.275,0.4){\footnotesize$10^2$}
      \put(-7.6625,0.4){\footnotesize$10^3$}
      \put(-7.05,0.4){\footnotesize$10^4$}
      \put(-6.25,0.4){\footnotesize$1$}
      \put(-5.08,0.4){\footnotesize$10$}
      \put(-3.9,0.4){\footnotesize$10^2$}
      \put(-3.15,0.4){\footnotesize$1$}
      \put(-1.95,0.4){\footnotesize$10$}
      \put(-0.75,0.4){\footnotesize$10^2$}
      \put(-9.4,3.5){\footnotesize$1$}
      \put(-8.65,3.5){\footnotesize$10$}
      \put(-7.8,3.5){\footnotesize$10^2$}
      \put(-7.05,3.5){\footnotesize$10^3$}
      \put(-6.3,3.5){\footnotesize$1$}
      \put(-5.75,3.5){\footnotesize$10$}
      \put(-5.125,3.5){\footnotesize$10^2$}
      \put(-4.5,3.5){\footnotesize$10^3$}
      \put(-3.9,3.5){\footnotesize$10^4$}
      \put(-3.15,3.5){\footnotesize$1$}
      \put(-1.95,3.5){\footnotesize$10$}
      \put(-0.75,3.5){\footnotesize$10^2$}
      \put(-9.85,3.7){\footnotesize$10^{-4}$}
      \put(-9.85,4.3125){\footnotesize$10^{-3}$}
      \put(-9.85,4.925){\footnotesize$10^{-2}$}
      \put(-9.85,5.5375){\footnotesize$10^{-1}$}
      \put(-9.6,6.15){\footnotesize$1$}
      \put(-6.72,3.7){\footnotesize$10^{-4}$}
      \put(-6.72,4.3125){\footnotesize$10^{-3}$}
      \put(-6.72,4,925){\footnotesize$10^{-2}$}
      \put(-6.72,5.5375){\footnotesize$10^{-1}$}
      \put(-6.4,6.15){\footnotesize$1$}
      \put(-3.59,3.7){\footnotesize$10^{-4}$}
      \put(-3.59,4.3125){\footnotesize$10^{-3}$}
      \put(-3.59,4.925){\footnotesize$10^{-2}$}
      \put(-3.59,5.5375){\footnotesize$10^{-1}$}
      \put(-3.3,6.15){\footnotesize$1$}
      \put(-9.85,0.6){\footnotesize$10^{-4}$}
      \put(-9.85,1.2){\footnotesize$10^{-3}$}
      \put(-9.85,1.8){\footnotesize$10^{-2}$}
      \put(-9.85,2.425){\footnotesize$10^{-1}$}
      \put(-9.6,3.05){\footnotesize$1$}
      \put(-6.72,0.75){\footnotesize$10^{-2}$}
      \put(-6.72,1.9){\footnotesize$10^{-1}$}
      \put(-6.4,3.05){\footnotesize$1$}
      \put(-3.59,0.6){\footnotesize$10^{-4}$}
      \put(-3.59,1.2){\footnotesize$10^{-3}$}
      \put(-3.59,1.8){\footnotesize$10^{-2}$}
      \put(-3.59,2.425){\footnotesize$10^{-1}$}
      \put(-3.3,3.05){\footnotesize$1$}
        \caption{ {\it Distributions of the cardinalities of hyperedges that are incident to a highly clustered node, and comparison with the corresponding distribution for generic nodes} 
         Comparison between the distributions  $W^\ast(\chi;\mathbf{I}_{\rm real})$ (blue circles)  and $W^\ast(\chi|C^{\rm q}=1;\mathbf{I}_{\rm real})$  (red squares) as defined in Eqs.~(\ref{eq:Wchix}) and (\ref{eq:WAstChiQ1}), respectively, for the six canonical real-world hypergraphs  considered in this Paper.         Panels represent  different real-world hypergraphs, as explained in the caption of Fig.~\ref{fig:undquadCdist}.   }
        \label{fig:chistar_quadC_scatter}
\end{figure*}

 In Fig.~\ref{fig:chistar_quadC_scatter}, we compare   the distribution  $W^\ast(\chi;\mathbf{I})$ with the related distribution $W^\ast(\chi|C^{\rm q}=1;\mathbf{I})$ defined  on nodes with a quad clustering coefficient equal to one.    The latter distribution is defined by 
 \begin{equation}
  W^\ast(\chi|C^{\rm q}=1;\mathbf{I}) =  \frac{ \sum^M_{k=1} W(k, \chi|C^{\rm q}=1;\mathbf{I})}{\sum^M_{k=1} \sum^{N-1}_{\chi=2} W(k, \chi|C^{\rm q}=1;\mathbf{I})}, \label{eq:WAstChiQ1}
   \end{equation}
   where
 \begin{equation}
  W(k, \chi|C^{\rm q}=1;\mathbf{I}) \equiv   \frac{\sum^N_{i=1}\sum^M_{\alpha=1}\delta_{C^{\rm q}_i(\mathbf{I}),1}  I_{i\alpha}\delta_{k,k_i(\mathbf{I})}\delta_{\chi,\chi_{\alpha}(\mathbf{I})}}{\sum^N_{i=1}\sum^M_{\alpha=1}\delta_{C^{\rm q}_i(\mathbf{I}),1} I_{i\alpha}}.
   \end{equation} 
 Interestingly, Fig.~\ref{fig:chistar_quadC_scatter}  reveals that  nodes with $C^{\rm q}_i(\mathbf{I})=1$ can have  a large cardinality $\chi \approx 2000$.  This highlights that the neighbourhood sets of highly clustered nodes can have a large number of quads, as they contain    hyperedges with large cardinality. Comparing Figs.~\ref{fig:kstar_quadC_scatter} and \ref{fig:chistar_quadC_scatter}, we observe that  support of the distribution $W^\ast(\chi;\mathbf{I}_{\rm real})$ is in most cases equal to the support of $W^\ast(\chi|C^{\rm q}=1)$, while   the support of the distribution $P(k^\ast;\mathbf{I}_{\rm real})$  is significantly smaller than the support of $P(k^\ast|C^{\rm q}=1)$.     Hence, highly clustered neighbourhoods  are  more biased towards low degree nodes than towards nodes of low cardinality, which is consistent with the formula (\ref{eq:qimax2}) for the maximal number of quads a node can have.

\section{Quad clustering coefficient for directed hypergraphs} \label{ch:directed_hypergraph}

In this Section we define a quad clustering coefficient for directed hypergraphs and we analyse its properties in real-world directed hypergraphs.
\subsection{Preliminaries on directed hypergraphs}
A directed hypergraph is a quadruplet $\H^{\leftrightarrow}=(\V, \W, \E^{\rm in}, \E^{\rm out})$ consisting of the set $\V$ of $\sizeV=|\V|$ nodes, the set  $\W$ of $\sizeW=|\W|$ hyperedges, and the sets $\E^{\rm in}\subset \V\times \W$ and  $\E^{\rm out}\subset \V\times \W$ of directed inlinks and outlinks, respectively.   Both inlinks and outlinks consist of pairs $(i,\alpha)$ with $i\in \V$ and $\alpha \in \W$, albeit the former represents links directed from a hyperedge to a vertex, while the latter represents links directed from a vertex to a hyperedge. 

We represent simple, directed, hypergraphs with a pair of incidence matrices $\mathbf{I}^{\leftrightarrow} \equiv (\mathbf{I}^{\rightarrow}, \mathbf{I}^{\leftarrow})$  defined by 
\begin{equation}
[\I^{\rightarrow}]_{i\alpha} \equiv \left\{\begin{array}{ccc} 1 &{\rm if}& (i,\alpha)\in \E^{\rm out} ,\\ 0 &{\rm if} & (i,\alpha)\notin \E^{\rm out}\end{array}\right.
\end{equation}
and 
\begin{equation}
[\I^{\leftarrow}]_{i\alpha} \equiv \left\{\begin{array}{ccc} 1 &{\rm if}& (i,\alpha)\in \E^{\rm in} ,\\ 0 &{\rm if} & (i,\alpha)\notin \E^{\rm in}.\end{array}\right.  
\end{equation}  

\begin{figure*}[t]
 \centering
 \setlength{\unitlength}{0.1\textwidth}
 \includegraphics[width=\textwidth]{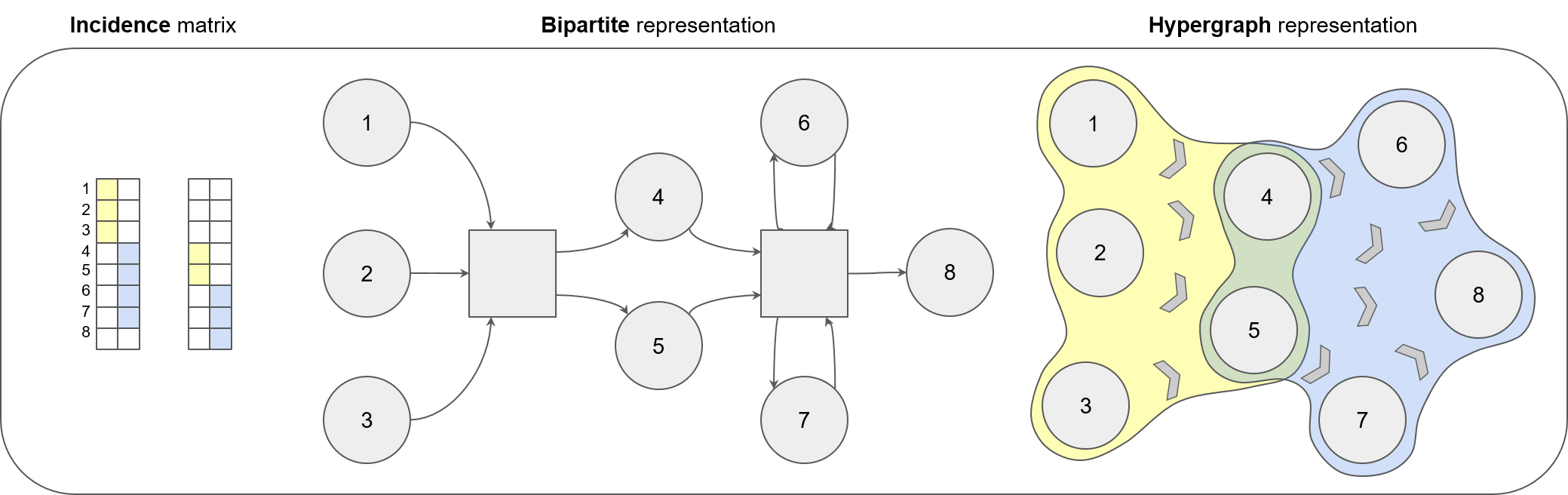}
  \put(-6.8,1.35){\small$\alpha$}
   \put(-4.9,1.35){\small$\beta$}
  \put(-2.5,2.35){\small$\alpha$}
   \put(-1.6,2.3){\small$\beta$}
  \put(-9.37,2.05){\footnotesize$\alpha$}
   \put(-9.23,2.05){\footnotesize$\beta$}
  \put(-9.3,0.75){\footnotesize$\mathbf{I}^{\rightarrow}$}
   \put(-8.7,0.75){\footnotesize$\mathbf{I}^{\leftarrow}$}
 \caption{{\it Representations of directed hypergraphs.}  The figure illustrates  with an example the three hypergraph representations, viz., with incidence matrices, as a bipartite graph, or as a graph with higher-order interactions.}
 \label{fig:dir_hypergraph}
\end{figure*}

Figure~\ref{fig:dir_hypergraph} illustrates different ways of representing hypergraphs with an example.

The {\it out-degree} and {\it in-degree} of node $i\in\V$ are defined by 
\begin{equation}
k^{\rm out}_i(\mathbf{I}^{\rightarrow})  \equiv \sum^\sizeW_{\alpha=1}I^{\rightarrow}_{i\alpha} \quad {\rm and} \quad  k^{\rm in}_i(\mathbf{I}^{\leftarrow})  \equiv \sum^\sizeW_{\alpha=1}I^{\leftarrow}_{i\alpha} , 
\end{equation}
and we also use the notations 
\begin{align}
\vec{k}^{\rm in}(\mathbf{I}^{\leftarrow}) &\equiv (k^{\rm in}_1(\mathbf{I}^{\leftarrow}), k^{\rm in}_2(\mathbf{I}^{\leftarrow}), \ldots, k^{\rm in}_N(\mathbf{I}^{\leftarrow}))
\end{align}
and 
\begin{align}
\vec{k}^{\rm out}(\mathbf{I}^{\rightarrow}) &\equiv (k^{\rm out}_1(\mathbf{I}^{\rightarrow}), k^{\rm out}_2(\mathbf{I}^{\rightarrow}), \ldots, k^{\rm out}_N(\mathbf{I}^{\rightarrow}))
\end{align}
for their sequences.   
Analogously, we define the  {\it out-cardinality} and {\it in-cardinality}  of hyperedge $\alpha\in \W$ by 
\begin{equation}
\chi^{\rm out}_\alpha(\mathbf{I}^{\leftarrow})  \equiv \sum^\sizeV_{i=1} I^{\leftarrow}_{i \alpha } \quad {\rm and} \quad 
\chi^{\rm in}_\alpha(\mathbf{I}^{\rightarrow})  \equiv \sum^\sizeV_{i=1} I^{\rightarrow}_{i \alpha },
\end{equation}      
and we also use the corresponding  sequences $\vec{\chi}^{\rm in}(\mathbf{I}^{\rightarrow})$ and  $\vec{\chi}^{\rm out}(\mathbf{I}^{\leftarrow})$.
In addition, we define  the {\it modified} {\it out-} and {\it in-cardinalities}
\begin{equation}
\chi^{\rm out}_{\alpha,i}(\mathbf{I}^{\leftarrow})  \equiv \sum^\sizeV_{\substack{j=1;\\j\neq i}} I^{\leftarrow}_{j \alpha } \quad {\rm and} \quad   \chi^{\rm in}_{\alpha,i}(\mathbf{I}^{\rightarrow})  \equiv \sum^\sizeV_{\substack{j=1;\\j\neq i}} I^{\rightarrow}_{j \alpha }
\end{equation}     
 excluding the stubs used to connect to a given node $i$.

Lastly, we define the set of hyperedges incident to the node $i$ as the union
\begin{equation}
\partial_{i}(\mathbf{I}^{\leftrightarrow})
\equiv \partial^{\rm out}_{i}(\mathbf{I}^{\rightarrow}) \cup\partial^{\rm in}_{i}(\mathbf{I}^{\leftarrow})
\end{equation}
of the two hyperedge neighbourhood sets $\partial^{\rm out}_{i}(\mathbf{I}^{\rightarrow})$ and $\partial^{\rm in}_{i}(\mathbf{I}^{\leftarrow})$
where
\begin{equation}
\partial^{\rm out}_{i}(\mathbf{I}^{\rightarrow})  \equiv \{\alpha\in \W| I^{\rightarrow}_{i\alpha}\neq0 \},
\end{equation}
and  
\begin{equation}
\partial^{\rm in}_{i}(\mathbf{I}^{\leftarrow})  \equiv \{\alpha\in \W| I^{\leftarrow}_{i\alpha}\neq0 \}.
\end{equation}

To each directed hypergraph  we can associate a 
projected, directed graph  
of order $N$, such that there exists a directed edge that points  from $i$ to $j$ in the projected graph whenever there exists a hyperedge $\alpha\in \W$ such that $(i,\alpha)\in \E^{\rm out}$ and $(j,\alpha)\in \E^{\rm in}$.   The adjacency matrix of the projected graph is given by  
\begin{equation}
A^{\rm proj}_{ij}(\mathbf{I}^{\leftrightarrow}) = \Theta\left(\sum^M_{\alpha=1}I^{\rightarrow}_{i\alpha}I^{\leftarrow}_{j\alpha}\right),
\end{equation}
for all $i,j\in \V$, where $\Theta(x) = 0$ if $x\leq 0$  and $\Theta(x)=1$ for $x>0$.       If $A^{\rm proj}_{ii}=0$ for all $i\in \V$, then we call the projected graph simple.

Note that there exists a one-to-one correspondence between simple, directed hypergraphs $\H^{\rm dir}$ and pairs $\mathbf{I}^{\leftrightarrow}$ of incidence matrices, while the mapping between $\H$ and $\mathbf{A}^{\rm proj}$ is not one-to-one, and hence the projected graph is a coarse-grained representation of the hypergraph.

\subsection{Clustering coefficient for directed graphs with pairwise interactions} 
We  review the definition of the pairwise clustering coefficient for directed graphs, as introduced in  Ref.~\cite{fagiolo2007clustering}.

Let $\mathbf{A}$ be the adjacency matrix of a simple, directed graph, such that $[\mathbf{A}]_{ij}=1$ whenever there exists a directed link that points from $i$ to $j$, and  $[\mathbf{A}]_{ij}=0$ whenever such a link is absent.  The directed clustering coefficient of node $i$ is defined  by~\cite{fagiolo2007clustering}
\begin{equation}
C^{\rm pi\leftrightarrow}_i(\mathbf{A}) \equiv \frac{T^{\leftrightarrow}_i(\mathbf{A})}{t^{\leftrightarrow}_{\rm max}(k^{\rm tot}_{i}(\mathbf{A}),k^{\leftrightarrow}_{i}(\mathbf{A}))} , \label{eq:sim_dirGraph}
\end{equation} 
where 
\begin{eqnarray}
T^{\leftrightarrow}_i(\mathbf{A}) &\equiv& \frac{1}{2}\left[\left(\mathbf{A}+\mathbf{A}^{\intercal}\right)^3\right]_{ii}\nonumber\\ 
&=&\frac{1}{2}\sum^N_{j=1}\sum^N_{h=1}(A_{ij}+A_{ji})(A_{ih}+A_{hi})(A_{jh}+A_{hj})
\end{eqnarray} 
counts the number of  directed triangles centered on  node $i$, 
and where  
\begin{equation}
t^{\leftrightarrow}_{\rm max}(k^{\rm tot}_{i}(\mathbf{A}),k^{\leftrightarrow}_{i}(\mathbf{A})) \equiv k^{\rm tot}_{i}(\mathbf{A})(k^{\rm tot}_{i}(\mathbf{A})-1)-2k^{\leftrightarrow}_{i}(\mathbf{A})
\end{equation} 
 is the maximum possible number of directed triangles incident to a node with a given total degree $k^{\rm tot}_{i}(\mathbf{A}) \equiv \sum^N_{j=1;j\neq i}(A_{ij} +A_{ji} )$, and a given degree of symmetric links $k^{\leftrightarrow}_{i}(\mathbf{A}) \equiv \sum^N_{j=1;j\neq i} A_{ji}A_{ij}$.   The denominator in the definition of  the pariwise clustering coefficient is independent of the directionality and the  symmetry (i.e., whether it is unidirectional or bidirection) of the links between  node $i$ and its neighbours.    Additionally, for simple and nondirected graphs ($A_{ij}=A_{ji}$,) the clustering coefficients in Eqs.~(\ref{eq:cGraph}) and (\ref{eq:sim_dirGraph}) are equal.

Following the example of pairwise clustering coefficients, we  define in the next Subsection a quad clustering coefficient for directed hypergraphs, which is an extension of the corresponding clustering coefficient for nondirected hypergraphs.

\subsection{Quad clustering coefficient for directed hypergraphs}
 We define  a quad clustering coefficient for directed hypergraphs.  Similarly to the pairwise clustering coefficient for directed graphs $C^{\rm pi\leftrightarrow}_{i}$, we require that the quad clustering coefficient counts the number of directed quads incident to the node $i$ of a hypergraph, and we require that for nondirected hypergraphs the directed quad clustering coefficient equals the  quad clustering coefficient defined in Eq.~(\ref{def:CQuad}). 

\begin{figure}
     \centering
     \setlength{\unitlength}{0.1\textwidth}
     \begin{subfigure}[b]{0.4\textwidth}
         \centering
         \includegraphics[width=\textwidth]{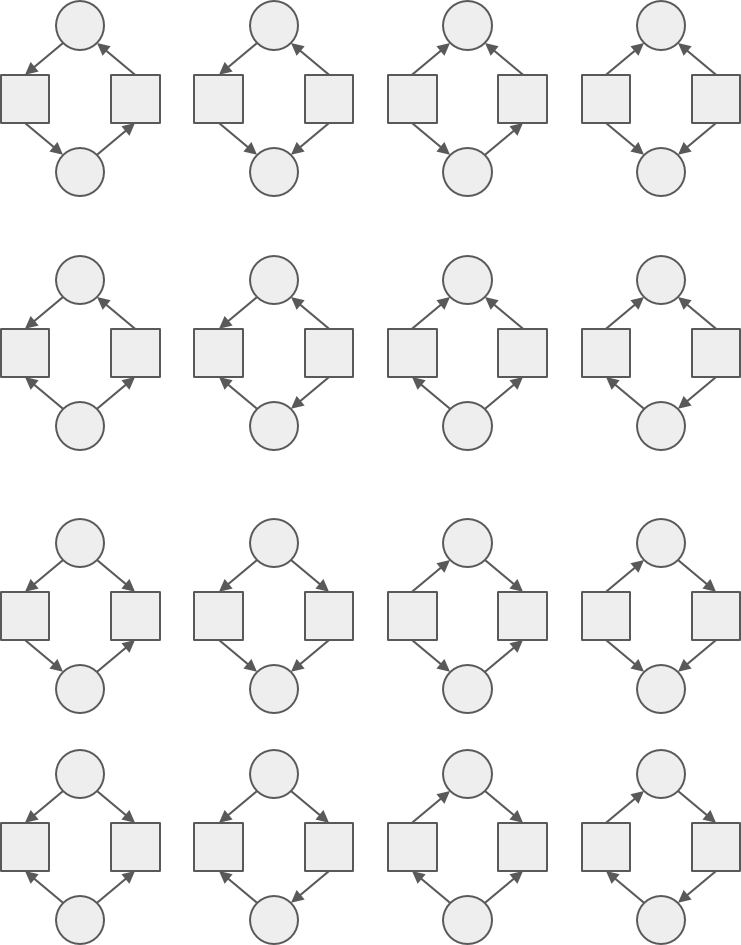}
         \label{fig:two_hyperedges2}
     \end{subfigure}
      \put(-4.27,5.2){\normalsize$(a)$}
      \put(-3.6,5.12){\scriptsize$i$}
      \put(-3.6,4.35){\scriptsize$j$}
      \put(-3.92,4.73){\scriptsize$\alpha$}
      \put(-3.3,4.73){\scriptsize$\beta$}
      \put(-2.55,5.12){\scriptsize$i$}
      \put(-2.55,4.35){\scriptsize$j$}
      \put(-2.87,4.73){\scriptsize$\alpha$}
      \put(-2.25,4.73){\scriptsize$\beta$}
      \put(-1.5,5.12){\scriptsize$i$}
      \put(-1.5,4.35){\scriptsize$j$}
      \put(-1.82,4.73){\scriptsize$\alpha$}
      \put(-1.2,4.73){\scriptsize$\beta$}
      \put(-0.46,5.12){\scriptsize$i$}
      \put(-0.46,4.35){\scriptsize$j$}
      \put(-0.78,4.73){\scriptsize$\alpha$}
      \put(-0.2,4.73){\scriptsize$\beta$}
      \put(-3.6,3.75){\scriptsize$i$}
      \put(-3.6,2.97){\scriptsize$j$}
      \put(-3.92,3.36){\scriptsize$\alpha$}
      \put(-3.3,3.36){\scriptsize$\beta$}
      \put(-2.55,3.75){\scriptsize$i$}
      \put(-2.55,2.97){\scriptsize$j$}
      \put(-2.87,3.36){\scriptsize$\alpha$}
      \put(-2.25,3.36){\scriptsize$\beta$}
      \put(-1.5,3.75){\scriptsize$i$}
      \put(-1.5,2.97){\scriptsize$j$}
      \put(-1.82,3.36){\scriptsize$\alpha$}
      \put(-1.2,3.36){\scriptsize$\beta$}
      \put(-0.46,3.75){\scriptsize$i$}
      \put(-0.46,2.97){\scriptsize$j$}
      \put(-0.78,3.36){\scriptsize$\alpha$}
      \put(-0.2,3.36){\scriptsize$\beta$}
      \put(-3.6,2.32){\scriptsize$i$}
      \put(-3.6,1.55){\scriptsize$j$}
      \put(-3.92,1.95){\scriptsize$\alpha$}
      \put(-3.3,1.95){\scriptsize$\beta$}
      \put(-2.55,2.32){\scriptsize$i$}
      \put(-2.55,1.55){\scriptsize$j$}
      \put(-2.87,1.95){\scriptsize$\alpha$}
      \put(-2.25,1.95){\scriptsize$\beta$}
      \put(-1.5,2.32){\scriptsize$i$}
      \put(-1.5,1.55){\scriptsize$j$}
      \put(-1.82,1.95){\scriptsize$\alpha$}
      \put(-1.2,1.95){\scriptsize$\beta$}
      \put(-0.46,2.32){\scriptsize$i$}
      \put(-0.46,1.55){\scriptsize$j$}
      \put(-0.78,1.95){\scriptsize$\alpha$}
      \put(-0.2,1.95){\scriptsize$\beta$}
      \put(-3.6,1.08){\scriptsize$i$}
      \put(-3.6,0.3){\scriptsize$j$}
      \put(-3.92,0.7){\scriptsize$\alpha$}
      \put(-3.3,0.7){\scriptsize$\beta$}
      \put(-2.55,1.08){\scriptsize$i$}
      \put(-2.55,0.3){\scriptsize$j$}
      \put(-2.87,0.7){\scriptsize$\alpha$}
      \put(-2.25,0.7){\scriptsize$\beta$}
      \put(-1.5,1.08){\scriptsize$i$}
      \put(-1.5,0.3){\scriptsize$j$}
      \put(-1.82,0.7){\scriptsize$\alpha$}
      \put(-1.2,0.7){\scriptsize$\beta$}
      \put(-0.46,1.08){\scriptsize$i$}
      \put(-0.46,0.3){\scriptsize$j$}
      \put(-0.78,0.7){\scriptsize$\alpha$}
      \put(-0.2,0.7){\scriptsize$\beta$}
     \hfill
     \begin{subfigure}[b]{0.35\textwidth}
         \centering
         \includegraphics[width=\textwidth]{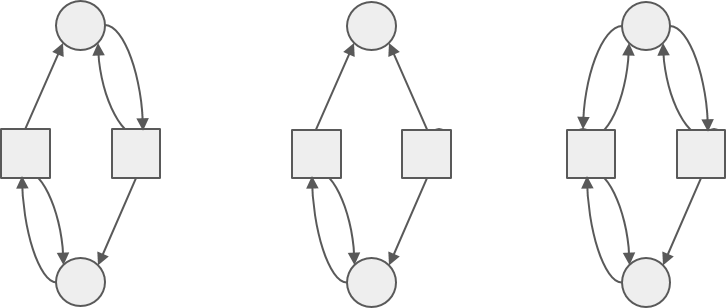}
     \end{subfigure}
      \put(-4.0,1.6){\normalsize$(b)$}
      \put(-3.13,1.31){\footnotesize$i$}
      \put(-2.6,1.6){\normalsize$(c)$}
      \put(-1.73,1.31){\footnotesize$i$}
      \put(-2.05,0.7){\footnotesize$\alpha$}
      \put(-1.5,0.7){\footnotesize$\beta$}
      \put(-0.43,1.31){\footnotesize$j$}
      \put(-0.71,0.7){\footnotesize$\gamma$}
      \put(-0.17,0.7){\footnotesize$\epsilon$}
 \caption{{\it Counting the number of directed quads incident to a node $i$.} $(a)$ The $16$ directed quads that contribute to $Q^{\rm q}_i(\mathbf{I})$.
$(b)$ Example graph with $C^{{\rm q}, \leftrightarrow}_i = 1$.   $(c)$  Two example graphs  with   $C^{{\rm q}, \leftrightarrow}_i = C^{{\rm q}, \leftrightarrow}_j = 1$.}
 \label{fig:dir_quads}
\end{figure}
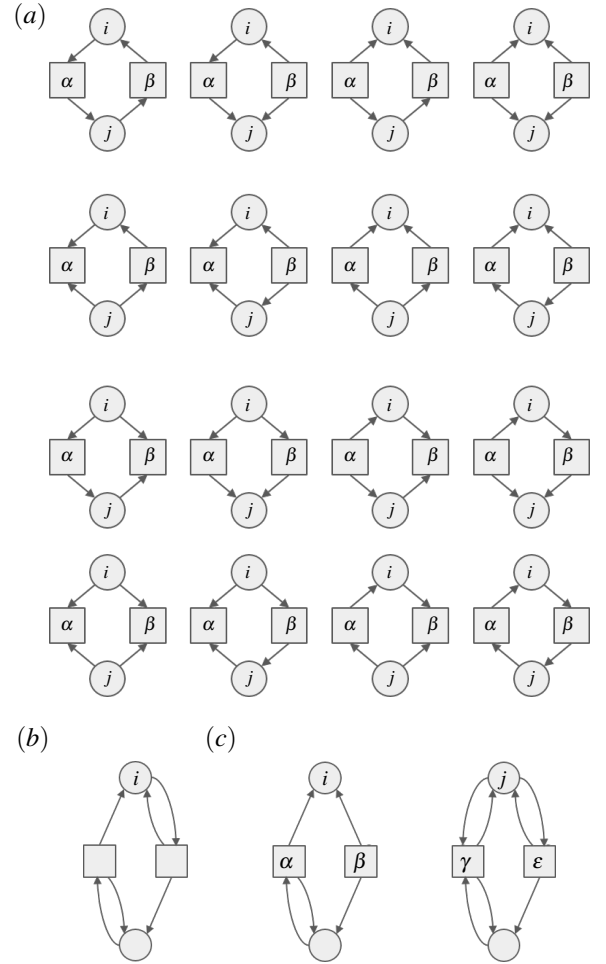

We define the  quad clustering coefficient $C^{\rm q \leftrightarrow}_i(\mathbf{I}^{\leftrightarrow})$ of a node $i$ in the directed hypergraph   represented by $\mathbf{I}^{\leftrightarrow}$, for which $\sum_{\alpha\in \partial_i(\mathbf{I}^{\leftrightarrow })}(\chi^{\rm in}_{\alpha, i} + \chi^{\rm out}_{\alpha, i})\geq 2$, as follows, 
\begin{equation}
C^{\rm q \leftrightarrow}_i(\mathbf{I}^{\leftrightarrow}) \equiv \frac{Q^{\leftrightarrow}_i(\mathbf{I}^{\leftrightarrow})}{q^{\leftrightarrow}_{\rm max}(\left\{ \mathcal{X}_{i\alpha}(\mathbf{I}^{\leftrightarrow}),I^{\leftrightarrow }_{i\alpha}\right\}_{\alpha\in \partial_i} )}, \label{def:dir_CQuad}
\end{equation}
where
\begin{eqnarray}
Q^{\leftrightarrow}_{i}(\mathbf{I}^{\leftrightarrow})
&\equiv&\sum^N_{j=1;j\neq i} \sum^M_{\alpha<\beta}I^{\leftrightarrow}_{i\alpha}I^{\leftrightarrow}_{j\alpha}I^{\leftrightarrow}_{i\beta}I^{\leftrightarrow}_{j\beta},
\label{eq:dir_quads}
\end{eqnarray} 
is the number of directed quads centred on the node $i$, and we have used the notation 
 $I^{\leftrightarrow}_{i\alpha} \equiv I^{\rightarrow}_{i\alpha}+I^{\leftarrow}_{i\alpha}$.   
    The denominator 
$q^{\leftrightarrow}_{\rm max}(\left\{ \mathcal{X}_{i\alpha}(\mathbf{I}^{\leftrightarrow}),I^{\leftrightarrow }_{i\alpha}\right\}_{\alpha\in \partial_i} )$   denotes the maximum possible number of directed quads incident to node $i$, given the  sets
\begin{equation}
\mathcal{X}_{i\alpha}(\mathbf{I}^{\leftrightarrow})\equiv\left\{\chi^{\rm in}_{\alpha, i}(\mathbf{I}^{\rightarrow}), \chi^{\rm out}_{\alpha,i}(\mathbf{I}^{\leftarrow})\right\} \label{eq:cardinalitysets}
\end{equation}
of modified {\it in-} and {\it out- cardinalities} of the hyperedges $\alpha\in \partial_i$, and the corresponding values of  $I^{\leftrightarrow}_{i\alpha}$.     We omit the explicit  mathematical expression for $q^{\leftrightarrow}_{\rm max}$ here, as it is elaborate, but it can be found in   Appendix~ \ref{app:clustering-norm}.    If  $\sum_{\alpha\in \partial_i(\mathbf{I}^{\leftrightarrow })}(\chi^{\rm in}_{\alpha, i} + \chi^{\rm out}_{\alpha, i})< 2$ then $C^{{\rm q} \leftrightarrow}(\mathbf{I})=0$.   To illustrate how quads are counted by 
 $Q^{\leftrightarrow}_{i}(\mathbf{I}^{\leftrightarrow})$,  consider the example in Panel (b) of Fig.~\ref{fig:dir_quads}.   In this case,   $Q^{\leftrightarrow}_{i}(\mathbf{I}^{\leftrightarrow})=4$, as the motif contains the four quads in the left column of Panel (a) of Fig~\ref{fig:dir_quads}.

Alternatively, we can express $Q^{\leftrightarrow}_{i}(\mathbf{I}^{\leftrightarrow})$  in terms of the number of closed paths of length $4$ (see Panel (a) of  Fig~\ref{fig:dir_quads} for all possible types of closed paths of length $4$) with the formula
\begin{multline}
Q^{\leftrightarrow}_{i}(\mathbf{I}^{\leftrightarrow})=\frac{1}{2}\left[\left(\mathbf{I}^{\leftrightarrow}\left(\mathbf{I}^{\leftrightarrow}\right)^{\intercal}\right)^{2}\right]_{ii}-\frac{1}{2}\left(\left[\mathbf{I}^{\leftrightarrow}\left(\mathbf{I}^{\leftrightarrow}\right)^{\intercal}\right]_{ii}\right)^{2}\\
-\frac{1}{2}\sum_{j;j\neq i}\left(\left[\mathbf{I}^{\leftrightarrow}\left(\mathbf{I}^{\leftrightarrow}\right)^{\intercal}\right]_{ij}\right)^{2}.
\end{multline}
The first term $\left[\left(\mathbf{I}^{\leftrightarrow}\left(\mathbf{I}^{\leftrightarrow}\right)^{\intercal}\right)^{2}\right]_{ii}$ counts the total  number of paths of length $4$ starting and ending in $i$. The second and third terms subtract off the contributions to the first term arising from paths returning to site $i$ via backtracking paths of length one and two, respectively.
The prefactor $1/2$ corrects for double counting arising from counting the same path with the opposite orientation.

Next we turn to the denominator of  the right-hand side of (\ref{def:dir_CQuad}). Similarly to the
pairwise, directed,  clustering coefficient $C^{\rm pi\leftrightarrow}_i(\mathbf{A})$,
the denominator $q^{\leftrightarrow}_{\rm max}(\left\{ \mathcal{X}_{i\alpha}(\mathbf{I}^{\leftrightarrow}),I^{\leftrightarrow }_{i\alpha}\right\}_{\alpha\in \partial_i} )$ normalizes the directed quad clustering coefficient $C^{\rm q \leftrightarrow}_i(\mathbf{I}^{\leftrightarrow})$ such that its value is 
independent of both  the directionality and  symmetry (i.e., unidirectional or bidirectional)
of the links that connect  node $i$ to its neighbouring hyperedges.   This means that if two nodes $i$ and $j$  have the same motif of inlinks, as shown in  Panel (c) of Fig~\ref{fig:dir_quads},  then the quad clustering coefficient of the two nodes, $C^{\rm q \leftrightarrow}_i$ and $C^{\rm q \leftrightarrow}_j$, must be the same, even if the motifs of outlinks are different.

Note that for nondirected hypergraphs  the  directed quad clustering coefficient, defined by Eq.~(\ref{def:dir_CQuad}), equals the  quad clustering coefficient for nondirected hypergraphs, defined by Eq.~(\ref{def:CQuad}) (see Appendix \ref{app:dir_undir_quadC}).

\subsection{Clustering in directed, realworld, hypergraphs} \label{ch:real_dirhyper}
In Sec.~\ref{ch:real_hypergraph} we found that the density of quads in  nondirected real-world hypergraphs is  large compared to the density of quads in  the configuration model. In this Section, we investigate whether an analogous  phenomenon can be observed in directed hypergraphs.    Specifically, we build  directed hypergraphs from three data sets related to the DNC-email network,  the English thesaurus, and the  Human metabolic pathway (see Appendix~\ref{app:data} for more detailed information about these data sets).

\begin{table}[b]
\caption{
Network characteristics of the real-world directed hypergraphs: number of nodes $N$ and hyperedges $M$, mean directed quad clustering coefficient $\overline{C}^{\rm q \leftrightarrow}(\textbf{I}^{\leftrightarrow}_{\rm real})$ and the average, mean directed quad clustering coefficient $\langle\overline{C}^{\rm q \leftrightarrow}(\textbf{I}^\leftrightarrow)\rangle$ of the corresponding configuration model.}\label{tb:dirqC_value}
\centering
\begin{tabular}{cccccccc}
\hline\hline
Dataset & $N$ & $M$ & $\overline{C}^{\rm q \leftrightarrow}(\textbf{I}^{\leftrightarrow}_{\rm real})$ & $\langle\overline{C}^{\rm q \leftrightarrow}(\textbf{I}^\leftrightarrow)\rangle$\\
\hline
DNC-email & 2,029 & 5,598 & 0.3419 & 0.0715\\
English thesaurus & 40,963 & 35,104 & 0.2371 & 0.0004\\
Metabolic pathways & 1,508 & 1,451 & 0.0684 & 0.0179\\
\hline\hline
\end{tabular}
\end{table}

In Table~\ref{tb:dirqC_value} we present  the mean quad clustering coefficient $\overline{C}^{\rm q \leftrightarrow}(\textbf{I}_{\rm real})\equiv\frac{1}{N}\sum^N_{i=1}C^{{\rm q \leftrightarrow}}_i(\textbf{I}_{\rm real})$ for the three real-world hypergraphs under study, and compare their values with the corresponding directed configuration models, which have  the  prescribed degree sequences $\vec{k}^{\rm in}(\mathbf{I}^{\leftarrow}_{\rm real})$ and $\vec{k}^{\rm out}(\mathbf{I}^{\rightarrow}_{\rm real})$,  and the prescribed  cardinality sequences $\vec{\chi}^{\rm in}(\mathbf{I}^{\rightarrow}_{\rm real})$ and $\vec{\chi}^{\rm out}(\mathbf{I}^{\leftarrow}_{\rm real})$. We observe that the real-world networks have significantly larger directe quad clustering coefficient, up to $500$   times larger than those of corresponding random models.  Hence, the density of directed quads in real-world directed hypergraphs is significantly higher than their density in the corresponding configuration models, consistent with earlier findings for nondirected hypergraphs.

\begin{figure*}[t]
     \centering
     \setlength{\unitlength}{0.1\textwidth}
     \hspace*{0.8cm}
     \includegraphics[width=\textwidth]{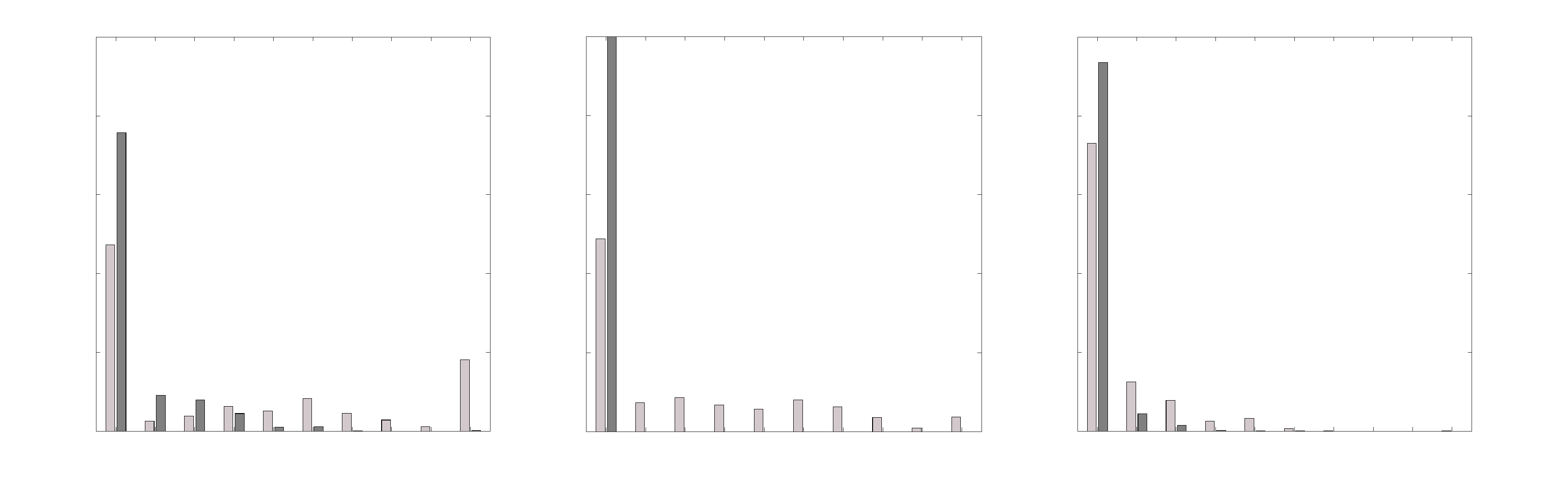}
      \put(-10.4,2.8){\small $(a)$}
      \put(-6.7,2.8){\small $(b)$}
      \put(-3.5,2.8){\small $(c)$}
      \put(-10.4,2.0){\small $P(C^{\rm q \leftrightarrow};\mathbf{I}^{\leftrightarrow}_{\rm real})$,}
      \put(-10.4,1.65){\small $\langle P(C^{\rm q \leftrightarrow};\mathbf{I}^{\leftrightarrow})\rangle$}
      \put(-5,-0.1){\small $C^{\rm q \leftrightarrow}$}
      \put(-9.35,0.15){\footnotesize$0.1$}
      \put(-8.62,0.15){\footnotesize$0.4$}
      \put(-7.86,0.15){\footnotesize$0.7$}
      \put(-7.04,0.15){\footnotesize$1$}
      \put(-6.21,0.15){\footnotesize$0.1$}
      \put(-5.48,0.15){\footnotesize$0.4$}
      \put(-4.72,0.15){\footnotesize$0.7$}
      \put(-3.9,0.15){\footnotesize$1$}
      \put(-3.08,0.15){\footnotesize$0.1$}
      \put(-2.34,0.15){\footnotesize$0.4$}
      \put(-1.57,0.15){\footnotesize$0.7$}
      \put(-0.77,0.15){\footnotesize$1$}
      \put(-9.55,0.35){\footnotesize$0$}
      \put(-9.7,0.8){\footnotesize$0.2$}
      \put(-9.7,1.3){\footnotesize$0.4$}
      \put(-9.7,1.8){\footnotesize$0.6$}
      \put(-9.7,2.3){\footnotesize$0.8$}
      \put(-9.55,2.8){\footnotesize$1$}
    \caption{{\it Distribution of quad clustering coefficients in directed hypergraphs.} The light grey histograms represent the distributions of the directed quad clustering coefficient measured in real-world hypergraphs. The grey bar graphs show the distributions of the directed quad clustering coefficient measured in the hypergraph configuration model that preserves the in-/out-degree and in-/out-cardinality sequences extracted from the real-world hypergraphs. Each plots are extracted from $(a)$ {\it DNC-email}, $(b)$ {\it English thesaurus}, $(c)$ {\it Human metabolic pathways}}
    \label{fig:dir_quadC_dist}
\end{figure*}
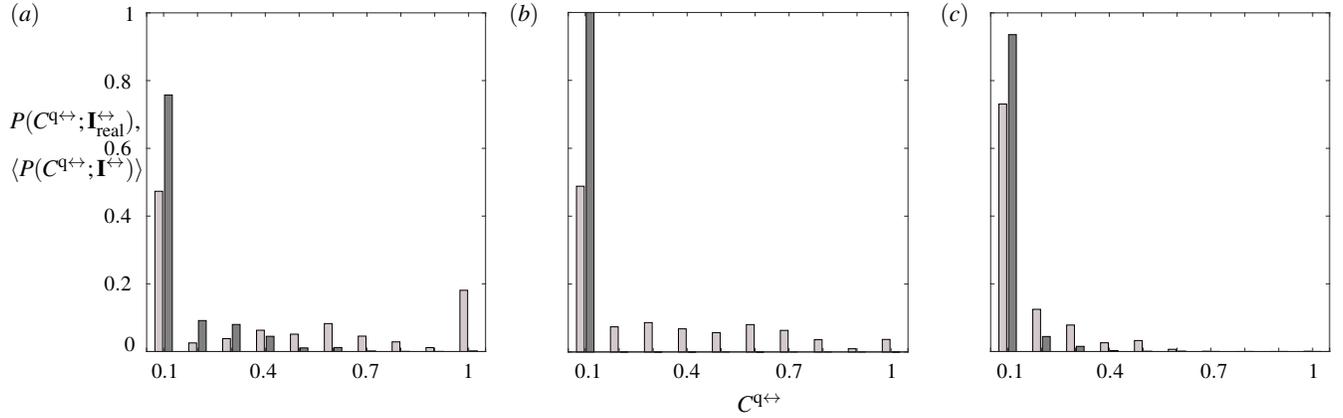

Furthermore, we determine the distribution of  directed, quad clustering coefficients in real-world hypergraphs defined by $P(C^{\rm q \leftrightarrow};\mathbf{I}^{\leftrightarrow}_{\rm real}) \equiv \frac{1}{N}\sum^N_{i=1}\delta(C^{\rm q \leftrightarrow}-C^{\rm q \leftrightarrow}_i(\mathbf{I}^{\leftrightarrow})_{\rm real}$, and present the results in Fig.~\ref{fig:dir_quadC_dist}.    Also in  directed real-world hypergraphs, we observe a   a peak at $C^{\rm q \leftrightarrow} \approx 1$ in the quad clustering distribution.   In the specific examples considered, the peak is most pronounced in the DNC-email hypergraph.

\section{Discussion} \label{ch:discussion}

We have introduced a clustering coefficient, called the quad clustering coefficient, that captures the multiplicity of interactions between neighbouring nodes in (non)directed hypergraphs with higher order interactions.     We have shown that for random hypergraphs the mean quad clustering coefficient has a value near zero, while for real-world networks it is one order of magnitude larger taking values ranging from $0.01$ to $0.34$, which is a smaller range than the one observed for pairwise clustering coefficients in real-world networks \cite{newman2003structure}; we note however that the distribution of  quad clustering coefficients is supported on the whole $[0,1]$ range of values.
Hence, the quad clustering coefficient describes a feature of real-world networks that is not captured by the current random hypergraph models.   

We have determined the average quad clustering coefficient in several random hypergraph models.    We have obtained exact expressions for models with fluctuating degrees and fixed cardinalities.    Our analysis shows that it  is significantly more difficult to deal with fluctuating cardinalities.   

Analysing the distribution of quad clustering coefficients in real-world networks we have  found that there exist a significant fraction of nodes that take its maximal value.  Analysing the topological properties of the neighbourhood sets of these highly clustered nodes we have found that they can exhibit large degrees, and their neighbouring nodes can have large cardinalities. 

The results of this paper show that 
the configuration model is not a good null model for real-world networks with higher order interactions.    This in itself is not a  surprising result, as the configuration model is also not a good model for networks without higher order interactions, see e.g.,~ discussions in Ref.~\cite{albert2002statistical}.    However, what is surprising is that the distribution of quad clustering coefficients exhibits a peak at its maximal value.     This result has, to the best of our knowledge, no counter part in systems without higher order interactions.  

This raises the question of what type of random hypergraph model can generate statistical properties similar to those observed in real-world networks with higher order interactions, see e.g., Ref.~\cite{ravasz2003hierarchical, barrat2000properties, watts1998collective} for related questions in networks without higher order interactions.  Another pertinent question concerns the implications of nodes with high quad clustering coefficients on dynamical processes, such as, percolation.  Since highly clustered nodes do not appear in random hypergraphs, they may play an important role in dynamical processes governed on real-world networks.

\begin{acknowledgments}
G.-G. Ha thanks D.-S. Lee, J.W. Lee, S.H. Lee, S.-W. Son, H.J. Park, M. Ha and N.W. Landry. This work was supported by the Engineering and Physical Sciences Research Council, part of the EPSRC DTP, Grant Ref No.: EP/V520019/1.
\end{acknowledgments}

\section*{Data Availability Statement}

We used the databases {\it NDC-substances}\cite{benson2018simplicial}, {\it Youtube}\cite{kunegis2013konect,mislove2009online}, {\it Food recipe}\cite{whats-cooking}, {\it Github}\cite{kunegis2013konect,Scott2009GitHub}, {\it Crime involvement}\cite{kunegis2013konect} and  {\it Wallmart}\cite{amburg2020clustering} as the real-world undirected hypergraph. And as a directed hypergraph, we used {\it DNC-email}\cite{kunegis2013konect}, {\it English thesaurus}\cite{ward2002moby} and {\it Human metabolic pathways}\cite{karp2019biocyc} database. And we implemented computation algorithms in Fortran to compute nondirected and directed quad clustering coefficients in a hypergraph, available from \url{https://github.com/Gyeong-GyunHa/qch}.

\appendix

\begin{widetext}
\section{Alternate expression for the denominator of the quad clustering coefficient }\label{app:C}

In this Section we show that $q_{\rm max}$, defined by Eq.~(\ref{eq:qimax1}), can also be expressed by Eq.~(\ref{eq:qimax2}).

We can express Eq.~(\ref{eq:qimax1}) 
\begin{equation}
\begin{split}  
2\:q_{\rm max}&=\sum_{\substack{\alpha,\beta;\\\chi_{\alpha}(\textbf{I})\leq\chi_{\beta}(\textbf{I})}}\chi_{\alpha}(\textbf{I})I_{i\alpha}I_{i\beta}-\sum_{\substack{\alpha,\beta;\\\chi_{\alpha}(\textbf{I})=\chi_{\beta}(\textbf{I})}}\chi_{\alpha}(\textbf{I})I_{i\alpha}I_{i\beta}+\sum_{\substack{\alpha,\beta;\\\chi_{\alpha}(\textbf{I})\geq\chi_{\beta}(\textbf{I})}}\chi_{\beta}(\textbf{I})I_{i\alpha}I_{i\beta}-\sum_{\alpha,\beta}I_{i\alpha}I_{i\beta}-\sum_{\alpha}(\chi_{\alpha}(\textbf{I})-1)I_{i\alpha}. \label{eq:intermediateQmax}
\end{split}
\end{equation}

To proceed, we introduce the  definitions
\begin{equation}
\begin{split}
\Omega_i(\textbf{I}) &\equiv \sum^M_{\gamma=1} I_{i\gamma} \chi_{\gamma}(\textbf{I})
\end{split}
\end{equation}
and
\begin{equation}
q_{i,\chi}(\vec{\chi}({\bf I});\mathbf{I}) \equiv\sum_{\alpha; \chi\leq \chi_{\alpha}(\textbf{I})}I_{i\alpha} = \sum_{\alpha}I_{i\alpha}\: \Theta(\chi_{\alpha}(\textbf{I})-\chi),
\end{equation}  
where  $\Theta(\chi_{\alpha}(\textbf{I})-\chi)$ is the Heaviside function defined below Eq.~(\ref{eq:AProj}). Using these definitions in Eq.~(\ref{eq:intermediateQmax}), yields
\begin{align}
2q_{\rm max}
&=\sum^M_{\alpha=1}\chi_{\alpha}(\mathbf{I})I_{i\alpha} q_{i,\chi_{\alpha}(\mathbf{I})}(\vec{\chi}({\bf I});\mathbf{I})-\sum^M_{\substack{\alpha,\beta=1;\\\chi_{\alpha}({\bf I})=\chi_{\beta}({\bf I})}}\chi_{\alpha}({\bf I})I_{i\alpha}I_{i\beta}+\sum^M_{\substack{\alpha,\beta=1;\\\chi_{\alpha}({\bf I})\geq\chi_{\beta}({\bf I})}}\chi_{\beta}({\bf I})I_{i\alpha}I_{i\beta}-k^2_{i}(\mathbf{I})-\Omega_i(\mathbf{I})+k_{i}(\mathbf{I})\\
&=2\sum^M_{\alpha=1}\chi_{\alpha}(\mathbf{I})I_{i\alpha} q_{i,\chi_{\alpha}(\mathbf{I})}(\vec{\chi}({\bf I});\mathbf{I})-\sum^M_{\alpha=1}\chi_{\alpha}(\mathbf{I})I_{i\alpha}\left[q_{i,\chi_{\alpha}(\mathbf{I})}(\vec{\chi}({\bf I});\mathbf{I})-q_{i,\chi_{\alpha}(\mathbf{I})}(\vec{\chi}({\bf I})-\vec{1};\mathbf{I})\right]-k^2_{i}(\mathbf{I})-\Omega_i(\mathbf{I})+k_{i}(\mathbf{I})\\
&=\sum^M_{\alpha=1}\chi_{\alpha}(\mathbf{I})I_{i\alpha}\left[q_{i,\chi_{\alpha}(\mathbf{I})}(\vec{\chi}({\bf I});\mathbf{I})+q_{i,\chi_{\alpha}(\mathbf{I})}(\vec{\chi}({\bf I})-\vec{1};\mathbf{I})\right]-k^2_{i}(\mathbf{I})-\Omega_i(\mathbf{I})+k_{i}(\mathbf{I})\\
&=\sum^M_{\alpha=1}\chi_{\alpha}(\mathbf{I})I_{i\alpha}\left[q_{i,\chi_{\alpha}(\mathbf{I})}(\vec{\chi}({\bf I});\mathbf{I})+q_{i,\chi_{\alpha}(\mathbf{I})}(\vec{\chi}({\bf I})-\vec{1};\mathbf{I})-1\right]+k_{i}(\mathbf{I})-k^2_{i}(\mathbf{I})\\
&=\sum^N_{\chi=1} \chi \sum_{\alpha} \delta_{\chi,\chi_{\alpha}(\mathbf{I})} I_{i\alpha}\left[q_{i,\chi_{\alpha}(\mathbf{I})}(\vec{\chi}({\bf I});\mathbf{I})+q_{i,\chi_{\alpha}(\mathbf{I})}(\vec{\chi}({\bf I})-\vec{1};\mathbf{I})-1\right]+k_{i}(\mathbf{I})-k^2_{i}(\mathbf{I}) \label{eq:before_apply_delta}\\
&=\sum^N_{\chi=1} \chi \left(\sum_{\alpha} \delta_{\chi,\chi_{\alpha}(\mathbf{I})} I_{i\alpha}\right)\left[q_{i,\chi}(\vec{\chi}({\bf I});\mathbf{I})+q_{i,\chi}(\vec{\chi}({\bf I})-\vec{1};\mathbf{I})-1\right]+k_{i}(\mathbf{I})-k^2_{i}(\mathbf{I})\label{eq:after_apply_delta}\\ 
&=\sum^N_{\chi=1} \chi\left[q_{i,\chi}(\vec{\chi}({\bf I});\mathbf{I})-q_{i,\chi}(\vec{\chi}({\bf I})-\vec{1};\mathbf{I})\right]\left[q_{i,\chi}(\vec{\chi}({\bf I});\mathbf{I})+q_{i,\chi}(\vec{\chi}({\bf I})-\vec{1};\mathbf{I})-1\right]+k_{i}(\mathbf{I})-k^2_{i}(\mathbf{I})\\
&=\sum^N_{\chi=1} \chi\left[\left(q_{i,\chi}(\vec{\chi}({\bf I});\mathbf{I})\right)^2-\left(q_{i,\chi}(\vec{\chi}({\bf I})-\vec{1};\mathbf{I})\right)^2-q_{i,\chi}(\vec{\chi}({\bf I});\mathbf{I})+q_{i,\chi}(\vec{\chi}({\bf I})-\vec{1};\mathbf{I})\right]+q_{i,1}(\vec{\chi}({\bf I});\mathbf{I})-\left(q_{i,1}(\vec{\chi}({\bf I});\mathbf{I})\right)^2\\
&=\sum_{\chi=2}^{N}q_{i,\chi}(\vec{\chi}({\bf I});\mathbf{I})\left[q_{i,\chi}(\vec{\chi}({\bf I});\mathbf{I})-1\right], \label{eq:qmaxIntermediate2}
\end{align} 
where $\vec{1}$ is the vector of $M$ entries, all equal to one.  Note that the passage from (\ref{eq:before_apply_delta}) to (\ref{eq:after_apply_delta}) we have used 

\begin{eqnarray}
&&\sum_{\alpha} \delta_{\chi,\chi_{\alpha}(\mathbf{I})} I_{i\alpha}\left[q_{i,\chi_{\alpha}(\mathbf{I})}(\vec{\chi}({\bf I});\mathbf{I})\right]
=\sum_{\alpha} \delta_{\chi,\chi_{\alpha}(\mathbf{I})} I_{i\alpha}\left[\sum_{\beta\neq \alpha} I_{i\beta}\Theta(\chi_\beta(\bI)-\chi_\alpha(\bI))\right]
\nonumber\\
&=&\sum_{\alpha} \delta_{\chi,\chi_{\alpha}(\mathbf{I})} I_{i\alpha}\left[\sum_{\beta\neq \alpha} I_{i\beta}\Theta(\chi_\beta(\bI)-\chi)\right]=\sum_{\alpha} \delta_{\chi,\chi_{\alpha}(\mathbf{I})} I_{i\alpha}\left[\sum_{\beta} I_{i\beta}\Theta(\chi_\beta(\bI)-\chi)\right]
\end{eqnarray}

Using the degrees $k_i(\mathbf{I};\chi)$ as defined in Eq.~(\ref{def:dec_k}),  we can express $q_{i,\chi}$ by
\begin{equation}
q_{i,\chi}(\vec{\chi}({\bf I});\mathbf{I})=\sum_{\lambda=\chi}^{N} k_i(\mathbf{I};\lambda), 
\end{equation}
and using this expression in Eq.~(\ref{eq:qmaxIntermediate2}) we find
\begin{align} 
q_{\rm max}(\left\{k_i(\chi)\right\}_{\chi\in \mathbb{N}})&=\frac{1}{2}\sum_{\chi=2}^{N}q_{i,\chi}(\vec{\chi}({\bf I});\mathbf{I})\left[q_{i,\chi}(\vec{\chi}({\bf I});\mathbf{I})-1\right]\nonumber\\
&=\frac{1}{2}\sum_{\chi=2}^{N}\left[\left(\sum_{\lambda=\chi}^{\infty} k_i(\mathbf{I};\lambda)\right)\left(\sum_{\lambda=\chi}^{\infty} k_i(\mathbf{I};\lambda)-1\right)\right]\nonumber\\
&=\frac{1}{2} \sum^{N}_{\chi=2} (\chi-1) k_{i}(\chi) \left( \sum^{\infty}_{\chi'=\chi}k_i(\chi')-1\right),
\end{align}
which is the equality (\ref{eq:qimax2}) in the main text, which we were meant to show. 

\section{Asymptotic expression of Lind's and Zhang's clustering coefficients for large  cardinalities} \label{app:Clustering_inf} 

We show that Eqs.~(\ref{eq:CLindCont}) and (\ref{eq:ZhangCont}) recover Eqs.~(\ref{def:CLind}) and (\ref{def:CZhang}) if $i$ is  connected to two hyperedes $\alpha$ and $\beta$ and if the  cardinality $\chi_{\alpha}\rightarrow\infty$ with the ratio  $r=\chi_{\beta}/\chi_{\alpha}>1$ fixed.

\subsection{Lind's clustering coefficient}
\noindent
If node $i$ is  connected two hyperedges $\alpha$ and $\beta$, Lind's clustering coefficient takes the form
\begin{align} 
C^{\rm Lind}_{i}=&
\frac{Q_i}
{(\chi_{\alpha}-1-Q_i)(\chi_{\beta}-1-Q_i)+Q_i}.
\end{align}
For large values of  $\chi_\alpha $, we can set $\chi_\beta = r\chi_\alpha$ and $Q_i = q\chi_\alpha$, which yields, after neglecting   leading order terms, the expression
\begin{align}
C^{\rm Lind}_{i} =\frac{q\chi_\alpha}{\chi^2_{\alpha}(1- q)(r-q)+\chi_{\alpha} q}=\frac{q}{\chi_{\alpha} (1-q)(r-q)+q}.
\end{align}

\subsection{Zhang's clustering coefficient}
\noindent
If node $i$ is only connected to hyperedges $\alpha$ and $\beta$, then Zhang's clustering coefficient takes the form
\begin{align} 
C^{\rm Zhang}_{i}=&
\frac{Q_i}{(\chi_{\alpha}-1)+(\chi_{\beta}-1)-Q_i},
\end{align}
For large values of $\chi_{\alpha}$, we can set $\chi_\beta = r\chi_\alpha$ and $Q_i = q\chi_\alpha$ as same as previous subsection. After neglecting   leading order terms, the expression

\begin{align} 
C^{\rm Zhang}_{i}=&\frac{Q_i}{(\chi_{\alpha}-1)(1+r)-Q_i}
=\frac{q}{1+r-q}.
\end{align}

\section{Explanation of the two configurations for $C^{\rm Lind}_i$ considered in the lower panel of Fig.~\ref{fig:clustering_landscape} }\label{app:QDetail} 
In the lower panel of Fig.~\ref{fig:clustering_landscape}  we consider motifs consisting of a central node $i$,  three hyperedges $\alpha$, $\beta$, and $\gamma$, and a  given number $Q_i(\mathbf{I})$ of quads.     There are different ways of assigning quads to a given node $i$ and three hyperedges, and this leads to different values of the Lind clustering coefficients $C^{{\rm Lind}}_i$, as shown in Fig.~\ref{fig:clustering_landscape}.   In this Appendix, we specify the two ways of assigning quads to  $i$ that have been considered in Fig.~\ref{fig:clustering_landscape} and which we call the uniform and the biased case.    Since there are three hyperedges, the different ways of assigning quads to these three hyperedges are fully determined by the numbers $q_{i\alpha\beta}(\mathbf{I})$, $q_{i\beta\gamma}(\mathbf{I})$, and $q_{i\alpha\gamma}(\mathbf{I})$ that denote the number of quads incident to node $i$ and two given hyperedges (see Eq.~(\ref{eq:qialphabeta}) for the definition).      
The example considered in Fig.~\ref{fig:clustering_landscape} has   cardinalities $\chi_{\alpha}=15$, $\chi_{\beta}=20$, and  $\chi_{\gamma}=25$, and therefore we focus on this case.

\subsection{Uniform case} \label{app:uniform}
In the uniform case, we assign uniformly and sequentially quads to the three hyperedges $\alpha$, $\beta$, and $\gamma$.   This gives 
\begin{equation}
 q_{i\alpha\beta}(Q_{i}(\mathbf{I})) = \sum^{13}_{a=0}\Theta(Q_{i}(\mathbf{I})-3a), \quad 
 q_{i\alpha\gamma}(Q_{i}(\mathbf{I})) = \sum^{13}_{a=0}\Theta(Q_{i}(\mathbf{I})-3a-2), 
 \end{equation} 
 and 
 \begin{equation}
 q_{i\beta\gamma}(Q_{i}(\mathbf{I})) = \sum^{13}_{a=0}\Theta(Q_{i}(\mathbf{I})-3a-1)+\sum^{46}_{b=42}\Theta(Q_{i}(\mathbf{I})-b),
\end{equation}
where  $\Theta(x)$ is the Heaviside function as defined below Eq.~(\ref{eq:AProj}).     We illustrate this configuration   in Panel (a) of Fig.~\ref{fig:app_common_nei} for the case of  $Q_i(\mathbf{I})=6$.

\begin{figure*}[t]
     \centering
     \setlength{\unitlength}{0.1\textwidth}
     \begin{subfigure}[b]{0.45\textwidth}
         \centering
         \includegraphics[width=0.6\textwidth]{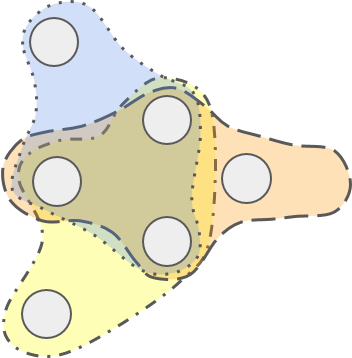}
     \end{subfigure}
      \put(-3.9,2.95){\normalsize$(a)$}
      \put(-3.2,1.3){\small $i$}
      \put(-3.6,0.7){\small $\alpha$}
      \put(-1.7,1.8){\small $\beta$}
      \put(-2.7,2.5){\small $\gamma$}
      \put(-3.0,2.25){\small $\times 22$}
      \put(-3.00,0.3){\small $\times 12$}
      \put(-1.45,1.34){\small $\times 17$}
     \hfill
     \begin{subfigure}[b]{0.45\textwidth}
         \centering
         \includegraphics[width=0.7\textwidth]{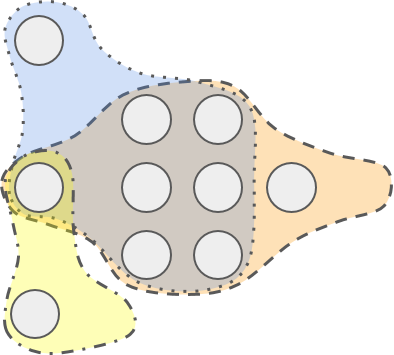}
     \end{subfigure}
      \put(-4.1,2.95){\normalsize$(b)$}
      \put(-3.55,1.3){\small $i$}
      \put(-3.9,0.7){\small $\alpha$}
      \put(-1.4,1.8){\small $\beta$}
      \put(-2.9,2.45){\small $\gamma$}
      \put(-3.3,2.27){\small $\times 18$}
      \put(-3.3,0.28){\small $\times 14$}
      \put(-1.2,1.3){\small $\times 13$}
      \hspace*{\fill}
        \caption{{\it Illustration of the   configurations of quads in the uniform and biased case as defined in Appendices~\ref{app:uniform} and \ref{app:biased}, respectively, for the case $Q_i = 6$.} The yellow shaded area bounded by a dash-dotted line  denotes  hyperedge $\alpha$; the blue shaded area bounded by  a dashed line represents  hyperedge $\beta$; and the orange shaded area with a dotted border represents  hyperedge $\gamma$.   Panel (a):  Three nodes, viz., $i$ and two other nodes, are incident to the three hyperedges $\alpha$, $\beta$, and $\gamma$, yielding $Q_i=6$.   Panel (b): Seven nodes, viz., $i$ and six other nodes, are incident to the two hyperedges $\gamma$ and $\beta$, yielding $Q_i=6$.   }
        \label{fig:app_common_nei}
\end{figure*}
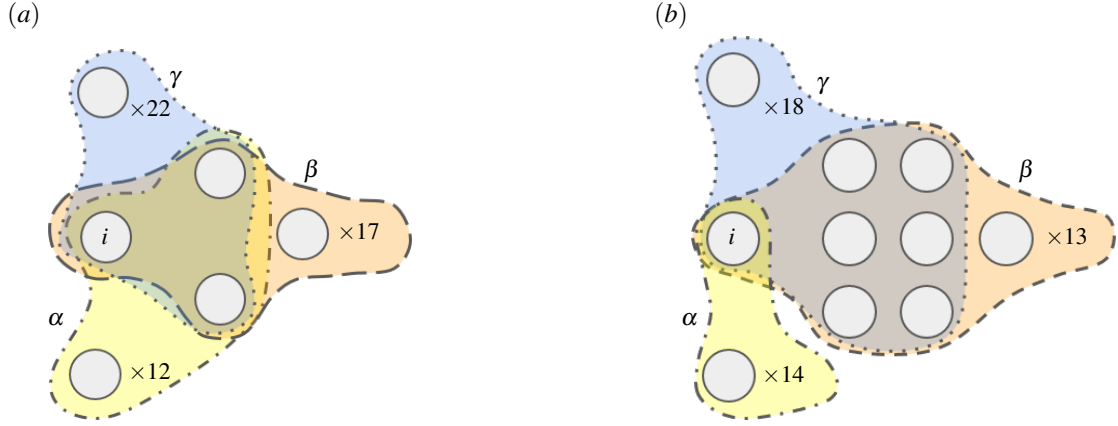

\subsection{Biased case} \label{app:biased}
This the opposing case where quads are fully assigned to one hyperedge, before assigning them to the other hyperedges.    In this case, we get 
\begin{equation}
\begin{split}
 q_{i\alpha\beta}(Q_{i}(\mathbf{I})) &= \sum^{28}_{a=24}\Theta(Q_{i}(\mathbf{I})-a)+\sum^{38}_{b=34}\Theta(Q_{i}(\mathbf{I})-b)+\sum^{46}_{c=43}\Theta(Q_{i}(\mathbf{I})-c),\\
 q_{i\alpha\gamma}(Q_{i}(\mathbf{I})) &= \sum^{23}_{a=19}\Theta(Q_{i}(\mathbf{I})-a)+\sum^{33}_{b=29}\Theta(Q_{i}(\mathbf{I})-b)+\sum^{42}_{c=39}\Theta(Q_{i}(\mathbf{I})-c),
 \end{split}
\end{equation}
 and 
\begin{equation}
\begin{split}
 q_{i\beta\gamma}(Q_{i}(\mathbf{I})) &= \sum^{18}_{a=0}\Theta(Q_{i}(\mathbf{I})-a).
\end{split}
\end{equation}
In  Panel (b) of Fig.~\ref{fig:app_common_nei}  we illustrate the biased case when $Q_i(\mathbf{I})=6$.

\section{Average quad clustering coefficent for random hypergraph models with regular cardinalities}
Building on random graph methods as developed in Refs.~\cite{coolen2017generating, newman2006structure, barabasi,dorogovtsev2022nature},   we derive in this Appendix the expressions (\ref{eq:clustering_finite_limit}), (\ref{eq:CAvkchi}) and (\ref{eq:CqkPresc}) for the average quad clustering coefficients of random hypergraph models with regular cardinalities.   In Appendix~\ref{app:chireg}, we derive Eq.~(\ref{eq:clustering_finite_limit}), and in Appendix~\ref{app:chiveckreg},   we derive Eq.~(\ref{eq:CqkPresc}).  Since (\ref{eq:CAvkchi}) is a special limiting case of (\ref{eq:CqkPresc}), we do not discuss it separately.

\subsection{$\chi$-regular ensemble}\label{app:chireg}
We derive the formula (\ref{eq:clustering_finite_limit}) for the average quad clustering coefficient of hypergraphs drawn from the ensemble $P_{\chi}(\mathbf{I})$ as defined in Eq.~(\ref{Eq:rand_model}).    

\subsubsection{Normalisation constant of $P_\chi$}
The normalisation constant in  Eq.~(\ref{Eq:rand_model}) is given by 
\begin{eqnarray} 
\mathcal{N}_{\chi} = \sum_{\mathbf{I}} \prod^M_{\gamma=1}\delta_{\chi,\chi_{\gamma}(\mathbf{I})}
=\left[\binom{N}{\chi}\right]^{M}
\label{eq:norm_coeff_chi},
\end{eqnarray}
as each hyperedge is connected to $\chi$ nodes that are randomly selected from the $N$ available options.  

\subsubsection{Average clustering coefficient}
Substituting the definition of the quad clustering coefficient, Eq.~(\ref{def:CQuad}), into the expression (\ref{eq:defensembleAverageCQ}) for the ensemble average clustering coefficient yields 
\begin{align} 
\mathcal{N}_{\chi}\langle C^{\rm q}_i(\textbf{I}) \rangle_{\chi}&= \frac{1}{\chi-1}\Bigg\langle  \sum^{\infty}_{q=2} \delta_{q,k_i({\bf I})}\delta_{\chi \vec{1},\vec{\chi}({\bf I})}
\frac{\sum_{\alpha,\beta,\alpha<\beta}\sum_{g \notin \left\{i\right\}}I_{g\alpha}I_{g\beta}I_{i\alpha}I_{i\beta}}
{\sum_{\alpha<\beta}I_{i\alpha}I_{i\beta}} \Bigg\rangle \nonumber\\
&=\sum^{\infty}_{q=2}\frac{2}{q(q-1)(\chi-1)}\sum_{\alpha,\beta,\alpha<\beta}\sum^N_{j=1;j\neq i}\Bigg\langle \delta_{q,k_i(\mathbf{I})}\delta_{\chi \vec{1},\vec{\chi}({\bf I})}I_{j\alpha}I_{j\beta}I_{i\alpha}I_{i\beta}\Bigg \rangle_{\chi} \nonumber \\ 
&=\sum^{\infty}_{q=2}\frac{2}{q(q-1)(\chi-1)}\int^{2\pi}_0\frac{d\hat{q}}{2\pi}e^{\text{i}\hat{q}q}   \int_{[0,2\pi]^M} \prod_{\xi=1}^{M}\frac{d\hat{\Xi}_{\xi}}{2\pi}e^{\text{i}\hat{\Xi}_{\xi}\chi}
\nonumber\\
&
\times \sum_{\alpha,\beta,\alpha<\beta}\sum^N_{j=1;j\neq i}\Bigg\langle e^{-\text{i}\hat{q}k_i(\mathbf{I})} \prod_{\xi'=1}^{M}e^{-\text{i}\hat{\Xi}_{\xi'}\sum_{o}I_{o\xi'}}I_{j\alpha}I_{j\beta}I_{i\alpha}I_{i\beta}\Bigg \rangle,
\label{Eq:clu_random1}
\end{align}
where we have used the notation 
\begin{equation}
\langle f(\mathbf{I}) \rangle \equiv \sum_{\mathbf{I}\in \left\{0,1\right\}^{NM}}f(\mathbf{I}).
\end{equation}

Performing the sum over all the entries $I_{j\alpha}$ of the incidence matrix $\mathbf{I}$ yields 
\begin{eqnarray} 
\mathcal{N}_{\chi}\langle C^{\rm q}_i(\textbf{I}) \rangle_{\chi}&=& \sum^{\infty}_{q=2}\frac{2}{q(q-1)(\chi-1)}\int^{2\pi}_0\frac{d\hat{q}}{2\pi}e^{\text{i}\hat{q}q}   \int_{[0,2\pi]^M} \prod_{\xi=1}^{M}\frac{d\hat{\Xi}_{\xi}}{2\pi}e^{\text{i}\hat{\Xi}_{\xi}\chi} 
\nonumber\\ 
&& \times\sum_{\alpha,\beta,\alpha<\beta}\sum^N_{j=1;j\neq i} e^{-2\text{i}\hat{q}}e^{-2\text{i}\hat{\Xi}_{\alpha}}e^{-2\text{i}\hat{\Xi}_{\beta}}\prod_{\gamma\notin\left\{\alpha,\beta\right\}}\left[e^{-\text{i}\hat{q}} e^{-\text{i}\hat{\Xi}_{\gamma}}+1\right]\nonumber\\
&& \times\left(e^{-\text{i}\hat{\Xi}_{\epsilon}}+1\right)^{N-1} \left(e^{-\text{i}\hat{\Xi}_{\alpha}}+1\right)^{N-2} \left(e^{-\text{i}\hat{\Xi}_{\beta}}+1\right)^{N-2} . \label{eq:intermediateReg}
\end{eqnarray}  
Expanding the power expressions in (\ref{eq:intermediateReg}) and integrating  over the $\hat{\Xi}_{\gamma}$ variables we get 
\begin{eqnarray} 
\mathcal{N}_{\chi}\langle C^{\rm q}_i(\textbf{I}) \rangle&=& \sum^{\infty}_{q=2}\frac{M(M-1)(N-1)}{q(q-1)(\chi-1)}\int^{2\pi}_0\frac{d\hat{q}}{2\pi}e^{\text{i}\hat{q}q}   e^{-2\text{i}\hat{q}} 
\left(\begin{array}{c}N-2 \\ \chi-2\end{array}\right)^2\left(\left(\begin{array}{c}N-1 \\ \chi-1\end{array}\right)e^{-\text{i}\hat{q}} + \left(\begin{array}{c}N-1 \\ \chi\end{array}\right)\right)^{M-2}. \label{eq:before2}
\end{eqnarray}
Further, expanding the power in (\ref{eq:before2}) and integrating over $\hat{q}$ reduces the expression into
\begin{eqnarray}  
 \mathcal{N}_{\chi}\langle C^{\rm q}_i(\textbf{I}) \rangle
&=&  \frac{\chi-1}{N-1}\left\{\left[\binom{N-1}{\chi-1}+\binom{N-1}{\chi}\right]^{M}-\left[\binom{N-1}{\chi}\right]^{M}-M\binom{N-1}{\chi-1}\binom{N-1}{\chi}^{M-1}\right\}. \label{eq:before}
\end{eqnarray} 
Lastly, dividing (\ref{eq:before}) by the normalisation constant (\ref{eq:norm_coeff_chi}) gives Eq.~(\ref{eq:clustering_finite_limit}), which we were meant to derive.

\subsection{$\chi$-regular with prescribed degree sequence}
\label{app:chiveckreg}
We derive the formula (\ref{eq:CqkPresc}) for the average quad clustering coefficient of the $\chi$-regular hypergraph ensemble  with a prescribed degree sequence $\vec{k}$, as defined in Eq.~(\ref{Eq:irregular_model}), in the limit $N\rightarrow \infty$ with fixed ratio 
\begin{equation}
 \mu \equiv  \frac{M}{N} = \frac{c}{ \chi} ,
\end{equation}
and where
\begin{equation}
c\equiv \frac{\sum^N_{j=1} k_j}{N}.
\end{equation}

The calculations are facilitated by  rewriting the expression for $P_{\vec{k},\chi}$ in the following form 
\begin{eqnarray} 
P_{\vec{k},\chi}(\textbf{I}) = \frac{1}{\mathcal{M}_{\vec{k},\chi}}\prod^N_{i=1} \prod^M_{\alpha=1}[p_{\ast}\delta_{I_{i\alpha},1}+(1-p_{\ast})\delta_{I_{i\alpha},0}]\prod^N_{j=1}\delta_{k_j,k_{j}(\mathbf{I})}\prod^M_{\alpha=1}\delta_{\chi,\chi_{\alpha}(\mathbf{I})}
\label{Eq:irregular_model2}
\end{eqnarray}
where $\mathcal{M}_{\vec{k},\chi}$ is the new normalisation constant that depends on the value of $p_{\ast}\in [0,1]$.   When $p^\ast=1/2$, we recover the expression Eq.~(\ref{Eq:irregular_model}).   Introducing a value $p^\ast\neq 1/2$ is a calculation trick that does not affect the average value of observables, such as $\langle C^q_i\rangle_{\vec{k},\chi}$, but it does alter the normalisation constant.   

In Appendix~\ref{app:IrregNorm}, we determine the normalisation constant $\mathcal{M}_{\vec{k},\chi}$, and in Appendix~\ref{app:IrregCQ} we calculate the average clustering coefficient.

\subsubsection{Normalisation constant of $P_{\vec{k},\chi}$}\label{app:IrregNorm}
From the definition of $\mathcal{M}_{\vec{k},\chi}$ as  the normalisation constant of $P_{\vec{k},\chi}$, as defined in  Eq.~(\ref{Eq:irregular_model2}),  it follows that  
\begin{align}
\mathcal{M}_{\vec{k},\chi} &= \sum_{\mathbf{I}} \prod_{i,\alpha}[p_{\ast}\delta_{I_{i\alpha},1}+(1-p_{\ast})\delta_{I_{i\alpha},0}]\prod^N_{j=1}\delta_{k_{j},k_{j}(\mathbf{I})}\prod^M_{\gamma=1}\delta_{\chi,\chi_{\gamma}(\mathbf{I})}\nonumber\\
&=\sum_{\mathbf{I}} \int_{[0,2\pi]^N}\prod_{j=1}^{N}\frac{d\hat{k}_{j}}{2\pi}e^{\text{i}\hat{k}_{j}k_{j}}\int_{[0,2\pi]^M} \prod_{\gamma=1}^{M}\frac{d\hat{\Xi}_{\gamma}}{2\pi}e^{\text{i}\hat{\Xi}_{\gamma}\chi}\prod_{i,\alpha}[p_{\ast}\delta_{I_{i\alpha},1}e^{-\text{i}\hat{k}_{i}}e^{-\text{i}\hat{\Xi}_{\alpha}}+(1-p_{\ast})\delta_{I_{i\alpha},0}], \label{eq:afterDir}
\end{align} 
where we have expressed the Kronecker delta functions  as integrals in order to get an expression that factorises in the $\mathbf{I}$ variables.  

Summing over the $\mathbf{I}$-variables we get 
\begin{eqnarray} 
\mathcal{M}_{\vec{k},\chi} &=&
\int\prod_{j=1}^{N}\frac{d\hat{k}_{j}}{2\pi}e^{\text{i}\hat{k}_{j}k_{j}}\int \prod_{\gamma=1}^{M}\frac{d\hat{\Xi}_{\gamma}}{2\pi}e^{\text{i}\hat{\Xi}_{\gamma}\chi} \prod_{i,\alpha}\left[p^\ast e^{-\text{i}\hat{k}_i-\text{i}\hat{\Xi}_{\alpha}} + 1-p^\ast\right]. 
\end{eqnarray}

We set  $p_\ast = \rho_\ast/N$ and take the limit 
$N\rightarrow \infty$ for fixed $M/N$  to obtain   \begin{equation}
 \prod^M_{\alpha=1}\left[1 + \frac{\rho_\ast}{N}\left(e^{-\text{i}\hat{k}_i-\text{i}\hat{\Xi}_{\alpha}}-1\right)\right] = \exp\left[\rho_\ast M\frac{ e^{-\text{i}\hat{k}_i}}{N} \frac{\sum^M_{\alpha=1}e^{-\text{i}\hat{\Xi}_{\alpha}}}{M}-\frac{\rho_\ast M}{N} + \mathcal{O}(1/N)\right],
\end{equation}
where $\mathcal{O}(1/N)$ represents a subleading order term that decays as $\sim 1/N$ for large valus of $N$.  The constant $\rho_\ast\in \mathbb{R}_+$ is an arbitrary constant that determines the normalisation constant but    disappears in the final expression of the average clustering coefficient.   
Identifying the term $\sum^M_{\alpha=1}e^{-\text{i}\hat{\Xi}_{\alpha}}$ in the exponent, and introducing the Dirac distribution
\begin{equation}
\int_{\mathbb{R}} d\iota \delta\left(\iota -\frac{\sum^M_{\alpha=1} e^{-\text{i}\hat{\Xi}_\alpha}}{M} \right) = 1
\end{equation}

yields 
\begin{align} 
\mathcal{M}_{\vec{k},\chi} 
&=\int_{\mathbb{R}^2} \frac{d\iota d\hat{\iota}}{2\pi} \prod_{j=1}^{N} \left(\int^{2\pi}_0\frac{d\hat{k}_{j}}{2\pi}  e^{\text{i}\hat{k}_{j}k_{j} + \rho_\ast M\iota e^{-\text{i}\hat{k}_{j}}/N} \right) \prod_{\gamma=1}^{M} \left(\int^{2\pi}_0 \frac{d\hat{\Xi}_{\gamma}}{2\pi}e^{\text{i}\hat{\Xi}_{\gamma}\chi + \text{i}\hat{\iota}e^{-\text{i}\hat{\Xi}_{\gamma}}/M}  \right)e^{-\text{i}\hat{\iota}\iota} e^{-\rho_\ast M+\mathcal{O}_N(1)}.\label{eq:NInt}
\end{align} 
We determine the integrals by expressing the exponentials in terms of their Taylor series, 
\begin{eqnarray} 
\int^{2\pi}_0\frac{d\hat{k}_{j}}{2\pi}  e^{\text{i}\hat{k}_{j}k_{j} +\rho_\ast  M\iota e^{-\text{i}\hat{k}_{j}}/N}  =  \int^{2\pi}_0\frac{d\hat{k}_{j}}{2\pi} e^{\text{i}\hat{k}_{j}k_{j}} \sum^{\infty}_{\ell=0} \left(\frac{\rho_\ast M\iota}{N}\right)^{\ell}\frac{e^{-\text{i}\ell \hat{k}_j}}{\ell!} = \frac{\left( \rho_\ast M\iota\right)^{k_{j}}}{N^{k_{j}}k_{j}!}, \label{eq:exp1}
\end{eqnarray}
and analogously we get
\begin{eqnarray}  
\int^{2\pi}_0 \frac{d\hat{\Xi}_{\gamma}}{2\pi}e^{\text{i}\hat{\Xi}_{\gamma}\chi + \text{i}\hat{\iota}e^{-\text{i}\hat{\Xi}_\gamma}/M}  = \frac{\left(\text{i}\hat{\iota}\right)^{\chi}}{M^{\chi}\chi!} . \label{eq:exp2}
\end{eqnarray}
Using the expressions (\ref{eq:exp1}) and (\ref{eq:exp2}) in Eq.~(\ref{eq:NInt}) gives 
\begin{align}
\mathcal{M}_{\vec{k},\chi} 
&= \int_{\mathbb{R}^2} \frac{d\iota d\hat{\iota}}{2\pi}\left(\prod^N_{j=1}\frac{1}{k_{j}!}\right)\left[\frac{\rho_\ast M\iota}{N}\right]^{\rho_\ast M} \left[\frac{\left(\text{i}\hat{\iota}\right)^{\chi}}{M^{\chi}\chi!} \right]^{M}e^{-\text{i}\hat{\iota}\iota-\rho_\ast M +\mathcal{O}_N(1)}.
\end{align} 
Using $M = \mu N$ and making the transformation $\hat{\iota}\rightarrow\mu N\hat{\iota}$, we get the saddle point integral
\begin{eqnarray} 
\mathcal{M}_{\vec{k},\chi} = \frac{\mu e^{-\rho_\ast M}N}{(\chi!)^M\prod_{j} k_{j}!}   \int_{\mathbb{R}^2} \frac{d\iota d\hat{\iota}}{2\pi}  e^{ N\Psi(\iota,\hat{\iota})+\mathcal{O}_N(1)}\label{eq:numeratorExp}
\end{eqnarray} 
with exponent 
\begin{eqnarray} 
\Psi(\iota,\hat{\iota}) = -\text{i}\mu\hat{\iota}\iota+c\log \rho_\ast \mu\iota+\mu\chi\log\text{i}\hat{\iota}. \label{eq:psi_func} 
\end{eqnarray}

In the limit of $N\rightarrow \infty$, the saddle point dominates, and we get the expression 
\begin{eqnarray} 
\mathcal{M}_{\vec{k},\chi} &=& \Phi_{\vec{k},\chi} e^{N\Psi(\iota^\ast,\hat{\iota}^\ast)+\mathcal{O}_N(1)},\label{eq:saddlePointEq1} 
\end{eqnarray} 
where $\iota^\ast$ and $\hat{\iota}^\ast$ solve the saddle point equation 
\begin{eqnarray} 
 \text{i}\mu\hat{\iota}^\ast\iota^\ast=c =\mu\chi,
\label{eq:prefactors_saddlePoint}
\end{eqnarray} 
and where 
\begin{eqnarray} 
\Phi_{\vec{k},\chi} = \frac{\mu e^{-\rho_\ast M}}{(\chi!)^M\prod_{j} k_{j}!}\frac{1}{\sqrt{{\rm det}\mathcal{H}}} \quad
\end{eqnarray}  
with $\mathcal{H}$ the Hessian of the function $\Psi$ evaluated at the saddle point.  
Using Eq.~(\ref{eq:prefactors_saddlePoint}) in (\ref{eq:saddlePointEq1}) we obtain the final expression 
\begin{eqnarray} 
\mathcal{M}_{\vec{k},\chi} &=& \Phi_{\vec{k},\chi} e^{ Nc(\log \rho_\ast c-1)+\mathcal{O}_N(1)}.
\label{eq:norm_coeffi}
\end{eqnarray} 
 As will become evident, the prefactor $\Phi_{\vec{k},\chi}$  cancels out with an identical prefactor that appears in the numerator of the derivation for $\langle C^{\rm q}_i(\mathbf{I})\rangle_{\vec{k}, \chi}$, which we do in the next Section.

\subsubsection{Average clustering coefficient}\label{app:IrregCQ}
Using the definition of the clustering coefficient, given by Eq.~(\ref{def:CQuad}), and the fact that in this model all cardinalities  are fixed to $\chi$, i.e.,  $\chi_\alpha=\chi$, we get 
\begin{eqnarray} 
\langle C^{\rm q}_{i}(\textbf{I}) \rangle_{\vec{k},\chi}
&=&\frac{2}{k_{i}(k_{i}-1)(\chi-1)}\left[1-\frac{\sum^N_{i=1}(\delta_{0,k_i}+\delta_{1,k_i})}{N}\right]
\nonumber\\ 
&& \times \frac{1}{\mathcal{M}_{\vec{k},\chi}} \sum_{\mathbf{I}}\sum_{\alpha,\beta,\alpha<\beta}\sum^N_{j=1;j\neq i}
\delta_{\vec{k},\vec{k}({\bf I})}\delta_{\chi \vec{1},\vec{\chi}({\bf I})}I_{j\alpha}I_{j\beta}I_{i\alpha}I_{i\beta}\prod^N_{g=1} \prod^M_{\epsilon=1}[p_{\ast}\delta_{I_{g\epsilon},1}+(1-p_{\ast}) \delta_{I_{g\epsilon},0}] . \label{eq:cqAvIrreg}
\end{eqnarray}
We represent the Kronecker delta functions in Eq.~(\ref{eq:cqAvIrreg}) as integrals, and then sum over the $\mathbf{I}$-variables, yielding
\begin{multline}
\mathcal{M}_{\vec{k},\chi}\langle C^{\rm q}_i(\textbf{I}) \rangle=\frac{2}{k_{i}(k_{i}-1)(\chi-1)}
\left[1-\frac{\sum^N_{i=1}(\delta_{0,k_i}+\delta_{1,k_i})}{N}\right]\int\prod_{n=1}^{N}\frac{d\hat{k}_{n}}{2\pi}e^{\text{i}\hat{k}_{n}k_{n}}\int \prod_{\xi=1}^{M}\frac{d\hat{\Xi}_{\xi}}{2\pi}e^{\text{i}\hat{\Xi}_{\xi}\chi}\\
\times\sum_{\alpha,\beta;\alpha<\beta}\sum^N_{j=1;j\neq i} (p_{\ast})^{4}e^{-2\text{i}\hat{k}_{i}}e^{-2\text{i}\hat{k}_{j}}e^{-2\text{i}\hat{\Xi}_{\alpha}}e^{-2\text{i}\hat{\Xi}_{\beta}}\prod_{(g,\epsilon)}'\left[p_{\ast}e^{-\text{i}\hat{k}_{g}} e^{-\text{i}\hat{\Xi}_{\epsilon}}+(1-p_{\ast})\right],
\end{multline}
where $\prod'_{(g,\epsilon)}$ is a product over all pairs $(g,\epsilon)\in \V\times \W$, but excluding  $\left\{(i,\alpha),(i,\beta),(j,\alpha),(j,\beta)\right\}$.
Setting $p^\ast = \rho_\ast/N$ and taking the limit $N\rightarrow \infty$, we get for all $g\notin \left\{i,j\right\}$ that 
\begin{eqnarray}
 \prod^M_{\epsilon=1}\left[\frac{\rho_\ast}{N}e^{-\text{i}\hat{k}_{g}} e^{-\text{i}\hat{\Xi}_{\epsilon}}+(1-\frac{\rho_\ast}{N})\right] = \exp\left[\rho_\ast M\frac{e^{-\text{i}\hat{k}_g}}{N} \frac{\sum^M_{\epsilon=1}e^{-\text{i}\hat{\Xi}_{\epsilon}}}{M}-\frac{\rho_\ast M}{N}+\mathcal{O}(M/N^2)\right]
\end{eqnarray}
and  for $g\in \left\{i,j\right\}$ we get 
\begin{eqnarray}
 \prod^M_{\substack{\epsilon=1;\\\epsilon\neq\{\alpha,\beta\}}}\left[\frac{\rho_\ast}{N}e^{-\text{i}\hat{k}_{g}} e^{-\text{i}\hat{\Xi}_{\epsilon}}+(1-\frac{\rho_\ast}{N})\right] = \exp\left[\rho_\ast M\frac{e^{-\text{i}\hat{k}_g}}{N} \frac{\sum_{\epsilon\notin\{\alpha,\beta\}}e^{-\text{i}\hat{\Xi}_{\epsilon}}}{M}-\frac{\rho_\ast(M-2)}{N}+\mathcal{O}(M/N^2)\right].
\end{eqnarray}
Introducing 
\begin{equation}
\int d\iota \delta\left(\iota - \frac{\sum^M_{\alpha=1}e^{-\text{i}\hat{\Xi}_\alpha}}{M} \right) = 1
\end{equation}
this simplifies into 
\begin{multline*}
\mathcal{M}_{\vec{k},\chi}\langle C^{\rm q}_i(\textbf{I}) \rangle
=\frac{2}{k_{i}(k_{i}-1)(\chi-1)}\left[1-\frac{\sum^N_{i=1}(\delta_{0,k_i}+\delta_{1,k_i})}{N}\right]\int \frac{d\iota d\hat{\iota}}{2\pi}\sum_{\substack{\alpha,\beta;\\\alpha<\beta}}\sum^N_{j=1;j\neq i}\prod_{n=1}^{N}\left(\int\frac{d\hat{k}_{n}}{2\pi}e^{\text{i}\hat{k}_{n}k_{n}+\rho_\ast M\iota e^{-\text{i}\hat{k}_{n}}/N}\right)\\
\times \prod_{\xi=1}^{M}\left(\int \frac{d\hat{\Xi}_{\xi}}{2\pi}e^{\text{i}\hat{\Xi}_{\xi}\chi+\text{i}\hat{\iota}e^{-\text{i}\hat{\Xi}_{\xi}}/M}\right)\left(\frac{\rho_\ast}{N}\right)^{4}e^{-2\text{i}\hat{k}_{i}}e^{-2\text{i}\hat{k}_{j}}e^{-2\text{i}\hat{\Xi}_{\alpha}}e^{-2\text{i}\hat{\Xi}_{\beta}}e^{-\rho_\ast M}e^{-\text{i}\hat{\iota}\iota+\mathcal{O}_N(1)}.
\end{multline*}
Integrating over $\hat{k}_n$ and $\hat{\Xi}_\xi$ and using the formula 
\begin{align*}
\int\frac{d\hat{k}_{n}}{2\pi}e^{\text{i}\hat{k}_{n}k_{n}+\rho_\ast M\iota e^{-\text{i}\hat{k}_{n}}/N}  &=  \int\frac{d\hat{k}_{n}}{2\pi}e^{\text{i}\hat{k}_{n}k_{n}}\sum^{\infty}_{\ell=0} \left(\frac{\rho_\ast M\iota}{N}\right)^{\ell}\frac{e^{-\text{i}\ell \hat{k}_n}}{\ell!} = \frac{\left(\rho_\ast M\iota\right)^{k_{n}}}{N^{k_{n}}k_{n}!}
\end{align*} 
yields 
\begin{multline*}
\mathcal{M}_{\vec{k},\chi}\langle C^{\rm q}_i(\textbf{I}) \rangle
=\frac{\mu N(\mu N-1)\chi^{2}(\chi-1)}{(\chi!)^M\prod_{j} k_{j}!}\left[1-\frac{\sum^N_{i=1}(\delta_{0,k_i}+\delta_{1,k_i})}{N}\right]\int \frac{d\iota d\hat{\iota}}{2\pi}\\
\times\sum^N_{j=1;j\neq i}\left(\frac{\rho_\ast}{N}\right)^{4}k_{j}(k_{j}-1)e^{-\rho_\ast M}e^{-\text{i}\hat{\iota}\iota}\left(\frac{\rho_\ast M\iota}{N}\right)^{\rho_\ast M-4}\left(\frac{\text{i}\hat{\iota}}{M} \right)^{M\chi-4}e^{\mathcal{O}_N(1)}.
\end{multline*}
After the transformation $\hat{\iota}\rightarrow\mu N\hat{\iota}$, we can write this expression as the following  saddle point integral 
\begin{align}
\mathcal{M}_{\vec{k},\chi}\langle C^{\rm q}_i(\textbf{I}) \rangle=\frac{\mu^{3}\chi^{2}(\chi-1)e^{-\rho_\ast M}}{N(\chi!)^M\prod_{j} k_{j}!}\left[1-\frac{\sum^N_{i=1}(\delta_{0,k_i}+\delta_{1,k_i})}{N}\right]\sum^N_{\substack{j=1;\\j\neq i}}k_{j}(k_{j}-1)\int \frac{d\iota d\hat{\iota}}{2\pi}\frac{e^{N\Psi(\iota,\hat{\iota})+\mathcal{O}_N(1)}}{\left(\mu\text{i}\hat{\iota}\iota\right)^{4}}
\end{align}
where $\Psi(\iota,\hat{\iota})$ is given by  Eq.~(\ref{eq:psi_func}).

In the limit of $N\rightarrow \infty$, the saddle point dominates.   However, since the exponent is identical to the one appearing in Eq.~(\ref{eq:numeratorExp}) for  $\mathcal{M}_{\vec{k},\chi}$, we get the simpler expression  
\begin{align}
\langle C^{\rm q}_i(\textbf{I}) \rangle&=\frac{\chi-1}{\left(N\mu\chi\right)^{2}}\left[1-\frac{\sum^N_{i=1}(\delta_{0,k_i}+\delta_{1,k_i})}{N}\right]\sum^N_{j=1;j\neq i}k_{j}(k_{j}-1)+\mathcal{O}\left(\frac{1}{N^2}\right)\nonumber\\
&= \frac{\chi-1}{\overline{k}^{2}N} 
\overline{ k(k-1)}\left(1-p_{\rm deg}(0)-p_{\rm deg}(1) \right)
+ \mathcal{O}\left(\frac{1}{N^2}\right),
\label{eq:derive_prescribed_k}
\end{align}
which is identical to Eq.~(\ref{eq:CqkPresc}) in the main text.  A comparison between  Eq.~(\ref{eq:derive_prescribed_k}) and the average quad clustering coefficient of large  numerically generated random graphs shows an excellent agreement (results not shown).  

If all terms of the degree sequence are equal (i.e., it is $(c,\chi)$-regular hypergraph), then Eq.~(\ref{eq:derive_prescribed_k}) becomes $\langle C^{\rm q}_i(\textbf{I}) \rangle=(c-1)(\chi-1)/(cN)+ \mathcal{O}\left(1/N^2\right)$.

\section{Average quad clustering coefficient for biregular cardinalities}\label{App:E}
We obtain the Eq.~(\ref{eq:biregAverage}) for the ensemble averaged   quad clustering coefficient of the model (\ref{Eq:rand_model2}) with biregular cardinalities.

\subsubsection{Normalisation constant of $P_{\chi_1,\chi_2}$}
By assumption, there  are $M_1$ hyperedges with cardinality $\chi_1$ and $M_2=M-M_1$ hyperedges with cardinality $\chi_2$.  The hyperedges and nodes  are connected randomly, given  their prescribed cardinalities.  
Therefore, the normalisation constant in  Eq.~(\ref{Eq:rand_model2}) is given by 
\begin{eqnarray} 
\mathcal{N}_{\chi_1,\chi_2} = \sum_{\mathbf{I}} \prod^{M_{1}}_{\alpha=1}\delta_{\chi_{1},\chi_{\alpha}(\mathbf{I})}\prod^{M}_{\beta=M_{1}+1}\delta_{\chi_{2},\chi_{\beta}(\mathbf{I})}= \left[\binom{N}{\chi_{1}}\right]^{M_{1}}\left[\binom{N}{\chi_{2}}\right]^{M_{2}}
\label{eq:norm_coeff_bireg}.
\end{eqnarray}

\subsubsection{Average clustering coefficient}

Using the definition of the clustering coefficient,   Eq.~(\ref{def:CQuad}), in the definition  of the average    clustering coefficient, Eq.~(\ref{eq:DefEnsemAvChi1Chi2}),  yields
\begin{multline}
\mathcal{N}_{\chi_1,\chi_2}\langle C_{q,i}(\textbf{I}) \rangle_{\chi_1,\chi_2}=\sum_{u=0}^{M_1}\sum_{v=0}^{M_{2}}\frac{2}{(\chi_{1}-1)(u+v)(u+v-1)+v(\chi_{2}-\chi_{1})(v-1)}\\
\times\sum_{\alpha,\beta,\alpha<\beta}\sum^N_{j=1;j\neq i} \sum_{\mathbf{I}}\delta_{u,k_i(\mathbf{I};\chi_{1})}\delta_{v,k_i(\mathbf{I};\chi_{2})}\prod_{\epsilon_{1}=1}^{M_{1}}\delta_{\chi_{1},\chi_{\epsilon_{1}}({\bf I})}\prod_{\epsilon_{2}=M_{1}+1}^{M}\delta_{\chi_{2},\chi_{\epsilon_{2}}({\bf I})}I_{j\alpha}I_{j\beta}I_{i\alpha}I_{i\beta}.
\end{multline}
Representing the Kronecker delta functions with integrals, we get 
\begin{multline}
\mathcal{N}_{\chi_1,\chi_2}\langle C^{\rm q}_i(\textbf{I}) \rangle_{\chi_1,\chi_2}=
\sum_{u=0}^{M_1}\sum_{v=0}^{M_{2}}\sum_{\vec{q}\in \mathbb{N}^N}\frac{2}{(\chi_{1}-1)(u+v)(u+v-1)+v(\chi_{2}-\chi_{1})(v-1)}\\
\times\int^{2\pi}_{0}\frac{d\hat{u}}{2\pi}e^{\text{i}\hat{u}u}\int^{2\pi}_{0}\frac{d\hat{v}}{2\pi}e^{\text{i}\hat{v}v}\int_{[0,2\pi]^N}\prod_{n=1}^{N}\frac{d\hat{q}_{n}}{2\pi}e^{\text{i}\hat{q}_{n}q_{n}}\int_{[0,2\pi]^{M_1}}\prod_{\xi=1}^{M_{1}}\frac{d\hat{\Xi}_{\xi}}{2\pi}e^{\text{i}\hat{\Xi}_{\xi}\chi_{1}}\int_{[0,2\pi]^{M_2}}\prod_{\xi=M_{1}+1}^{M}\frac{d\hat{\Xi}_{\xi}}{2\pi}e^{\text{i}\hat{\Xi}_{\xi}\chi_{2}}\\
\times\sum_{\alpha,\beta,\alpha<\beta}\sum^N_{j=1;j\neq i}\sum_{\mathbf{I}} e^{-\text{i}\hat{u}\sum_{\gamma}I_{i\gamma}\delta_{\chi_{1},\chi_{\gamma}(\mathbf{I})}}e^{-\text{i}\hat{v}\sum_{\gamma}I_{i\gamma}\delta_{\chi_{2},\chi_{\gamma}(\mathbf{I})}}
\prod_{n'=1}^{N}e^{-\text{i}\hat{q}_{n'}\sum_{\mu}I_{n'\mu}} \prod_{\xi'=1}^{M}e^{-\text{i}\hat{\Xi}_{\xi'}\sum_{o}I_{o\xi'}}I_{j\alpha}I_{j\beta}I_{i\alpha}I_{i\beta} .
\end{multline}

Summing over the $\mathbf{I}$ variables, and subsequently integrating over the $\hat{q}_n$ and $\hat{\Xi}_{\xi}$ variables, we get the expression 
\begin{multline}
\mathcal{N}_{\chi_1,\chi_2}\langle C^{\rm q}_i(\textbf{I}) \rangle_{\chi_1,\chi_2} =\sum_{u=0}^{M_1}\sum_{v=0}^{M_{2}}\frac{2(N-1)}{(\chi_{1}-1)(u+v)(u+v-1)+v(\chi_{2}-\chi_{1})(v-1)}\int^{2\pi}_{0}\frac{d\hat{u}}{2\pi}e^{\text{i}\hat{u}u}\int^{2\pi}_{0}\frac{d\hat{v}}{2\pi}e^{\text{i}\hat{v}v}\\
\times \left[(p_{\ast})^{\chi_{a}}(1-p_{\ast})^{N-\chi_{a}}\right]^{M_{1}}\left[(p_{\ast})^{\chi_{2}}(1-p_{\ast})^{N-\chi_{2}}\right]^{M_{2}}\\
\times \Biggl\{\biggl[\binom{M_{1}}{2}\left(\binom{N-2}{\chi_{1}-2}e^{-\hat{u}}\right)^{2}\left(\binom{N-1}{\chi_{1}-1}e^{-\text{i}\hat{u}}+\binom{N-1}{\chi_{1}}\right)^{M_{1}-2}\left(\binom{N-1}{\chi_{2}-1}e^{-\text{i}\hat{v}}+\binom{N-1}{\chi_{2}}\right)^{M_{2}}\\
+\binom{M_{1}}{1}\binom{M_{2}}{1}\binom{N-2}{\chi_{1}-2}e^{-\hat{u}}\binom{N-2}{\chi_{2}-2}e^{-\hat{v}}\left(\binom{N-1}{\chi_{1}-1}e^{-\text{i}\hat{u}}+\binom{N-1}{\chi_{1}}\right)^{M_{1}-1}\left(\binom{N-1}{\chi_{2}-1}e^{-\text{i}\hat{v}}+\binom{N-1}{\chi_{2}}\right)^{M_{2}-1}\\
+\binom{M_{2}}{2}\left(\binom{N-2}{\chi_{2}-2}e^{-\hat{v}}\right)^{2}\left(\binom{N-1}{\chi_{1}-1}e^{-\text{i}\hat{u}}+\binom{N-1}{\chi_{1}}\right)^{M_{1}}\left(\binom{N-1}{\chi_{2}-1}e^{-\text{i}\hat{v}}+\binom{N-1}{\chi_{2}}\right)^{M_{2}-2}\biggr]\Biggr\}.
\end{multline}
Lastly, integrating over the variables $\hat{u}$ and $\hat{v}$ yields
\begin{multline}
\mathcal{N}_{\chi_1,\chi_2}\langle C^{\rm q}_i(\textbf{I}) \rangle_{\chi_1,\chi_2} =\sum_{u=2}^{M_1}\sum_{v=0}^{M_{2}}\frac{2(N-1)\left[(p_{\ast})^{\chi_{1}}(1-p_{\ast})^{N-\chi_{1}}\right]^{M_{1}}\left[(p_{\ast})^{\chi_{2}}(1-p_{\ast})^{N-\chi_{2}}\right]^{M_{2}}}{(\chi_{1}-1)(u+v)(u+v-1)+v(\chi_{2}-\chi_{1})(v-1)}\\
\times\binom{M_{1}}{2}\left[\binom{N-2}{\chi_{1}-2}\right]^{2}\binom{M_{1}-2}{u-2}\left[\binom{N-1}{\chi_{1}-1}\right]^{u-2}\left[\binom{N-1}{\chi_{1}}\right]^{M_{1}-u}\binom{M_{2}}{v}\left[\binom{N-1}{\chi_{2}-1}\right]^{v}\left[\binom{N-1}{\chi_{2}}\right]^{M_{2}-v}\\
+\sum_{u=1}^{M_1}\sum_{v=1}^{M_{2}}\frac{2(N-1)\left[(p_{\ast})^{\chi_{1}}(1-p_{\ast})^{N-\chi_{1}}\right]^{M_{1}}\left[(p_{\ast})^{\chi_{2}}(1-p_{\ast})^{N-\chi_{2}}\right]^{M_{2}}}{(\chi_{1}-1)(u+v)(u+v-1)+v(\chi_{2}-\chi_{1})(v-1)}\\
\times M_{1}M_{2}\binom{N-2}{\chi_{1}-2}\binom{N-2}{\chi_{2}-2}\binom{M_{1}-1}{u-1}\left[\binom{N-1}{\chi_{1}-1}\right]^{u-1}\left[\binom{N-1}{\chi_{1}}\right]^{M_{1}-u}\binom{M_{2}-1}{v-1}\left[\binom{N-1}{\chi_{2}-1}\right]^{v-1}\left[\binom{N-1}{\chi_{2}}\right]^{M_{2}-v}\\
+\sum_{u=0}^{M_1}\sum_{v=2}^{M_{2}}\frac{2(N-1)\left[(p_{\ast})^{\chi_{1}}(1-p_{\ast})^{N-\chi_{1}}\right]^{M_{1}}\left[(p_{\ast})^{\chi_{2}}(1-p_{\ast})^{N-\chi_{2}}\right]^{M_{2}}}{(\chi_{1}-1)(u+v)(u+v-1)+v(\chi_{2}-\chi_{1})(v-1)}\\
\times\binom{M_{2}}{2}\left[\binom{N-2}{\chi_{2}-2}\right]^{2}\binom{M_{1}}{u}\left[\binom{N-1}{\chi_{1}-1}\right]^{u}\left[\binom{N-1}{\chi_{1}}\right]^{M_{1}-u}\binom{M_{2}-2}{v-2}\left[\binom{N-1}{\chi_{2}-1}\right]^{v-2}\left[\binom{N-1}{\chi_{2}}\right]^{M_{2}-v} ,\label{eq:finalBiregExpr}
\end{multline}
which gives the  Eqs.~(\ref{eq:biregAverage}-\ref{eq:phiuv}) in the main text after  dividing by the expression (\ref{eq:norm_coeff_bireg}) for the normalisation constant.

\section{Datasets for real-world hypergraphs}\label{app:data}
In Secs.~\ref{ch:real_hypergraph} and \ref{ch:directed_hypergraph} of this Paper,   
we have considered six nondirected hypergraphs.  These are: 
   \begin{enumerate}
     \item {\it NDC-substances}\cite{benson2018simplicial}: The nodes are substances, and the hyperedges are commercial drugs registered in by the U.S. Food and Drug Administration in the National Drug Code (NDC).   A node is linked to a hyperedge whenever the corresponding substance is used to synthesise the drug.  
     \item {\it Youtube}\cite{kunegis2013konect,mislove2009online}: Nodes represent YouTube users and  hyperedges represent Youtube channels with paid subscription.   A user is linked to a hyperedge when the user pays for the  membership service.
     \item {\it Food recipe}\cite{whats-cooking}: Nodes are ingredients and  hyperedges are recipes for food dishes. 
     \item {\it Github}\cite{kunegis2013konect,Scott2009GitHub}: Nodes are  GitHub users and  hyperedges are GitHub projects.  A node is linked to a hyperedge whenever the corresponding user contributes to the GitHub project.   
     \item {\it Crime involvement}\cite{kunegis2013konect}: The nodes are  suspects, and the hyperedges are  crime cases.   Nodes are linked to hyperedges whenever the corresponding suspects are involved with the crime investigation.  
     \item  {\it Wallmart}\cite{amburg2020clustering}: Nodes are products sold by Walmart, and the  hyperedges represent  purchase orders.  Nodes are linked to hyperedges whenever the corresponding products are  part of the purchased order.  
   \end{enumerate}

In Sec.~\ref{ch:real_dirhyper}, we have considered three directed hypergraphs: 
   \begin{enumerate}
     \item {\it DNC-email}\cite{kunegis2013konect}:  Nodes are  users sending and receiving emails and  hyperedges are  emails that are part of the 2016 Democratic National Committee (DNC) email leak.  Hyperedges are directed from the sender to its recipients.    Since an email always has  a single sender, all hyperedges have an  in-cardinality equal to one.   
     \item {\it Human metabolic pathways}\cite{karp2019biocyc}: Nodes represent metabolic compounds in the human metabolism, and  hyperedges are  metabolic reactions.   A hyperedge is directed from the reactants towards the products of the metabolic reaction, and metabolic reactions with very small rates are omitted, yielding a directed hypergraph.
     \item {\it English thesaurus}\cite{ward2002moby}: Nodes are English words and  hyperedges represent synonym relations between words.   Hyperedges are directed from a root word to target words.  Since not all words occur as root words, the hypergraph is directed.    The in-cardinality of each hyperedge equals to one.   
   \end{enumerate}

\section{Configuration model for hypergraphs} \label{app:conf_model}

We describe the algorithm used to generate a single instance from the configuration model for  hypergraphs.    There are two type of configuration models: the microcanonical ensemble that specifies the degree $\vec{k}(\mathbf{I})$ and cardinality  $\vec{\chi}(\mathbf{I})$ sequences and the canonical ensemble that specifies the distributions $P(k)$ and $P(\chi)$ for  the degrees and cardinalities of nodes and hyperedges, respectively.   In the microcanonical ensemble links are generated randomly between nodes and hyperedges given the specified sequences, while in the canonical ensemble we first generate these sequences, and then generate the links.

In Section \ref{ch:real_hypergraph} and \ref{ch:directed_hypergraph}, we use a micro-canonical ensemble with the  number of nodes $N$, hyperedges $M$, degree sequence $\vec{k}$,  and the cardinality sequence $\vec{\chi}$ as given by the  real-world hypergraph under study.  The links between the nodes and hyperedges are generated as follows.   We associate a number of stubs to the nodes and hyperedges of the graph corresponding to their degrees and cardinalities.   Subsequently,  
we randomly connect the stubs of nodes with those of  hyperedges with the additional constraints that there are no multiple links connecting the same pair of nodes and hyperedges.  
The upper Panel of Fig.~\ref{fig:hypergraph configuration model} shows an example of this process for the case of  $\vec{k}=(1,1,1,2,2,1,1,1)$ and $\vec{\chi}=(5,5)$.   An analogous process applies to directed   hypergraphs and is illustrated in the lower Panel of Fig.~\ref{fig:hypergraph configuration model}  for  $\vec{k}^{\rm in}_{i}=(0,0,0,1,1,1,1,1)$, $\vec{k}^{\rm out}=(1,1,1,1,1,1,1,0)$, $\vec{\chi}^{\rm in}=\{3,4\}$ and $\vec{\chi}^{\rm out}=\{2,3\}$.

\begin{figure}[h]
 \centering
 \includegraphics[width=\textwidth]{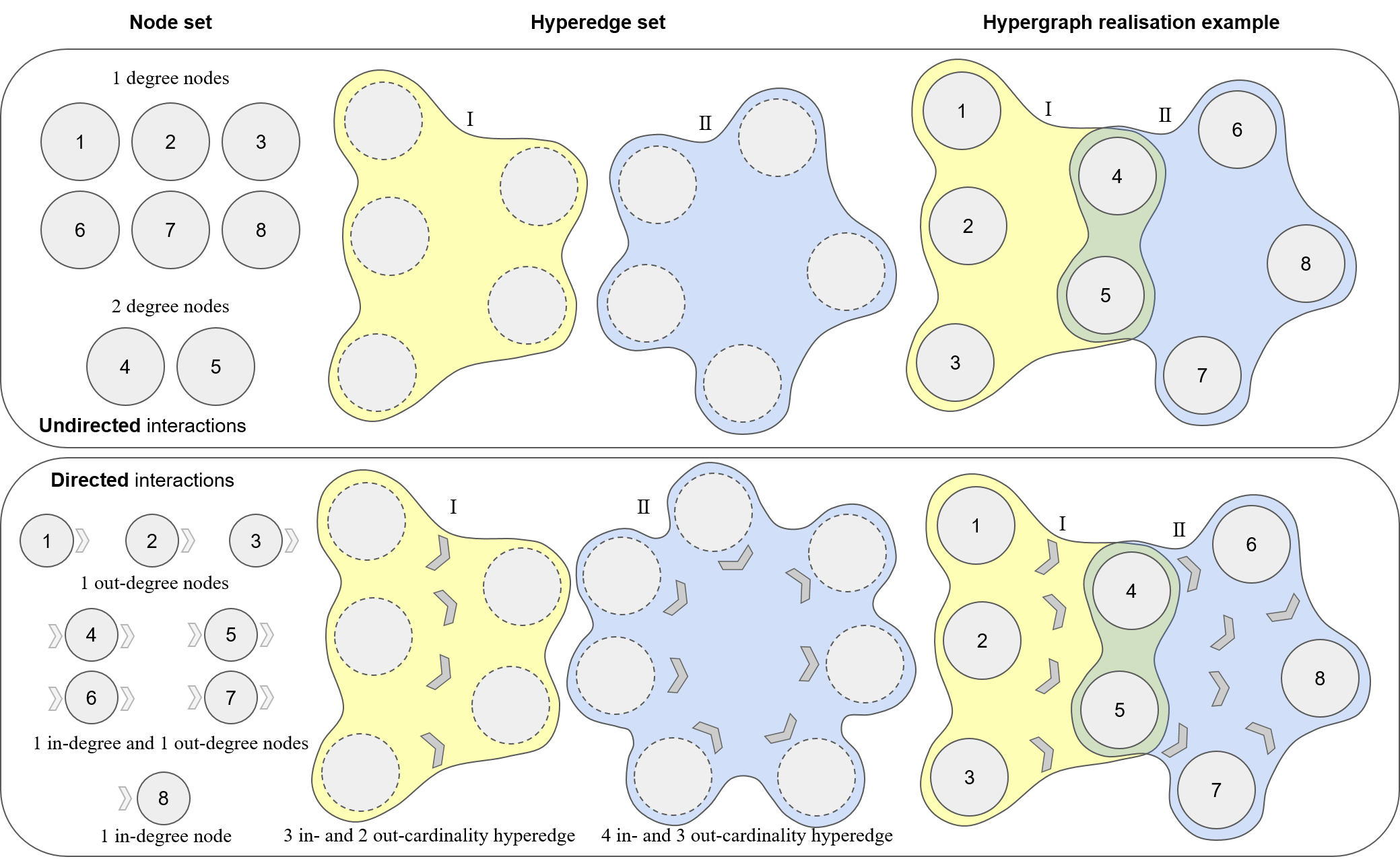}
 \caption{\it Example of two sets of nodes and hyperedges with their stubs (one nondirected and another one directed), and an example of two  random hypergraphs generated from these initial configurations. }
 \label{fig:hypergraph configuration model}
\end{figure}

\section{Denominator of    the quad clustering coefficient for a directed hypergraphs} \label{app:clustering-norm}

In this Appendix, we provide an explicit  expression for  the maximum number of directed quads, $q^{\leftrightarrow}_{\rm max}(\left\{ \mathcal{X}_{i\alpha}(\mathbf{I}^{\leftrightarrow}),I^{\leftrightarrow }_{i\alpha}\right\}_{\alpha\in \partial_i} )$,  that a node $i$ can have when it is  connected to a set of hyperedges $\alpha \in \partial_i$  for which (i)
the cardinalities of the hyperedges $\alpha$  are     given by those in the sets $\mathcal{X}_{i\alpha} = \left\{\chi^{\rm in}_{\alpha, i}(\mathbf{I}^{\rightarrow}), \chi^{\rm out}_{\alpha,i}(\mathbf{I}^{\leftarrow})\right\} $  and  (ii) the symmetry of the  links connecting node $i$ to the hyperedges $\alpha$ are   determined by the values    $I^{\leftrightarrow }_{i\alpha}$.       The expression for 
$q^{\leftrightarrow}_{\rm max}$ can be decomposed into 
\begin{eqnarray}
q^{\leftrightarrow}_{\rm max}(\left\{ \mathcal{X}_{i\alpha}(\mathbf{I}^{\leftrightarrow}),I^{\leftrightarrow }_{i\alpha}\right\}_{\alpha\in \partial_i} )\equiv\sum_{\alpha,\beta;\alpha<\beta}\left(\textbf{I}^{\leftrightarrow}\right)_{i\alpha}\left(\textbf{I}^{\leftrightarrow}\right)_{i\beta}\mathcal{W}\left(\mathcal{X}_{i\alpha}(\mathbf{I}^{\leftrightarrow}),\mathcal{X}_{i\beta}(\mathbf{I}^{\leftrightarrow})\right),
\end{eqnarray} 
where $\mathcal{W}$ is an integer valued function that is independent of the symmetry of the links $(i,\alpha)$ and $(i,\beta)$ (as determined by $\textbf{I}^{\leftrightarrow}_{i\alpha}$ and   $\textbf{I}^{\leftrightarrow}_{i\beta}$).    In what follows we specify $\mathcal{W}\left(\mathcal{X}_{i\alpha}, \mathcal{X}_{i\beta}\right)$ for the four possible scenarios, viz., (1)  $\chi^{\rm in}_{\alpha, i} = \chi^{\rm out}_{\alpha, i}$ and  $\chi^{\rm in}_{\beta, i} = \chi^{\rm out}_{\beta, i}$ (see Appendix~\ref{App:F1}); (2) $\chi^{\rm in}_{\alpha, i} = \chi^{\rm out}_{\alpha, i}$ and  $\chi^{\rm in}_{\beta, i} \neq \chi^{\rm out}_{\beta, i}$, or $\chi^{\rm in}_{\alpha, i} \neq \chi^{\rm out}_{\alpha, i}$ and  $\chi^{\rm in}_{\beta, i} = \chi^{\rm out}_{\beta, i}$ (see Appendix~\ref{App:F2});  (3) $\chi^{\rm in}_{\alpha, i} \neq \chi^{\rm out}_{\alpha, i}$ and  $\chi^{\rm in}_{\beta, i} \neq \chi^{\rm out}_{\beta, i}$.  Additionally,  $|\mathcal{X}_{i\alpha}\cup \mathcal{X}_{i\beta}| \neq 4$ (see Appendix~\ref{App:F3}); (4) $\chi^{\rm in}_{\alpha, i} \neq \chi^{\rm out}_{\alpha, i}$ and  $\chi^{\rm in}_{\beta, i} \neq \chi^{\rm out}_{\beta, i}$.  Additionally,  $|\mathcal{X}_{i\alpha}\cup \mathcal{X}_{i\beta}| = 4$ (see Appendix~\ref{App:F4}).

\subsection{$\chi^{\rm in}_{\alpha, i} = \chi^{\rm out}_{\alpha, i}$ and  $\chi^{\rm in}_{\beta, i} = \chi^{\rm out}_{\beta, i}$}\label{App:F1}

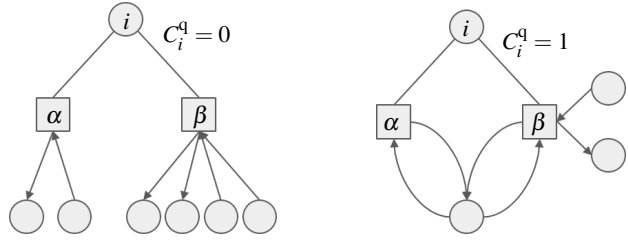
\begin{figure}[h!]
     \centering
     \setlength{\unitlength}{0.1\textwidth}
     \includegraphics[scale=0.54]{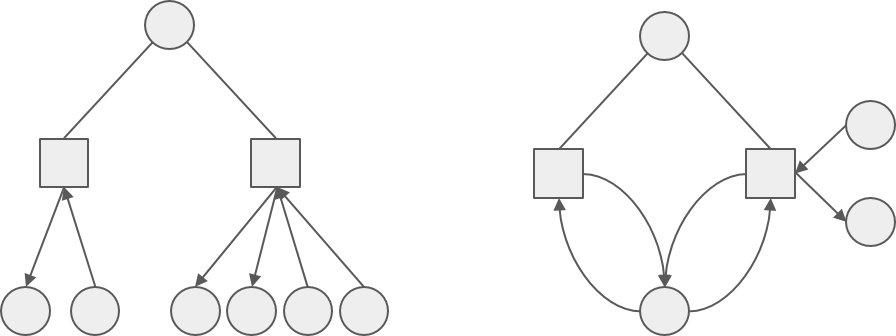}
      \put(-3.725,1.52){\small$i$}
      \put(-3.425,1.42){\small$C^{\rm q}_i=0$}
      \put(-4.3,0.825){\small$\alpha$}
      \put(-3.21,0.825){\small$\beta$}
      \put(-1.21,1.47){\small$i$}
      \put(-0.91,1.37){\small$C^{\rm q}_i=1$}
      \put(-1.8,0.77){\small$\alpha$}
      \put(-0.7,0.77){\small$\beta$}
 \caption{Two motifs consisting of a node $i$ linked with  two hyperedges $\alpha$ and $\beta$ and for which  $\chi^{\rm in}_{\alpha, i} = \chi^{\rm out}_{\alpha, i}=1$ and  $\chi^{\rm in}_{\beta, i} = \chi^{\rm out}_{\beta, i}=2$, corresponding with Appendix~\ref{App:F1}.   Left panel shows an example with $C^{\rm q}_i=0$ and the right panel has $C^{\rm q}_i=1$.}
 \label{fig:clustering_case1}
\end{figure}

In this case
\begin{align}
\mathcal{W}\left(\mathcal{X}_{i\alpha}(\mathbf{I}^{\leftrightarrow}),\mathcal{X}_{i\beta}(\mathbf{I}^{\leftrightarrow})\right)&\equiv4\min\left(\mathcal{X}_{i\alpha}(\mathbf{I}^{\leftrightarrow})\cup\mathcal{X}_{i\beta}(\mathbf{I}^{\leftrightarrow})\right).
 \label{eq:clustering_case1}
\end{align}

Figure~\ref{fig:clustering_case1} shows two examples, one for which  $C^{\rm q}_i=0$ and another one for which  $C^{\rm q}_i=1$.

\subsection{$\chi^{\rm in}_{\alpha, i} = \chi^{\rm out}_{\alpha, i}$ and  $\chi^{\rm in}_{\beta, i} \neq \chi^{\rm out}_{\beta, i}$}\label{App:F2}
We define the minimum cardinality by $\chi_{\rm min} \equiv \min\left(\left\{\chi^{\rm in}_{\alpha, i}, \chi^{\rm in}_{\beta, i} ,  \chi^{\rm out}_{\beta, i}\right\}\right)$ and the maximum value by $\chi_{\rm max} \equiv \max\left(\left\{\chi^{\rm in}_{\alpha, i}, \chi^{\rm in}_{\beta, i} ,  \chi^{\rm out}_{\beta, i}\right\}\right)$.   In case the three values $\chi^{\rm in}_{\alpha, i}$,  $\chi^{\rm in}_{\beta, i}$ and   $\chi^{\rm out}_{\beta, i}$ are distinct, we use the notation 
 $ \chi_{\rm med}$ for the median value.   Using this notation, we can express 
\begin{align}
\mathcal{W}\left(\mathcal{X}_{i\alpha},\mathcal{X}_{i\beta}\right)\equiv\begin{cases}
			4\chi_{\rm min}, & \text{if $\min\left(\mathcal{X}_{i\alpha}\cup  \mathcal{X}_{i\beta}\right)=\chi^{\rm in}_{\alpha, i}$,}\\[10pt]
			2\chi_{\rm min}+2\chi_{\rm med}, & \text{if $\min\left(\mathcal{X}_{i\alpha}\cup  \mathcal{X}_{i\beta}\right)\neq\chi^{\rm in}_{\alpha, i}$ and $\left|\mathcal{X}_{i\alpha}\cup\mathcal{X}_{i\beta}\right|=3$},\\[10pt]
			2\chi_{\rm min}+2\chi_{\rm max}, & \text{otherwise}.
		 \end{cases}
\label{eq:Wcase2}
\end{align}

Fig.~\ref{fig:clustering_case2} shows  examples with $C^{\rm q}_i=0$ and $C^{\rm q}_i=1$ for each the three above cases.

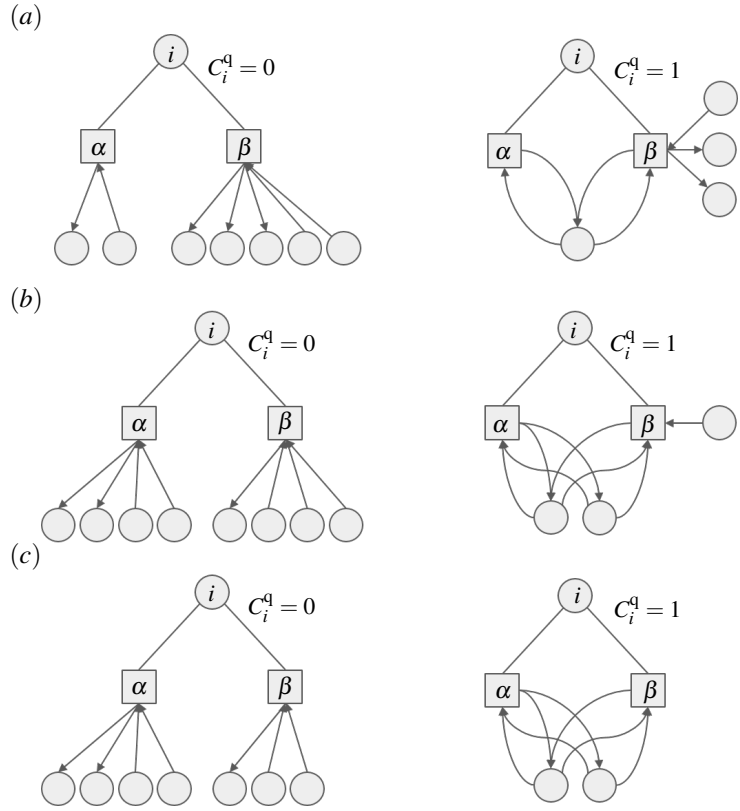
\begin{figure}[h!]
     \centering
     \setlength{\unitlength}{0.1\textwidth}
     \includegraphics[scale=0.54]{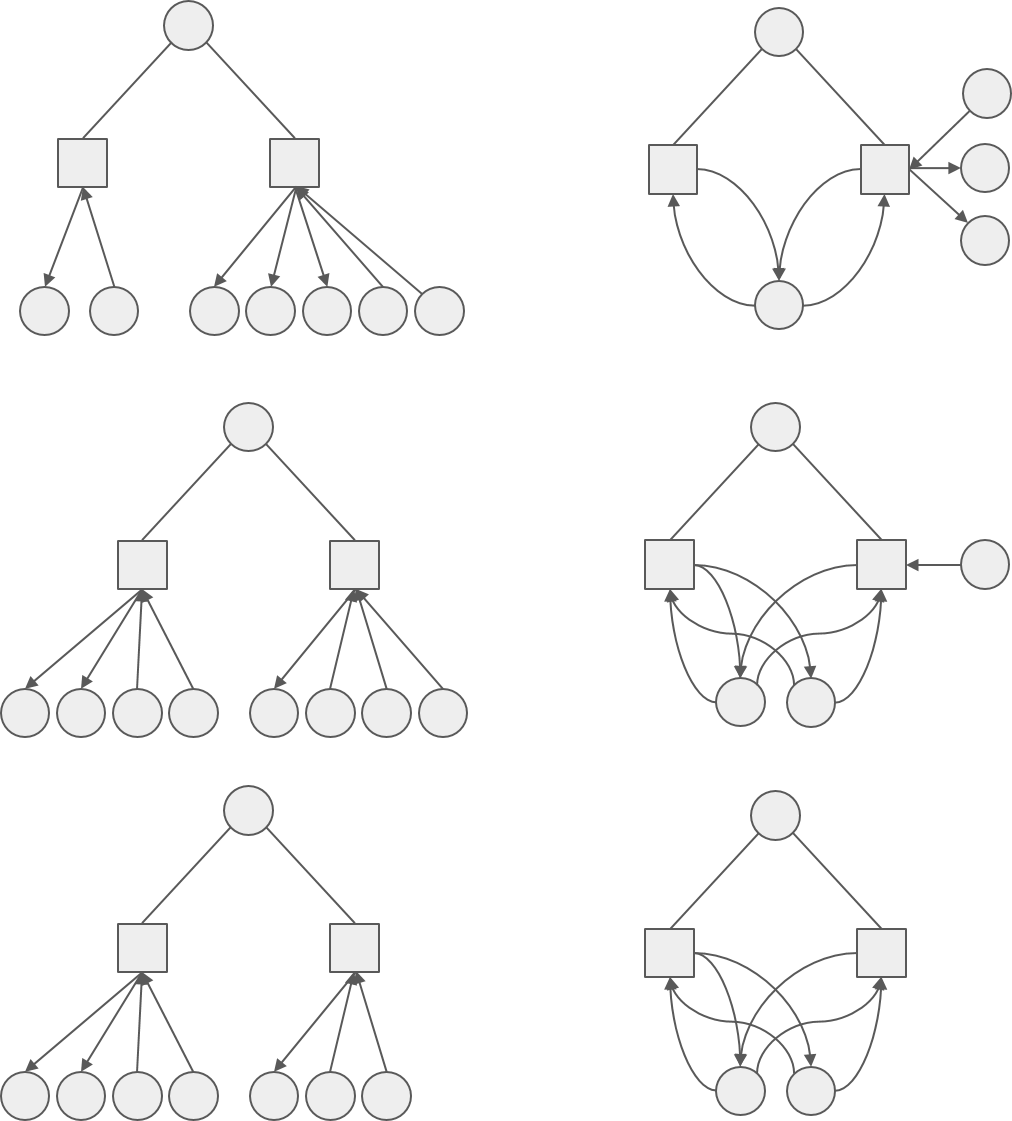}
      \put(-5.4,5.8){\normalsize$(a)$}
      \put(-4.225,5.52){\small$i$}
      \put(-3.925,5.42){\small$C^{\rm q}_i=0$}
      \put(-4.8,4.825){\small$\alpha$}
      \put(-3.71,4.825){\small$\beta$}
      \put(-1.21,5.5){\small$i$}
      \put(-0.91,5.4){\small$C^{\rm q}_i=1$}
      \put(-1.8,4.8){\small$\alpha$}
      \put(-0.7,4.8){\small$\beta$}
      \put(-5.4,3.7){\normalsize$(b)$}
      \put(-3.925,3.47){\small$i$}
      \put(-3.625,3.37){\small$C^{\rm q}_i=0$}
      \put(-4.5,2.775){\small$\alpha$}
      \put(-3.41,2.795){\small$\beta$}
      \put(-1.24,3.47){\small$i$}
      \put(-0.94,3.37){\small$C^{\rm q}_i=1$}
      \put(-1.82,2.77){\small$\alpha$}
      \put(-0.72,2.77){\small$\beta$}
      \put(-5.4,1.8){\normalsize$(c)$}
      \put(-3.925,1.52){\small$i$}
      \put(-3.625,1.42){\small$C^{\rm q}_i=0$}
      \put(-4.5,0.825){\small$\alpha$}
      \put(-3.41,0.825){\small$\beta$}
      \put(-1.21,1.5){\small$i$}
      \put(-0.91,1.4){\small$C^{\rm q}_i=1$}
      \put(-1.8,0.8){\small$\alpha$}
      \put(-0.7,0.8){\small$\beta$}
 \caption{ Motifs consisting of a node $i$ linked with  two hyperedges $\alpha$ and $\beta$ and for which $\chi^{\rm in}_{\alpha, i} = \chi^{\rm out}_{\alpha, i}$  and  $\chi^{\rm in}_{\beta, i} \neq \chi^{\rm out}_{\beta, i}$, corresponding with Appendix~\ref{App:F2}.   Left panels show examples with $C^{\rm q}_i=0$ and for the right panels  $C^{\rm q}_i=1$.  Panels (a)-(c) correspond with the different cases in Eq.~(\ref{eq:Wcase2}).  Panel (a): $\chi^{\rm in}_{\alpha, i} = \chi^{\rm out}_{\alpha, i} = 1$,  $\chi^{\rm in}_{\beta, i} = 2$, and $\chi^{\rm out}_{\beta, i} = 3$. Panel (b): $\chi^{\rm in}_{\alpha, i} = \chi^{\rm out}_{\alpha, i} = 2$, $\chi^{\rm in}_{\beta, i} = 3$, and $\chi^{\rm out}_{\beta, i} = 1$.  Panel (c): $\chi^{\rm in}_{\alpha, i} = \chi^{\rm out}_{\alpha, i} = 2$, $\chi^{\rm in}_{\beta, i} = 2$, and $\chi^{\rm out}_{\beta, i} = 1$.  }
 \label{fig:clustering_case2}
\end{figure}

For the case with  $\chi^{\rm in}_{\alpha, i} \neq \chi^{\rm out}_{\alpha, i}$ and  $\chi^{\rm in}_{\beta, i} = \chi^{\rm out}_{\beta, i}$ an analogous expression applies with the two indices $\alpha$ and $\beta$ swapped.

\subsection{$\chi^{\rm in}_{\alpha, i} \neq \chi^{\rm out}_{\alpha, i}$,  $\chi^{\rm in}_{\beta, i} \neq \chi^{\rm out}_{\beta, i}$, and  $|\mathcal{X}_{i\alpha}\cup \mathcal{X}_{i\beta}| \neq 4$.}\label{App:F3}
As in Appendix~\ref{App:F2}, we use the notation 
 $\chi_{\rm min} \equiv {\rm min}\left(\left\{\chi^{\rm in}_{\alpha, i}, \chi^{\rm out}_{\alpha, i}, \chi^{\rm in}_{\beta, i} ,  \chi^{\rm out}_{\beta, i}\right\}\right)$ and $\chi_{\rm max} \equiv {\rm max}\left(\left\{\chi^{\rm in}_{\alpha, i}, \chi^{\rm out}_{\alpha, i}, \chi^{\rm in}_{\beta, i} ,  \chi^{\rm out}_{\beta, i}\right\}\right)$.   In addition,  if $\left|\mathcal{X}_{i\alpha}\cup\mathcal{X}_{i\beta}\right|=3$, then a third medican value exists, which we denote by $\chi_{\rm med}$.    Using this notation , we get
\begin{align}
\mathcal{W}\left(\mathcal{X}_{i\alpha},\mathcal{X}_{i\beta}\right)\equiv\begin{cases}
			3\chi_{\rm min}+\chi_{\rm max}, & \text{if $\left|\mathcal{X}_{i\alpha}\cup\mathcal{X}_{i\beta}\right|=2$},\\[10pt]
			3\chi_{\rm min}+\chi_{\rm med}, & \text{if $\left|\mathcal{X}_{i\alpha}\cup\mathcal{X}_{i\beta}\right|=3$ and $\min\left(\mathcal{X}_{i\alpha}\right)=\min\left(\mathcal{X}_{i\beta}\right)$,}\\[10pt]
			2\chi_{\rm min}+2\chi_{\rm med}, & \text{if $\left|\mathcal{X}_{i\alpha}\cup\mathcal{X}_{i\beta}\right|=3$ and either $\max\left(\mathcal{X}_{i\alpha}\right)=\min\left(\mathcal{X}_{i\beta}\right)$ or $\max\left(\mathcal{X}_{i\beta}\right)=\min\left(\mathcal{X}_{i\alpha}\right)$,}\\[10pt]
			2\chi_{\rm min}+\chi_{\rm max}+\chi_{\rm med}, & \text{if $\left|\mathcal{X}_{i\alpha}\cup\mathcal{X}_{i\beta}\right|=3$ and $\max\left(\mathcal{X}_{i\alpha}\right)=\max\left(\mathcal{X}_{i\beta}\right)$.}
		 \end{cases}
\label{eq:Wcase3}
\end{align}
Fig.~\ref{fig:clustering_case3} shows  examples with $C^{\rm q}_i=0$ or  $C^{\rm q}_i=1$ for each of the four cases mentioned in formula (\ref{eq:Wcase3}).

\begin{figure}[h!]
     \centering
     \setlength{\unitlength}{0.1\textwidth}
     \includegraphics[scale=0.54]{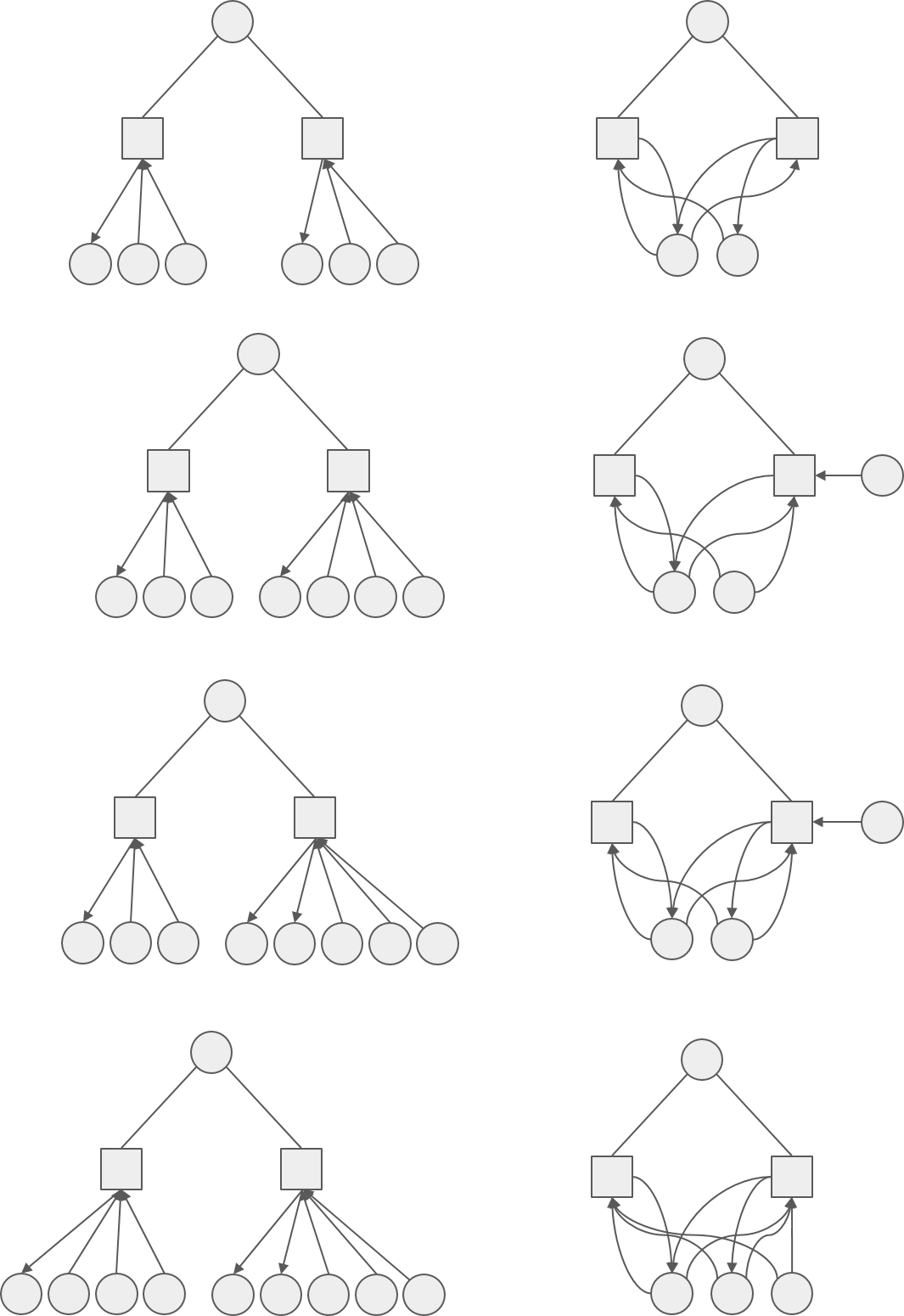}
      \put(-5.4,7.9){\normalsize$(a)$}
      \put(-4.05,7.72){\small$i$}
      \put(-3.75,7.62){\small$C^{\rm q}_i=0$}
      \put(-4.65,7.025){\small$\alpha$}
      \put(-3.525,7.025){\small$\beta$}
      \put(-1.21,7.7){\small$i$}
      \put(-0.91,7.6){\small$C^{\rm q}_i=1$}
      \put(-1.8,7){\small$\alpha$}
      \put(-0.7,7){\small$\beta$}
      \put(-5.4,5.8){\normalsize$(b)$}
      \put(-3.9,5.72){\small$i$}
      \put(-3.6,5.62){\small$C^{\rm q}_i=0$}
      \put(-4.5,5.025){\small$\alpha$}
      \put(-3.4,5.025){\small$\beta$}
      \put(-1.21,5.7){\small$i$}
      \put(-0.91,5.6){\small$C^{\rm q}_i=1$}
      \put(-1.8,5){\small$\alpha$}
      \put(-0.7,5){\small$\beta$}
      \put(-5.4,3.7){\normalsize$(c)$}
      \put(-4.1,3.63){\small$i$}
      \put(-3.8,3.53){\small$C^{\rm q}_i=0$}
      \put(-4.7,2.925){\small$\alpha$}
      \put(-3.6,2.925){\small$\beta$}
      \put(-1.24,3.6){\small$i$}
      \put(-0.94,3.5){\small$C^{\rm q}_i=1$}
      \put(-1.82,2.9){\small$\alpha$}
      \put(-0.72,2.9){\small$\beta$}
      \put(-5.4,1.8){\normalsize$(d)$}
      \put(-4.18,1.52){\small$i$}
      \put(-3.88,1.42){\small$C^{\rm q}_i=0$}
      \put(-4.78,0.825){\small$\alpha$}
      \put(-3.68,0.825){\small$\beta$}
      \put(-1.235,1.475){\small$i$}
      \put(-0.935,1.375){\small$C^{\rm q}_i=1$}
      \put(-1.825,0.775){\small$\alpha$}
      \put(-0.725,0.775){\small$\beta$}
 \caption{Motifs consisting of a node $i$ linked with  two hyperedges $\alpha$ and $\beta$ and for which $\chi^{\rm in}_{\alpha, i} \neq \chi^{\rm out}_{\alpha, i}$,  $\chi^{\rm in}_{\beta, i} \neq \chi^{\rm out}_{\beta, i}$, and  $|\mathcal{X}_{i\alpha}\cup \mathcal{X}_{i\beta}| \neq4$, corresponding with Appendix~\ref{App:F3}. Left panels show examples with $C^{\rm q}_i=0$   and the right panels have  $C^{\rm q}_i=1$.   Left panels show examples with $C^{\rm q}_i=0$ and for the right panels  $C^{\rm q}_i=1$.  Panels (a)-(d) correspond with the different cases in Eq.~(\ref{eq:Wcase3}).  Panel (a): $\chi^{\rm in}_{\alpha, i} = 2$, $\chi^{\rm out}_{\alpha, i} = 1$,  $\chi^{\rm in}_{\beta, i} = 2$, and $\chi^{\rm out}_{\beta, i} = 1$.  Panel (b): $\chi^{\rm in}_{\alpha, i} = 2$, $\chi^{\rm out}_{\alpha, i} = 1$,  $\chi^{\rm in}_{\beta, i} = 3$, and $\chi^{\rm out}_{\beta, i} = 1$. Panel (c): $\chi^{\rm in}_{\alpha, i} = 2$, $\chi^{\rm out}_{\alpha, i} = 1$,  $\chi^{\rm in}_{\beta, i} = 3$, and $\chi^{\rm out}_{\beta, i} = 2$. Panel (d): $\chi^{\rm in}_{\alpha, i} = 3$, $\chi^{\rm out}_{\alpha, i} = 1$,  $\chi^{\rm in}_{\beta, i} = 3$, and $\chi^{\rm out}_{\beta, i} = 2$.}
 \label{fig:clustering_case3}
\end{figure}
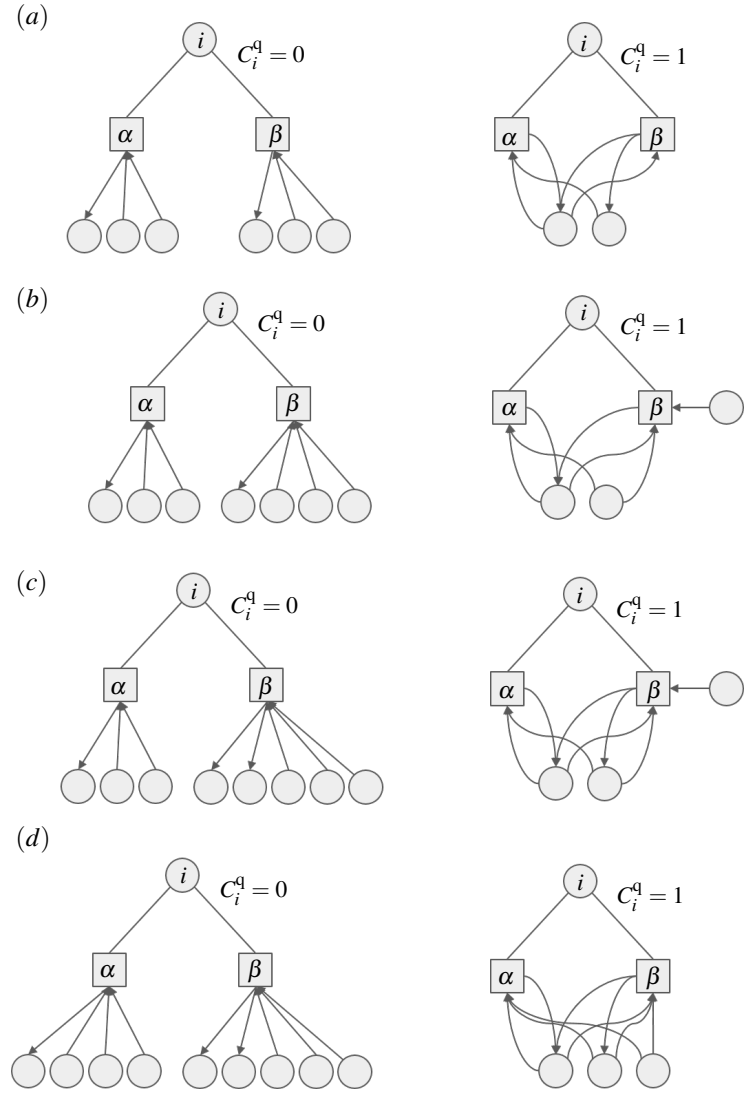

\subsection{$\chi^{\rm in}_{\alpha, i} \neq \chi^{\rm out}_{\alpha, i}$,  $\chi^{\rm in}_{\beta, i} \neq \chi^{\rm out}_{\beta, i}$, and  $|\mathcal{X}_{i\alpha}\cup \mathcal{X}_{i\beta}| =4$.}\label{App:F4}
In this case, the four cardinalities in the set  $\left\{\chi^{\rm in}_{\alpha, i}, \chi^{\rm out}_{\alpha, i}, \chi^{\rm in}_{\beta, i} ,  \chi^{\rm out}_{\beta, i}\right\}$ are all different.   We order them from small to large and use the notation $\chi_{\rm smallest}<\chi_{\rm small}<\chi_{\rm large}<\chi_{\rm largest}$ so that    $\chi_{\rm smallest} \equiv {\rm min}\left(\left\{\chi^{\rm in}_{\alpha, i}, \chi^{\rm out}_{\alpha, i}, \chi^{\rm in}_{\beta, i} ,  \chi^{\rm out}_{\beta, i}\right\}\right)$, and so forth.    The expression for $\mathcal{W}$ takes then the form 
\begin{align}
\mathcal{W}\left(\mathcal{X}_{i\alpha},\mathcal{X}_{i\beta}\right)\equiv\begin{cases}
			2\chi_{\rm smallest}+2\chi_{\rm small}, & \text{if $\max\left(\mathcal{X}_{i\alpha}\right)<\min\left(\mathcal{X}_{i\beta}\right)$ or $\min\left(\mathcal{X}_{i\alpha}\right)>\max\left(\mathcal{X}_{i\beta}\right)$},\\
			2\chi_{\rm smallest}+\chi_{\rm large}+\chi_{\rm small}, & \text{otherwise.}
		 \end{cases}
\label{eq:Wcase4_rewritten}
\end{align}
Fig.~\ref{fig:clustering_case4} shows the examples of $C^{\rm q}_i=0$ and $C^{\rm q}_i=1$ for both cases in Eq.~(\ref{eq:Wcase4_rewritten}).  

\begin{figure}[h!]
     \centering
     \setlength{\unitlength}{0.1\textwidth}
     \includegraphics[scale=0.54]{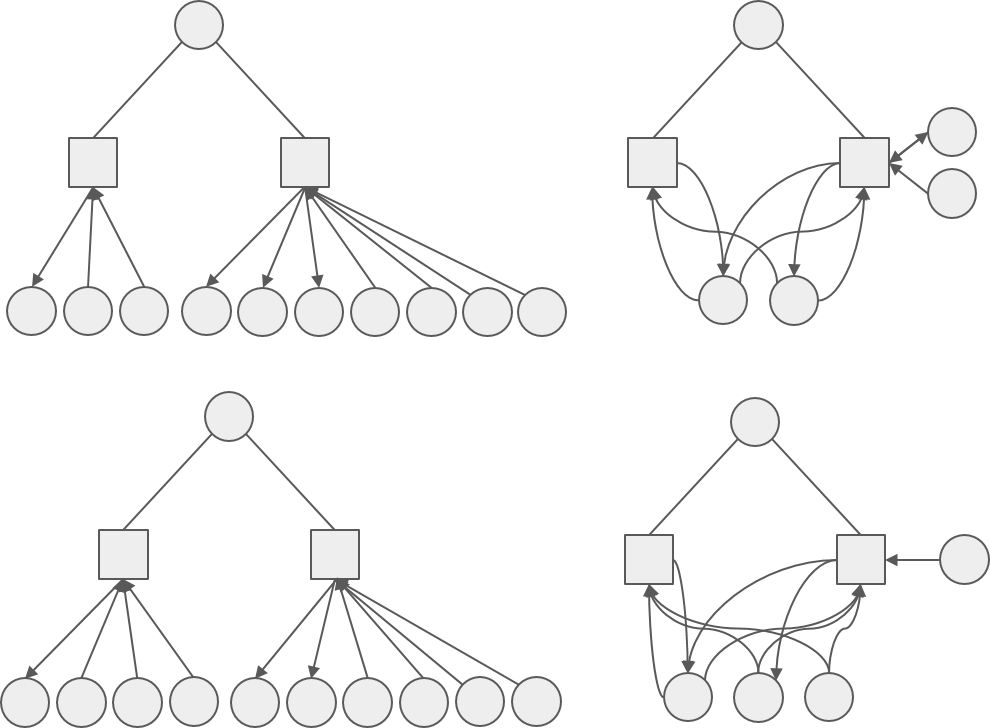}
      \put(-5.8,3.7){\normalsize$(a)$}
      \put(-4.07,3.52){\small$i$}
      \put(-3.77,3.42){\small$C^{\rm q}_i=0$}
      \put(-4.65,2.815){\small$\alpha$}
      \put(-3.55,2.815){\small$\beta$}
      \put(-1.2,3.52){\small$i$}
      \put(-0.9,3.42){\small$C^{\rm q}_i=1$}
      \put(-1.78,2.82){\small$\alpha$}
      \put(-0.68,2.82){\small$\beta$}
      \put(-5.8,1.7){\normalsize$(b)$}
      \put(-3.9,1.52){\small$i$}
      \put(-3.6,1.42){\small$C^{\rm q}_i=0$}
      \put(-4.5,0.825){\small$\alpha$}
      \put(-3.4,0.825){\small$\beta$}
      \put(-1.21,1.5){\small$i$}
      \put(-0.91,1.4){\small$C^{\rm q}_i=1$}
      \put(-1.8,0.8){\small$\alpha$}
      \put(-0.7,0.8){\small$\beta$}
 \caption{ Motifs consisting of a node $i$ linked with  two hyperedges $\alpha$ and $\beta$ and for which  $\chi^{\rm in}_{\alpha, i} \neq \chi^{\rm out}_{\alpha, i}$,  $\chi^{\rm in}_{\beta, i} \neq \chi^{\rm out}_{\beta, i}$, and  $|\mathcal{X}_{i\alpha}\cup \mathcal{X}_{i\beta}| =4$, corresponding with Appendix~\ref{App:F4}.   Left panels show examples with $C^{\rm q}_i=0$   and the right panels have  $C^{\rm q}_i=1$.    Panels (a)-(b) correspond with the two different cases in Eq.~(\ref{eq:Wcase4_rewritten}).   Panel (a): $\chi^{\rm in}_{\alpha, i} = 2$, $\chi^{\rm out}_{\alpha, i} = 1$,  $\chi^{\rm in}_{\beta, i} = 4$, and $\chi^{\rm out}_{\beta, i} = 3$. Panel (b): $\chi^{\rm in}_{\alpha, i} = 3$, $\chi^{\rm out}_{\alpha, i} = 1$,  $\chi^{\rm in}_{\beta, i} = 4$, and $\chi^{\rm out}_{\beta, i} = 2$. }
 \label{fig:clustering_case4}
\end{figure}
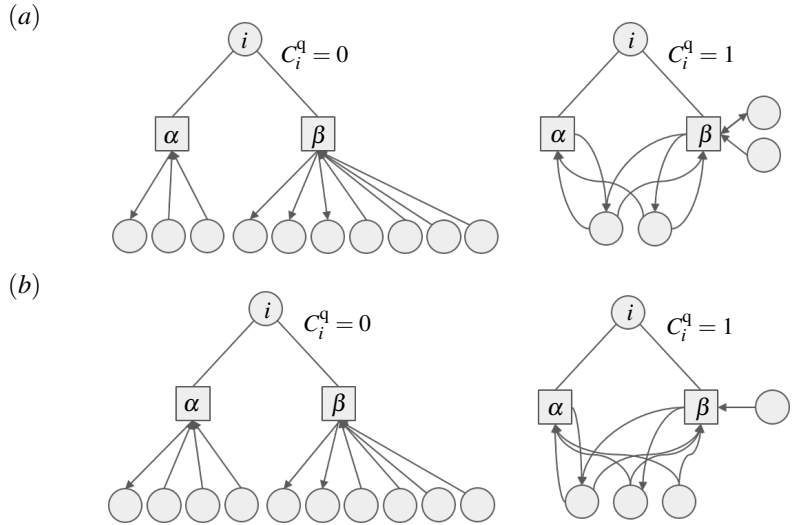

\section{For nondirected hypergraphs $C^{{\rm q}\leftrightarrow}(\mathbf{I}^{\leftrightarrow}) = C^{\rm q}(\mathbf{I})$} \label{app:dir_undir_quadC}
For a nondirected hypergraph $\mathbf{I}^{\rightarrow}=\mathbf{I}^{\leftarrow} = \mathbf{I}$, such that (\ref{def:dir_CQuad}) reads
 \begin{eqnarray}
\begin{split}
C^{\rm q \leftrightarrow}_i(\mathbf{I}^{\leftrightarrow})
&=4\frac{\sum_{j;j\neq i} \sum_{\alpha,\beta;\alpha<\beta}I_{i\alpha}  I_{j\alpha}  I_{i\beta} I_{j\beta}}
{\sum_{\alpha,\beta;\alpha<\beta} I_{i\alpha} I_{i\beta} \mathcal{W}\left(\mathcal{X}_{i\alpha},\mathcal{X}_{i\beta}\right)}.
\end{split}
\end{eqnarray}
Furthermore, as the hypergraph is nondirected,  $\chi^{\rm in}_{\alpha, i} = \chi^{\rm out}_{\alpha, i}$ and  $\chi^{\rm in}_{\beta, i} = \chi^{\rm out}_{\beta, i}$, and hence the case of  Appendix~\ref{App:F1} applies for $\mathcal{W}$ and its expression is given by Eq.~(\ref{eq:clustering_case1}).  Using this formula we obtain  
\begin{eqnarray}
\begin{split}
C^{\rm q \leftrightarrow}_i(\mathbf{I})
&=\frac{4\sum_{j;j\neq i} \sum_{\alpha,\beta;\alpha<\beta}I_{i\alpha}I_{j\alpha}I_{i\beta}I_{j\beta}}
{\sum_{\alpha,\beta;\alpha<\beta}\left(I_{i\alpha}I_{i\beta}\right) 4 \min\left\{(\sum_{j,i\neq j}I_{j\alpha}),(\sum_{j,i\neq j}I_{j\beta})\right\}} \label{eq:G2}
\end{split}
\end{eqnarray}
Eliminating the $4$ from both the numerator and the denominator in the right-hand side of (\ref{eq:G2}), we recover  the quad clustering coefficient $C^{\rm q}_{i}$ as defined in Eq.~(\ref{def:CQuad}), which completes the derivation.

\end{widetext}

\bibliography{aipsamp}

\providecommand{\noopsort}[1]{}\providecommand{\singleletter}[1]{#1}%
\begin{thebibliography}{35}%
\makeatletter
\providecommand \@ifxundefined [1]{%
 \@ifx{#1\undefined}
}%
\providecommand \@ifnum [1]{%
 \ifnum #1\expandafter \@firstoftwo
 \else \expandafter \@secondoftwo
 \fi
}%
\providecommand \@ifx [1]{%
 \ifx #1\expandafter \@firstoftwo
 \else \expandafter \@secondoftwo
 \fi
}%
\providecommand \natexlab [1]{#1}%
\providecommand \enquote  [1]{``#1''}%
\providecommand \bibnamefont  [1]{#1}%
\providecommand \bibfnamefont [1]{#1}%
\providecommand \citenamefont [1]{#1}%
\providecommand \href@noop [0]{\@secondoftwo}%
\providecommand \href [0]{\begingroup \@sanitize@url \@href}%
\providecommand \@href[1]{\@@startlink{#1}\@@href}%
\providecommand \@@href[1]{\endgroup#1\@@endlink}%
\providecommand \@sanitize@url [0]{\catcode `\\12\catcode `\$12\catcode
  `\&12\catcode `\#12\catcode `\^12\catcode `\_12\catcode `\%12\relax}%
\providecommand \@@startlink[1]{}%
\providecommand \@@endlink[0]{}%
\providecommand \url  [0]{\begingroup\@sanitize@url \@url }%
\providecommand \@url [1]{\endgroup\@href {#1}{\urlprefix }}%
\providecommand \urlprefix  [0]{URL }%
\providecommand \Eprint [0]{\href }%
\providecommand \doibase [0]{http://dx.doi.org/}%
\providecommand \selectlanguage [0]{\@gobble}%
\providecommand \bibinfo  [0]{\@secondoftwo}%
\providecommand \bibfield  [0]{\@secondoftwo}%
\providecommand \translation [1]{[#1]}%
\providecommand \BibitemOpen [0]{}%
\providecommand \bibitemStop [0]{}%
\providecommand \bibitemNoStop [0]{.\EOS\space}%
\providecommand \EOS [0]{\spacefactor3000\relax}%
\providecommand \BibitemShut  [1]{\csname bibitem#1\endcsname}%
\let\auto@bib@innerbib\@empty
\bibitem [{\citenamefont {Newman}, \citenamefont {Barab{\'a}si},\ and\
  \citenamefont {Watts}(2006)}]{newman2006structure}%
  \BibitemOpen
  \bibfield  {author} {\bibinfo {author} {\bibfnamefont {M.~E.}\ \bibnamefont
  {Newman}}, \bibinfo {author} {\bibfnamefont {A.-L.~E.}\ \bibnamefont
  {Barab{\'a}si}}, \ and\ \bibinfo {author} {\bibfnamefont {D.~J.}\
  \bibnamefont {Watts}},\ }\href@noop {} {\emph {\bibinfo {title} {The
  structure and dynamics of networks.}}}\ (\bibinfo  {publisher} {Princeton
  university press},\ \bibinfo {year} {2006})\BibitemShut {NoStop}%
\bibitem [{\citenamefont {Barabási}\ and\ \citenamefont
  {Pósfai}(2016)}]{barabasi}%
  \BibitemOpen
  \bibfield  {author} {\bibinfo {author} {\bibfnamefont {A.-L.}\ \bibnamefont
  {Barabási}}\ and\ \bibinfo {author} {\bibfnamefont {M.}~\bibnamefont
  {Pósfai}},\ }\href@noop {} {\emph {\bibinfo {title} {Network Science}}}\
  (\bibinfo  {publisher} {Cambridge University Press},\ \bibinfo {year}
  {2016})\BibitemShut {NoStop}%
\bibitem [{\citenamefont {Dorogovtsev}\ and\ \citenamefont
  {Mendes}(2022)}]{dorogovtsev2022nature}%
  \BibitemOpen
  \bibfield  {author} {\bibinfo {author} {\bibfnamefont {S.~N.}\ \bibnamefont
  {Dorogovtsev}}\ and\ \bibinfo {author} {\bibfnamefont {J.~F.}\ \bibnamefont
  {Mendes}},\ }\href@noop {} {\emph {\bibinfo {title} {The nature of complex
  networks}}}\ (\bibinfo  {publisher} {Oxford University Press},\ \bibinfo
  {year} {2022})\BibitemShut {NoStop}%
\bibitem [{\citenamefont {Battiston}\ \emph {et~al.}(2020)\citenamefont
  {Battiston}, \citenamefont {Cencetti}, \citenamefont {Iacopini},
  \citenamefont {Latora}, \citenamefont {Lucas}, \citenamefont {Patania},
  \citenamefont {Young},\ and\ \citenamefont {Petri}}]{battiston2020networks}%
  \BibitemOpen
  \bibfield  {author} {\bibinfo {author} {\bibfnamefont {F.}~\bibnamefont
  {Battiston}}, \bibinfo {author} {\bibfnamefont {G.}~\bibnamefont {Cencetti}},
  \bibinfo {author} {\bibfnamefont {I.}~\bibnamefont {Iacopini}}, \bibinfo
  {author} {\bibfnamefont {V.}~\bibnamefont {Latora}}, \bibinfo {author}
  {\bibfnamefont {M.}~\bibnamefont {Lucas}}, \bibinfo {author} {\bibfnamefont
  {A.}~\bibnamefont {Patania}}, \bibinfo {author} {\bibfnamefont {J.-G.}\
  \bibnamefont {Young}}, \ and\ \bibinfo {author} {\bibfnamefont
  {G.}~\bibnamefont {Petri}},\ }\bibfield  {title} {\enquote {\bibinfo {title}
  {Networks beyond pairwise interactions: structure and dynamics},}\
  }\href@noop {} {\bibfield  {journal} {\bibinfo  {journal} {Physics Reports}\
  }\textbf {\bibinfo {volume} {874}},\ \bibinfo {pages} {1--92} (\bibinfo
  {year} {2020})}\BibitemShut {NoStop}%
\bibitem [{\citenamefont {Watts}\ and\ \citenamefont
  {Strogatz}(1998)}]{watts1998collective}%
  \BibitemOpen
  \bibfield  {author} {\bibinfo {author} {\bibfnamefont {D.~J.}\ \bibnamefont
  {Watts}}\ and\ \bibinfo {author} {\bibfnamefont {S.~H.}\ \bibnamefont
  {Strogatz}},\ }\bibfield  {title} {\enquote {\bibinfo {title} {Collective
  dynamics of ‘small-world’networks},}\ }\href@noop {} {\bibfield
  {journal} {\bibinfo  {journal} {nature}\ }\textbf {\bibinfo {volume} {393}},\
  \bibinfo {pages} {440--442} (\bibinfo {year} {1998})}\BibitemShut {NoStop}%
\bibitem [{\citenamefont {Albert}\ and\ \citenamefont
  {Barab{\'a}si}(2002)}]{albert2002statistical}%
  \BibitemOpen
  \bibfield  {author} {\bibinfo {author} {\bibfnamefont {R.}~\bibnamefont
  {Albert}}\ and\ \bibinfo {author} {\bibfnamefont {A.-L.}\ \bibnamefont
  {Barab{\'a}si}},\ }\bibfield  {title} {\enquote {\bibinfo {title}
  {Statistical mechanics of complex networks},}\ }\href@noop {} {\bibfield
  {journal} {\bibinfo  {journal} {Reviews of modern physics}\ }\textbf
  {\bibinfo {volume} {74}},\ \bibinfo {pages} {47} (\bibinfo {year}
  {2002})}\BibitemShut {NoStop}%
\bibitem [{\citenamefont {Ravasz}\ and\ \citenamefont
  {Barab{\'a}si}(2003)}]{ravasz2003hierarchical}%
  \BibitemOpen
  \bibfield  {author} {\bibinfo {author} {\bibfnamefont {E.}~\bibnamefont
  {Ravasz}}\ and\ \bibinfo {author} {\bibfnamefont {A.-L.}\ \bibnamefont
  {Barab{\'a}si}},\ }\bibfield  {title} {\enquote {\bibinfo {title}
  {Hierarchical organization in complex networks},}\ }\href@noop {} {\bibfield
  {journal} {\bibinfo  {journal} {Physical review E}\ }\textbf {\bibinfo
  {volume} {67}},\ \bibinfo {pages} {026112} (\bibinfo {year}
  {2003})}\BibitemShut {NoStop}%
\bibitem [{\citenamefont {Barrat}\ and\ \citenamefont
  {Weigt}(2000)}]{barrat2000properties}%
  \BibitemOpen
  \bibfield  {author} {\bibinfo {author} {\bibfnamefont {A.}~\bibnamefont
  {Barrat}}\ and\ \bibinfo {author} {\bibfnamefont {M.}~\bibnamefont {Weigt}},\
  }\bibfield  {title} {\enquote {\bibinfo {title} {On the properties of
  small-world network models},}\ }\href@noop {} {\bibfield  {journal} {\bibinfo
   {journal} {The European Physical Journal B-Condensed Matter and Complex
  Systems}\ }\textbf {\bibinfo {volume} {13}},\ \bibinfo {pages} {547--560}
  (\bibinfo {year} {2000})}\BibitemShut {NoStop}%
\bibitem [{\citenamefont {Opsahl}(2013)}]{opsahl2013triadic}%
  \BibitemOpen
  \bibfield  {author} {\bibinfo {author} {\bibfnamefont {T.}~\bibnamefont
  {Opsahl}},\ }\bibfield  {title} {\enquote {\bibinfo {title} {Triadic closure
  in two-mode networks: Redefining the global and local clustering
  coefficients},}\ }\href@noop {} {\bibfield  {journal} {\bibinfo  {journal}
  {Social networks}\ }\textbf {\bibinfo {volume} {35}},\ \bibinfo {pages}
  {159--167} (\bibinfo {year} {2013})}\BibitemShut {NoStop}%
\bibitem [{\citenamefont {Brunson}(2015)}]{brunson2015triadic}%
  \BibitemOpen
  \bibfield  {author} {\bibinfo {author} {\bibfnamefont {J.~C.}\ \bibnamefont
  {Brunson}},\ }\bibfield  {title} {\enquote {\bibinfo {title} {Triadic
  analysis of affiliation networks},}\ }\href@noop {} {\bibfield  {journal}
  {\bibinfo  {journal} {Network Science}\ }\textbf {\bibinfo {volume} {3}},\
  \bibinfo {pages} {480--508} (\bibinfo {year} {2015})}\BibitemShut {NoStop}%
\bibitem [{\citenamefont {Kartun-Giles}\ and\ \citenamefont
  {Bianconi}(2019)}]{kartun2019beyond}%
  \BibitemOpen
  \bibfield  {author} {\bibinfo {author} {\bibfnamefont {A.~P.}\ \bibnamefont
  {Kartun-Giles}}\ and\ \bibinfo {author} {\bibfnamefont {G.}~\bibnamefont
  {Bianconi}},\ }\bibfield  {title} {\enquote {\bibinfo {title} {Beyond the
  clustering coefficient: A topological analysis of node neighbourhoods in
  complex networks},}\ }\href@noop {} {\bibfield  {journal} {\bibinfo
  {journal} {Chaos, Solitons \& Fractals: X}\ }\textbf {\bibinfo {volume}
  {1}},\ \bibinfo {pages} {100004} (\bibinfo {year} {2019})}\BibitemShut
  {NoStop}%
\bibitem [{\citenamefont {Serrano}\ and\ \citenamefont
  {G{\'o}mez}(2020)}]{serrano2020centrality}%
  \BibitemOpen
  \bibfield  {author} {\bibinfo {author} {\bibfnamefont {D.~H.}\ \bibnamefont
  {Serrano}}\ and\ \bibinfo {author} {\bibfnamefont {D.~S.}\ \bibnamefont
  {G{\'o}mez}},\ }\bibfield  {title} {\enquote {\bibinfo {title} {Centrality
  measures in simplicial complexes: Applications of topological data analysis
  to network science},}\ }\href@noop {} {\bibfield  {journal} {\bibinfo
  {journal} {Applied Mathematics and Computation}\ }\textbf {\bibinfo {volume}
  {382}},\ \bibinfo {pages} {125331} (\bibinfo {year} {2020})}\BibitemShut
  {NoStop}%
\bibitem [{\citenamefont {Yin}, \citenamefont {Benson},\ and\ \citenamefont
  {Leskovec}(2018)}]{yin2018higher}%
  \BibitemOpen
  \bibfield  {author} {\bibinfo {author} {\bibfnamefont {H.}~\bibnamefont
  {Yin}}, \bibinfo {author} {\bibfnamefont {A.~R.}\ \bibnamefont {Benson}}, \
  and\ \bibinfo {author} {\bibfnamefont {J.}~\bibnamefont {Leskovec}},\
  }\bibfield  {title} {\enquote {\bibinfo {title} {Higher-order clustering in
  networks},}\ }\href@noop {} {\bibfield  {journal} {\bibinfo  {journal}
  {Physical Review E}\ }\textbf {\bibinfo {volume} {97}},\ \bibinfo {pages}
  {052306} (\bibinfo {year} {2018})}\BibitemShut {NoStop}%
\bibitem [{\citenamefont {Lind}, \citenamefont {Gonz{\'a}lez},\ and\
  \citenamefont {Herrmann}(2005)}]{lind2005cycles}%
  \BibitemOpen
  \bibfield  {author} {\bibinfo {author} {\bibfnamefont {P.~G.}\ \bibnamefont
  {Lind}}, \bibinfo {author} {\bibfnamefont {M.~C.}\ \bibnamefont
  {Gonz{\'a}lez}}, \ and\ \bibinfo {author} {\bibfnamefont {H.~J.}\
  \bibnamefont {Herrmann}},\ }\bibfield  {title} {\enquote {\bibinfo {title}
  {Cycles and clustering in bipartite networks},}\ }\href@noop {} {\bibfield
  {journal} {\bibinfo  {journal} {Physical review E}\ }\textbf {\bibinfo
  {volume} {72}},\ \bibinfo {pages} {056127} (\bibinfo {year}
  {2005})}\BibitemShut {NoStop}%
\bibitem [{\citenamefont {Zhang}\ \emph {et~al.}(2008)\citenamefont {Zhang},
  \citenamefont {Wang}, \citenamefont {Li}, \citenamefont {Li}, \citenamefont
  {Di},\ and\ \citenamefont {Fan}}]{zhang2008clustering}%
  \BibitemOpen
  \bibfield  {author} {\bibinfo {author} {\bibfnamefont {P.}~\bibnamefont
  {Zhang}}, \bibinfo {author} {\bibfnamefont {J.}~\bibnamefont {Wang}},
  \bibinfo {author} {\bibfnamefont {X.}~\bibnamefont {Li}}, \bibinfo {author}
  {\bibfnamefont {M.}~\bibnamefont {Li}}, \bibinfo {author} {\bibfnamefont
  {Z.}~\bibnamefont {Di}}, \ and\ \bibinfo {author} {\bibfnamefont
  {Y.}~\bibnamefont {Fan}},\ }\bibfield  {title} {\enquote {\bibinfo {title}
  {Clustering coefficient and community structure of bipartite networks},}\
  }\href@noop {} {\bibfield  {journal} {\bibinfo  {journal} {Physica A:
  Statistical Mechanics and its Applications}\ }\textbf {\bibinfo {volume}
  {387}},\ \bibinfo {pages} {6869--6875} (\bibinfo {year} {2008})}\BibitemShut
  {NoStop}%
\bibitem [{\citenamefont {Kitsak}\ and\ \citenamefont
  {Krioukov}(2011)}]{kitsak2011hidden}%
  \BibitemOpen
  \bibfield  {author} {\bibinfo {author} {\bibfnamefont {M.}~\bibnamefont
  {Kitsak}}\ and\ \bibinfo {author} {\bibfnamefont {D.}~\bibnamefont
  {Krioukov}},\ }\bibfield  {title} {\enquote {\bibinfo {title} {Hidden
  variables in bipartite networks},}\ }\href@noop {} {\bibfield  {journal}
  {\bibinfo  {journal} {Physical Review E}\ }\textbf {\bibinfo {volume} {84}},\
  \bibinfo {pages} {026114} (\bibinfo {year} {2011})}\BibitemShut {NoStop}%
\bibitem [{\citenamefont {Jeong}\ and\ \citenamefont
  {Yu}(2022)}]{jeong2022effects}%
  \BibitemOpen
  \bibfield  {author} {\bibinfo {author} {\bibfnamefont {W.}~\bibnamefont
  {Jeong}}\ and\ \bibinfo {author} {\bibfnamefont {U.}~\bibnamefont {Yu}},\
  }\bibfield  {title} {\enquote {\bibinfo {title} {Effects of quadrilateral
  clustering on complex contagion},}\ }\href@noop {} {\bibfield  {journal}
  {\bibinfo  {journal} {Chaos, Solitons \& Fractals}\ }\textbf {\bibinfo
  {volume} {165}},\ \bibinfo {pages} {112784} (\bibinfo {year}
  {2022})}\BibitemShut {NoStop}%
\bibitem [{\citenamefont {Jia}, \citenamefont {Gabrys},\ and\ \citenamefont
  {Musial}(2021)}]{jia2021measuring}%
  \BibitemOpen
  \bibfield  {author} {\bibinfo {author} {\bibfnamefont {M.}~\bibnamefont
  {Jia}}, \bibinfo {author} {\bibfnamefont {B.}~\bibnamefont {Gabrys}}, \ and\
  \bibinfo {author} {\bibfnamefont {K.}~\bibnamefont {Musial}},\ }\bibfield
  {title} {\enquote {\bibinfo {title} {Measuring quadrangle formation in
  complex networks},}\ }\href@noop {} {\bibfield  {journal} {\bibinfo
  {journal} {IEEE Transactions on Network Science and Engineering}\ }\textbf
  {\bibinfo {volume} {9}},\ \bibinfo {pages} {538--551} (\bibinfo {year}
  {2021})}\BibitemShut {NoStop}%
\bibitem [{\citenamefont {Aksoy}, \citenamefont {Kolda},\ and\ \citenamefont
  {Pinar}(2017)}]{aksoy2017measuring}%
  \BibitemOpen
  \bibfield  {author} {\bibinfo {author} {\bibfnamefont {S.~G.}\ \bibnamefont
  {Aksoy}}, \bibinfo {author} {\bibfnamefont {T.~G.}\ \bibnamefont {Kolda}}, \
  and\ \bibinfo {author} {\bibfnamefont {A.}~\bibnamefont {Pinar}},\ }\bibfield
   {title} {\enquote {\bibinfo {title} {Measuring and modeling bipartite graphs
  with community structure},}\ }\href@noop {} {\bibfield  {journal} {\bibinfo
  {journal} {Journal of Complex Networks}\ }\textbf {\bibinfo {volume} {5}},\
  \bibinfo {pages} {581--603} (\bibinfo {year} {2017})}\BibitemShut {NoStop}%
\bibitem [{\citenamefont {Malizia}\ \emph {et~al.}(2023)\citenamefont
  {Malizia}, \citenamefont {Lamata-Ot{\'\i}n}, \citenamefont {Frasca},
  \citenamefont {Latora},\ and\ \citenamefont
  {G{\'o}mez-Garde{\~n}es}}]{malizia2023hyperedge}%
  \BibitemOpen
  \bibfield  {author} {\bibinfo {author} {\bibfnamefont {F.}~\bibnamefont
  {Malizia}}, \bibinfo {author} {\bibfnamefont {S.}~\bibnamefont
  {Lamata-Ot{\'\i}n}}, \bibinfo {author} {\bibfnamefont {M.}~\bibnamefont
  {Frasca}}, \bibinfo {author} {\bibfnamefont {V.}~\bibnamefont {Latora}}, \
  and\ \bibinfo {author} {\bibfnamefont {J.}~\bibnamefont
  {G{\'o}mez-Garde{\~n}es}},\ }\bibfield  {title} {\enquote {\bibinfo {title}
  {Hyperedge overlap drives explosive collective behaviors in systems with
  higher-order interactions},}\ }\href@noop {} {\bibfield  {journal} {\bibinfo
  {journal} {arXiv preprint arXiv:2307.03519}\ } (\bibinfo {year}
  {2023})}\BibitemShut {NoStop}%
\bibitem [{\citenamefont {Lee}, \citenamefont {Choe},\ and\ \citenamefont
  {Shin}(2021)}]{lee2021hyperedges}%
  \BibitemOpen
  \bibfield  {author} {\bibinfo {author} {\bibfnamefont {G.}~\bibnamefont
  {Lee}}, \bibinfo {author} {\bibfnamefont {M.}~\bibnamefont {Choe}}, \ and\
  \bibinfo {author} {\bibfnamefont {K.}~\bibnamefont {Shin}},\ }\bibfield
  {title} {\enquote {\bibinfo {title} {How do hyperedges overlap in real-world
  hypergraphs?-patterns, measures, and generators},}\ }in\ \href@noop {} {\emph
  {\bibinfo {booktitle} {Proceedings of the web conference 2021}}}\ (\bibinfo
  {year} {2021})\ pp.\ \bibinfo {pages} {3396--3407}\BibitemShut {NoStop}%
\bibitem [{\citenamefont {Jaccard}(1901)}]{jaccard1901etude}%
  \BibitemOpen
  \bibfield  {author} {\bibinfo {author} {\bibfnamefont {P.}~\bibnamefont
  {Jaccard}},\ }\bibfield  {title} {\enquote {\bibinfo {title} {{\'E}tude
  comparative de la distribution florale dans une portion des alpes et des
  jura},}\ }\href@noop {} {\bibfield  {journal} {\bibinfo  {journal} {Bull Soc
  Vaudoise Sci Nat}\ }\textbf {\bibinfo {volume} {37}},\ \bibinfo {pages}
  {547--579} (\bibinfo {year} {1901})}\BibitemShut {NoStop}%
\bibitem [{\citenamefont {Bianconi}\ and\ \citenamefont
  {Marsili}(2005)}]{bianconi2005loops}%
  \BibitemOpen
  \bibfield  {author} {\bibinfo {author} {\bibfnamefont {G.}~\bibnamefont
  {Bianconi}}\ and\ \bibinfo {author} {\bibfnamefont {M.}~\bibnamefont
  {Marsili}},\ }\bibfield  {title} {\enquote {\bibinfo {title} {Loops of any
  size and hamilton cycles in random scale-free networks},}\ }\href@noop {}
  {\bibfield  {journal} {\bibinfo  {journal} {Journal of Statistical Mechanics:
  Theory and Experiment}\ }\textbf {\bibinfo {volume} {2005}},\ \bibinfo
  {pages} {P06005} (\bibinfo {year} {2005})}\BibitemShut {NoStop}%
\bibitem [{\citenamefont {Newman}, \citenamefont {Strogatz},\ and\
  \citenamefont {Watts}(2001)}]{Newman2001}%
  \BibitemOpen
  \bibfield  {author} {\bibinfo {author} {\bibfnamefont {M.~E.~J.}\
  \bibnamefont {Newman}}, \bibinfo {author} {\bibfnamefont {S.~H.}\
  \bibnamefont {Strogatz}}, \ and\ \bibinfo {author} {\bibfnamefont {D.~J.}\
  \bibnamefont {Watts}},\ }\bibfield  {title} {\enquote {\bibinfo {title}
  {Random graphs with arbitrary degree distributions and their applications},}\
  }\href {\doibase 10.1103/PhysRevE.64.026118} {\bibfield  {journal} {\bibinfo
  {journal} {Phys. Rev. E}\ }\textbf {\bibinfo {volume} {64}},\ \bibinfo
  {pages} {026118} (\bibinfo {year} {2001})}\BibitemShut {NoStop}%
\bibitem [{\citenamefont {Coolen}, \citenamefont {Annibale},\ and\
  \citenamefont {Roberts}(2017)}]{coolen2017generating}%
  \BibitemOpen
  \bibfield  {author} {\bibinfo {author} {\bibfnamefont {T.}~\bibnamefont
  {Coolen}}, \bibinfo {author} {\bibfnamefont {A.}~\bibnamefont {Annibale}}, \
  and\ \bibinfo {author} {\bibfnamefont {E.}~\bibnamefont {Roberts}},\
  }\href@noop {} {\emph {\bibinfo {title} {Generating random networks and
  graphs}}}\ (\bibinfo  {publisher} {Oxford university press},\ \bibinfo {year}
  {2017})\BibitemShut {NoStop}%
\bibitem [{\citenamefont {Fagiolo}(2007)}]{fagiolo2007clustering}%
  \BibitemOpen
  \bibfield  {author} {\bibinfo {author} {\bibfnamefont {G.}~\bibnamefont
  {Fagiolo}},\ }\bibfield  {title} {\enquote {\bibinfo {title} {Clustering in
  complex directed networks},}\ }\href@noop {} {\bibfield  {journal} {\bibinfo
  {journal} {Physical Review E}\ }\textbf {\bibinfo {volume} {76}},\ \bibinfo
  {pages} {026107} (\bibinfo {year} {2007})}\BibitemShut {NoStop}%
\bibitem [{\citenamefont {Newman}(2003)}]{newman2003structure}%
  \BibitemOpen
  \bibfield  {author} {\bibinfo {author} {\bibfnamefont {M.~E.}\ \bibnamefont
  {Newman}},\ }\bibfield  {title} {\enquote {\bibinfo {title} {The structure
  and function of complex networks},}\ }\href@noop {} {\bibfield  {journal}
  {\bibinfo  {journal} {SIAM review}\ }\textbf {\bibinfo {volume} {45}},\
  \bibinfo {pages} {167--256} (\bibinfo {year} {2003})}\BibitemShut {NoStop}%
\bibitem [{\citenamefont {Benson}\ \emph {et~al.}(2018)\citenamefont {Benson},
  \citenamefont {Abebe}, \citenamefont {Schaub}, \citenamefont {Jadbabaie},\
  and\ \citenamefont {Kleinberg}}]{benson2018simplicial}%
  \BibitemOpen
  \bibfield  {author} {\bibinfo {author} {\bibfnamefont {A.~R.}\ \bibnamefont
  {Benson}}, \bibinfo {author} {\bibfnamefont {R.}~\bibnamefont {Abebe}},
  \bibinfo {author} {\bibfnamefont {M.~T.}\ \bibnamefont {Schaub}}, \bibinfo
  {author} {\bibfnamefont {A.}~\bibnamefont {Jadbabaie}}, \ and\ \bibinfo
  {author} {\bibfnamefont {J.}~\bibnamefont {Kleinberg}},\ }\bibfield  {title}
  {\enquote {\bibinfo {title} {Simplicial closure and higher-order link
  prediction},}\ }\href@noop {} {\bibfield  {journal} {\bibinfo  {journal}
  {Proceedings of the National Academy of Sciences}\ }\textbf {\bibinfo
  {volume} {115}},\ \bibinfo {pages} {E11221--E11230} (\bibinfo {year}
  {2018})}\BibitemShut {NoStop}%
\bibitem [{\citenamefont {Kunegis}(2013)}]{kunegis2013konect}%
  \BibitemOpen
  \bibfield  {author} {\bibinfo {author} {\bibfnamefont {J.}~\bibnamefont
  {Kunegis}},\ }\bibfield  {title} {\enquote {\bibinfo {title} {Konect: the
  koblenz network collection},}\ }in\ \href@noop {} {\emph {\bibinfo
  {booktitle} {Proceedings of the 22nd international conference on world wide
  web}}}\ (\bibinfo {year} {2013})\ pp.\ \bibinfo {pages}
  {1343--1350}\BibitemShut {NoStop}%
\bibitem [{\citenamefont {Mislove}(2009)}]{mislove2009online}%
  \BibitemOpen
  \bibfield  {author} {\bibinfo {author} {\bibfnamefont {A.~E.}\ \bibnamefont
  {Mislove}},\ }\href@noop {} {\emph {\bibinfo {title} {Online social networks:
  measurement, analysis, and applications to distributed information
  systems}}}\ (\bibinfo  {publisher} {Rice University},\ \bibinfo {year}
  {2009})\BibitemShut {NoStop}%
\bibitem [{\citenamefont {Kan}(2015)}]{whats-cooking}%
  \BibitemOpen
  \bibfield  {author} {\bibinfo {author} {\bibfnamefont {W.}~\bibnamefont
  {Kan}},\ }\href {https://kaggle.com/competitions/whats-cooking} {\enquote
  {\bibinfo {title} {What's cooking?}}\ } (\bibinfo {year} {2015})\BibitemShut
  {NoStop}%
\bibitem [{\citenamefont {{Scott Chacon}}(2009)}]{Scott2009GitHub}%
  \BibitemOpen
  \bibfield  {author} {\bibinfo {author} {\bibnamefont {{Scott Chacon}}},\
  }\href@noop {} {\enquote {\bibinfo {title} {The 2009 github contest},}\
  }\bibinfo {howpublished}
  {\url{https://github.com/blog/466-the-2009-github-contest}} (\bibinfo {year}
  {2009}),\ \bibinfo {note} {[Online; accessed June-2023]}\BibitemShut
  {NoStop}%
\bibitem [{\citenamefont {Amburg}, \citenamefont {Veldt},\ and\ \citenamefont
  {Benson}(2020)}]{amburg2020clustering}%
  \BibitemOpen
  \bibfield  {author} {\bibinfo {author} {\bibfnamefont {I.}~\bibnamefont
  {Amburg}}, \bibinfo {author} {\bibfnamefont {N.}~\bibnamefont {Veldt}}, \
  and\ \bibinfo {author} {\bibfnamefont {A.}~\bibnamefont {Benson}},\
  }\bibfield  {title} {\enquote {\bibinfo {title} {Clustering in graphs and
  hypergraphs with categorical edge labels},}\ }in\ \href@noop {} {\emph
  {\bibinfo {booktitle} {Proceedings of The Web Conference 2020}}}\ (\bibinfo
  {year} {2020})\ pp.\ \bibinfo {pages} {706--717}\BibitemShut {NoStop}%
\bibitem [{\citenamefont {Ward}(2002)}]{ward2002moby}%
  \BibitemOpen
  \bibfield  {author} {\bibinfo {author} {\bibfnamefont {G.}~\bibnamefont
  {Ward}},\ }\bibfield  {title} {\enquote {\bibinfo {title} {Moby thesaurus
  ii},}\ }\href@noop {} {\bibfield  {journal} {\bibinfo  {journal} {Project
  Gutenberg Literary Archive Foundation}\ } (\bibinfo {year}
  {2002})}\BibitemShut {NoStop}%
\bibitem [{\citenamefont {Karp}\ \emph {et~al.}(2019)\citenamefont {Karp},
  \citenamefont {Billington}, \citenamefont {Caspi}, \citenamefont {Fulcher},
  \citenamefont {Latendresse}, \citenamefont {Kothari}, \citenamefont
  {Keseler}, \citenamefont {Krummenacker}, \citenamefont {Midford},\ and\
  \citenamefont {Ong}}]{karp2019biocyc}%
  \BibitemOpen
  \bibfield  {author} {\bibinfo {author} {\bibfnamefont {P.~D.}\ \bibnamefont
  {Karp}}, \bibinfo {author} {\bibfnamefont {R.}~\bibnamefont {Billington}},
  \bibinfo {author} {\bibfnamefont {R.}~\bibnamefont {Caspi}}, \bibinfo
  {author} {\bibfnamefont {C.~A.}\ \bibnamefont {Fulcher}}, \bibinfo {author}
  {\bibfnamefont {M.}~\bibnamefont {Latendresse}}, \bibinfo {author}
  {\bibfnamefont {A.}~\bibnamefont {Kothari}}, \bibinfo {author} {\bibfnamefont
  {I.~M.}\ \bibnamefont {Keseler}}, \bibinfo {author} {\bibfnamefont
  {M.}~\bibnamefont {Krummenacker}}, \bibinfo {author} {\bibfnamefont {P.~E.}\
  \bibnamefont {Midford}}, \ and\ \bibinfo {author} {\bibfnamefont
  {Q.}~\bibnamefont {Ong}},\ }\bibfield  {title} {\enquote {\bibinfo {title}
  {The biocyc collection of microbial genomes and metabolic pathways},}\
  }\href@noop {} {\bibfield  {journal} {\bibinfo  {journal} {Briefings in
  bioinformatics}\ }\textbf {\bibinfo {volume} {20}},\ \bibinfo {pages}
  {1085--1093} (\bibinfo {year} {2019})}\BibitemShut {NoStop}%
\end{thebibliography}%

\end{document}